\documentclass[preprint,nofootinbib,
preprintnumbers,
amsmath,amssymb]{revtex4}

\usepackage{graphicx,epsfig}
\usepackage{simplewick}
\usepackage{slashed}
\usepackage{color}
\usepackage{setspace}
\usepackage{hyperref}
\usepackage{empheq}

\def\simg{{\ \lower-1.2pt\vbox{\hbox{\rlap{$>$}\lower6pt\vbox{\hbox{$\sim$}}}}\ }}
\def\siml{{\ \lower-1.2pt\vbox{\hbox{\rlap{$<$}\lower6pt\vbox{\hbox{$\sim$}}}}\ }}

\newcommand{\eq}[1]{Eq.~\eqref{#1}}
\newcommand{\eqs}[2]{Eqs.~\eqref{#1} and \eqref{#2}}

\newcommand{\Sec}[1]{Sec.~\ref{#1}}

   \graphicspath{{./figs/}}
\usepackage{color}

\usepackage{relsize}
\newcommand{\babar}{{\mbox{\slshape B\kern-0.1em{\smaller A}\kern-0.1em
            B\kern-0.1em{\smaller A\kern-0.2em R}}}
\def\MSbar{\relax\ifmmode\overline                        %%%%%%%%%
            {\rm MS}\else{$\overline{\rm MS}${ }}\fi}     %%%%%%%%%
           }                                              %%%%%%%%%
\def\MSbar{\relax\ifmmode\overline                        %%%%%%%%%
            {\rm MS}\else{$\overline{\rm MS}${ }}\fi}     %%%%%%%%%
\def\1{\hbox{{1}\kern-.25em\hbox{l}}}
 \date{\today}
\def\be{\begin{equation}}
\def\ee{\end{equation}}
\def\bea{\begin{eqnarray}}
\def\eea{\end{eqnarray}}
\def\bear{\begin{array}}
\def\eear{\end{array}}

\def\als{\alpha_{s}}
\def\al{\alpha}
\def\nn{\nonumber}

\newcommand{\MS}{\overline{\rm MS}}
\newcommand{\RS}{\rm RS}

\def\lQ{\Lambda_{\rm QCD}}

\begin{document}
\title{Superasymptotic and hyperasymptotic approximation to the operator product expansion}

\author{Cesar Ayala}
\affiliation{Department of Physics, Universidad T{\'e}cnica Federico
Santa Mar{\'\i}a (UTFSM),  
Casilla 110-V, 
Valpara{\'\i}so, Chile}
\author{Xabier Lobregat}
\author{Antonio Pineda}
\affiliation{Grup de F\'\i sica Te\`orica, Dept. F\'\i sica and IFAE-BIST, Universitat Aut\`onoma de Barcelona,\\ 
E-08193 Bellaterra (Barcelona), Spain}

\date{\today}

\begin{abstract}
Given an observable and its operator product expansion (OPE), we present expressions that carefully disentangle truncated sums of the perturbative series in powers of $\al$ from the non-perturbative (NP) corrections. This splitting is done with NP power accuracy. Analytic control of the splitting is achieved and the organization of the different terms is done along an super/hyper-asymptotic expansion. As a test we apply the methods to the static potential in the large $\beta_0$ approximation. We see the superasymptotic and hyperasymptotic structure of the observable in full glory. 
\end{abstract}

\maketitle
\tableofcontents

\vfill
\newpage

\section{Introduction}

Non-perturbative (NP) effects are dominant for QCD phenomena with characteristic energy of 
${\cal O}(\lQ)$. Consequently, the absence of analytic tools for dealing with NP effects in QCD makes impossible to produce quantitative semi-analytic predictions in terms of $\lQ$ and renormalized quark masses for most low energy observables.

On the other hand, there are observables for which their perturbative expansions in powers of $\alpha$ are reasonable approximations. 
This typically happens when there is a large scale, generically referred as $Q$ ($\gg \lQ$), in the process. In principle, 
it is then possible to perform perturbative calculations up to any
finite order in $\alpha$. Nevertheless, such perturbative expansions are expected to be asymptotic and divergent. 
Such divergent behavior is not arbitrary. Besides the perturbative series in powers of $\alpha$, one also expects the observable to depend on,
non-analytic, NP, functions of order $e^{-A\frac{2\pi}{\beta_0\al(Q)}} \sim (\lQ/Q)^A$. These NP effects and the perturbative series in powers of $\alpha$ are not 
independent of each other. Indeed the former determines the late-term behavior of the later. Leaving aside instantons, that we will neglect in what follows (as they yield smaller NP corrections than those we consider in this paper), such relation 
can be quantified using the operator product expansion (OPE) of the observable for large $Q$. The allowed operators determine the allowed corrections in powers of $\lQ$ (up to logarithms), and, therefore, the large order behavior of the perturbative expansion, since the latter can be related with singularities in the Borel plane (located in the positive real axis), which mix with the NP corrections. To these singularities (and the associated asymptotic perturbative expansion) we generically refer to as infrared renormalons \cite{tHooft:1977xjm}. 

On a more general scenario one can consider more than one large scale: $Q_1 \gg Q_2 \gg \lQ$. Then the use of the OPE and the factorization between the different scales makes the perturbative expansions associated with each scale to be asymptotic. In some cases one has renormalon singularities associated with the scales $Q_1$ and $Q_2$ that cancel among themselves. This is indeed the case for the leading renormalon singularity of the pole mass and the static potential, as first found in \cite{Pineda:1998id}, and later in \cite{Hoang:1998nz,Beneke:1998rk}. We name these renormalon singularities spurious. 

So, in general, we want to:
\begin{enumerate}
\item
Predict observables with $e^{-A\frac{2 \pi }{\beta_0 \alpha(Q)}}$ precision.
\item
Avoid spurious renormalon problems.
\end{enumerate}
In this paper we focus on 1), though our results will be relevant for 2) too. 

Besides its intrinsic theoretical interest, the asymptotic behavior of perturbative expansions in QCD is starting to be seen in a series of observables, in particular, in heavy quark physics. In this case, in order to handle the renormalon problem associated with the pole mass, different threshold masses have been introduced \cite{Bigi:1994em,Beneke:1998rk,Pineda:2001zq,Lee:2003hh,Hoang:2009yr,Brambilla:2017hcq}. Some of these threshold masses introduce (explicitly or implicitly) a scale $\nu_f$ that acts as an infrared cutoff. Such infrared cutoff kills the renormalon behavior of the perturbative series producing a convergent perturbative series and introducing a linear power-like dependence in $\nu_f$. In practice these threshold masses work quite well. The error associated to the fact that we have this linear cutoff is typically small (see, for instance, \cite{Ayala:2014yxa,Ayala:2016sdn,Peset:2018ria}). Still, it is not optimal conceptually\footnote{In the same way that there is nothing conceptually wrong in using cutoff regularization in perturbative computations, but regularizations that kill spurious power-like divergences, like dimensional regularization, and preserve more symmetries are much more convenient.}. Other of these  threshold masses use approximate expressions for the Borel transform of the pole mass that partially incorporate the renormalon singularities in the Borel plane. The inverse of the Borel transform (which we will name Borel sum or Borel integral in the following) is then ill defined. This requires using some prescription to regulate the Borel integral. In this last case the perturbative series is typically abandoned and one directly works with the Borel integral expression. In this approach it is not quantified what is the error made by using (the unavoidably) approximated expressions for the Borel transform. 

This discussion leads us to consider an alternative method that is also often used to tame the asymptotic behavior of the perturbative series: truncating the perturbative sum at the minimal term. In mathematical literature, such approximation is often named the {\it superasymptotic} approximation of the original function (see \cite{BerryandHowls}), which is a name we will also use in the following. This procedure has long since been used (see \cite{Dingle},
 or \cite{LeGuillou:1990nq}, for references), mainly in the context of solutions to one-dimensional differential equations. Nevertheless, in that context, renormalons do not show up, nor it does the issue of scheme/scale dependence. 
 
In the context of four dimensional quantum gauge field theories, truncation of the perturbative sum in different formulations or using approximated expressions for the Borel integrals has also been considered since the early days of OPE/renormalon analyses to determine observables with NP accuracy (see for instance, \cite{DiGiacomo:1981lcx,LeGuillou:1990nq,Mueller:1993pa,Neubert:1994vb,Altarelli:1994vz,Ball:1995ni,Martinelli:1996pk,Broadhurst:2000yc}). However, it was not possible to make quantitative analyses beyond the large-$\beta_0$ approximation, since the existing perturbative series were only known to low orders. More recently, perturbative expansions have been obtained to high enough orders for some observables in the lattice scheme \cite{Bauer:2011ws,Bali:2013pla,Bali:2013qla,Bali:2014fea}. This has allowed us to quantitatively use perturbative sums truncated at the minimal term and successfully determine the gluon condensate and $\bar \Lambda$ in the quenched approximation \cite{Bali:2014sja}. This success motivates us to try to improve this approach, and to revisit with it observables already computed in the $\MS$ scheme, even if only few coefficients are known, since in the $\MS$ scheme (and in particular in heavy quark physics) renormalon dominance shows up at relatively low orders. 
 
Whereas, by construction, the superasymptotic approximation does not explicitly introduce the factorization scale $\nu_f$, the dependence on the renormalization scale $\nu$ remains to be assessed. Therefore, to push this method forward we need to get a quantitative understanding of the error on the truncation of the sum and of its remaining scheme and scale dependence. Similarly, 
the NP power corrections are potentially dependent on how the divergent perturbative series is regulated and on the renormalization scheme/scale used to define the strong coupling: $\al_X(\mu)$. A major point of this paper is to be able to control (in an analytic way) the dependence of the power corrections in this generalized scheme dependence. 
We will only then be able to add NP power corrections to the pertubative series in a systematic way, since the mixing between the perturbative series and the leading NP terms (or between the perturbative series associated to the scales $Q_1$ and $Q_2$) makes impossible to determine them independently. An unambiguous definition of the NP power corrections requires defining the perturbative series with power accuracy. Such combined expansion of perturbative series and NP terms will be called {\it hyperasymptotic} expansion as in \cite{BerryandHowls}. Organizing the computation in this way allows us to precisely state the parametric accuracy of the result at each step.

The mixing between perturbative and NP 
effects may hinder estimating the {\it real} size of the
NP effects. This happens when using threshold masses. In this case the problem is not severe. A more extreme example of this problem appears in lattice regularization. The gluon condensate is, up to a factor, the expectation value of the plaquette: 
\be
\langle G^2 \rangle_{\rm latt} =\frac{36}{\pi^2}C_G^{-1}(\al)
\frac{1}{a^4}\langle P \rangle_{MC}(\al) \simeq \frac{36}{\pi^2}\frac{1}{a^4}p_0 \al
,\ee 
where $p_0=4\pi/3$. For $\beta=3/(2\pi\al)=6.65$ we have $ \langle G^2 \rangle_{\rm latt} \sim 3.3 \times 10^4 \;  r_0^{-4}$, whereas the NP gluon condensate is $\sim 3.2 \; r_0^{-4}$ \cite{Bali:2014sja}. We see that the perturbative contribution overwhelms the NP contribution by orders of magnitude. Therefore, it is convenient to devise schemes where one has extracted as much information as
possible from perturbation theory in such a way that the remaining
NP object has a minimal mixing with perturbation theory. This
scheme would provide a natural place to estimate the real size of the NP corrections without the distortion due to perturbative
effects. We believe that in this scheme one could
get a better understanding of the real structure (size) of the NP effects. This could be important, once the precision increases and to set a standard for the future. 

It is also our aim to relate the hyperasymptotic expansion with the previously mentioned methods used to handle the pole mass renormalon. This will allow us to parametrically quantify the error those methods have, in particular those using approximate expressions for the Borel transform. 

Finally, it is also worth mentioning that truncating the perturbative series at the minimal term can be motivated in the context of factorization of scales and effective field theories, where one wants to factor out the physics associated to $Q$ from the physics associated with $\lQ$. The point is that in $n$-loop diagrams, new scales are effectively generated. These scales are proportional to $Q$, but are modulated by small factors $\sim e^{-\frac{n}{k}}$, where $k$ is an integer. The dominant contribution to the $n$-loop diagram is not then due to $Q$ but to $Qe^{-\frac{n}{k_{\rm min}}}$, 
where $k_{\rm min}$ is the smallest possible $k$ for the process at hand. For the case of the pole mass $k_{\rm min}=1$ and $Q=m$. In this case, for small $n$, we still have that $m e^{-n} \gg \lQ$. Nevertheless, for $n \sim \frac{2 \pi}{\beta_0  \alpha}$, we have $me^{-n} \sim \lQ$. Doing perturbation theory for 
$ n \simg \frac{2 \pi}{\beta_0  \alpha}$ would simply mean treating $m e^{-n}$ as much bigger than $\lQ$, which is incorrect.  

The structure of the paper will be as follows. In Sec. \ref{Sec:Scheme} we discuss the general case when there are no ultraviolet renormalons. In Sec. \ref{Sec:Vs} we discuss the QCD static potential in the large $\beta_0$ approximation. We use this quantity as a toy-model NP observable to test our methods. The inclusion of ultraviolet renormalons and real QCD examples will be discussed in followup papers. 

\section{Determination of power corrections and summation scheme dependence}
\label{Sec:Scheme}

The generic form of the OPE of a dimensionless observable is the following:
\be
{\rm Observable}(\frac{Q}{\lQ})=S(\alpha_X(Q))+\sum_d C_{O,d}(\alpha_X(Q)) \frac{\langle O_d \rangle}{Q^d}
\,.
\ee
For observables that live in the Euclidean (like the Adler function or the plaquette), $O_d$  generically represents a local operator, but not necessarily so if the OPE is applied to EFTs in the Minkowski (as it could be the case for the $B$ meson mass). In any case, the expectation values $\langle O_d \rangle$, are of order $\lQ^d$ (up to some anomalous dimension). On the other hand 
$S(\alpha_X(Q))$ can be computed as a Taylor expansion in powers of $\alpha_X(Q)$. This series is assumed to be asymptotic. 
Up to some anomalous dimension, $C_{O,d}(\alpha_X(Q))$ can also be computed as a Taylor expansion in powers of 
$\al_X(Q)$ and the generated series is also assumed to be asymptotic (we will not explicitly ellaborate much on this fact though, since this leads us to consider subleading corrections in the OPE expansion, which can be handled in an analogous way). Then, the observable is often represented in the following way:
\bea
\label{Observable}
&&
{\rm Observable}(\frac{Q}{\lQ})=\sum_{n=0}^{\infty}p^{(X)}_n \alpha_X^{n+1}(Q)+\left(K+\sum_{n=0}^{\infty}p^{(X,d)}_n \alpha_X^{n+1}(Q)\right)\alpha_X^{\gamma}(Q)\frac{\Lambda_X^{d}}{Q^d}+\cdots
\\
&&
\qquad
=
\sum_{n=0}^{\infty}p^{(X)}_n \alpha_X^{n+1}(Q)+\left(K'+\sum_{n=0}^{\infty}p^{`(X,d)}_n \alpha_X^{n+1}(Q)\right)\alpha_X^{\gamma-db}(Q)e^{-d\frac{2\pi}{\beta_0\alpha_X(Q)}}+\cdots
\nn
\\
\nn
&&
\qquad
=
\sum_{n=0}^{\infty}p^{(X)}_n(\frac{\mu}{Q}) \alpha_X^{n+1}(\mu)+
\left(K'+\sum_{n=0}^{\infty}p^{`(X,d)}_n(\frac{\mu}{Q}) \alpha_X^{n+1}(\mu)\right)\alpha_X^{\gamma-db}(\mu)\frac{\mu^d}{Q^d}e^{-d\frac{2\pi}{\beta_0\alpha_X(\mu)}}+\cdots
\,,
\eea
where the dots stand for terms suppressed by higher powers of $\lQ/Q$, $\beta_0=\frac{11}{3}C_A-\frac{4}{3}T_F N_f$, $\gamma$ is the anomalous dimension of the operator $O_d$, and
\be
\label{eq:betafun}
\Lambda_X=\mu\exp\left\{-\left[\frac{2\pi}{\beta_0\alpha_X(\mu)}
+b
\ln\left(\frac12 \frac{\beta_0\alpha_X(\mu)}{2\pi}\right)
+\sum_{j\geq 1}
s^{(X)}_j\,(-b)^j\!\left(\frac{\beta_0\alpha_X(\mu)}{2\pi}\right)^{\!j}\right]\right\}
\,,
\ee
with
\begin{equation}
b=\frac{\beta_1}{2\beta_0^2}\,,\quad
s_1^{(X)}=\frac{\beta_1^2-\beta_0\beta^{(X)}_2}{4b\beta_0^4}\,,\quad
s_2^{(X)}=\frac{\beta_1^3-2\beta_0\beta_1\beta^{(X)}_2+\beta_0^2\beta^{(X)}_3}{16b^2\beta_0^6}\,,
\end{equation}
and so on. Obviously, the three equalities in \eq{Observable} are symbolic representations of the observable, as the perturbative series are asymptotic.  We need to define first the perturbative series  with $\lQ$ power accuracy. This definition of the perturbative series will, in turn, define unambiguously the NP power correction. In realistic cases, the only information that we will have of the OPE of the observable will be:
\begin{enumerate}
\item
The exact knowledge of the coefficients $p_n$ up to $n=N$, where $N \gg 1$ is large enough such that $p_n$ is well approximated by its asymptotic behavior\footnote{Otherwise the perturbative expression is not accurate enough (in principle) to be sensitive to NP corrections and it does not make much sense the consideration of NP power corrections, which is the aim of this paper.};
\item
The knowledge of the structure of the leading NP power corrections: values of $d$, $\gamma$ and the very first few terms of $p^{(d)}_n$;
\item
The knowledge of the asymptotic behavior of $p_n$ (which relies on the previous item and demanding consistency to the OPE):
\be
p^{(as)}_n(\frac{\mu}{Q})=Z^{X}_{O_d}\frac{\mu^d}{Q^d} \frac{\Gamma(1+db-\gamma+n)}{\Gamma(1+db-\gamma)}\left(\frac{\beta_0}{2\pi d}\right)^n\left[
1+{\cal O}\left(\frac{1}{n}\right)\right]
\,;
\ee
\item
The knowledge of the $\mu$ dependence of $p_n$ and $p^{`(d)}_n$ dictated by the renormalization group invariance. In realistic cases, only to some order.
\end{enumerate}
Therefore, we will devise definition methods that only use this information. This naturally leads us to consider perturbative series truncated at the minimal term $N^*$ (or close by):
\be
\label{Nop}
(N^*+bd-\gamma)\frac{\beta_0 \alpha_X(\mu)}{2\pi d}= e^{-\frac{1}{2(N^*+bd-\gamma)}+{\cal O}(\frac{1}{N^{*2}})} \rightarrow 
N^*=\frac{d 2 \pi}{\beta_0 \al_X(\mu)}-\frac{1}{2}-db+\gamma+{\cal O}(\al_X(\mu))
\,.
\ee
Note that $N^*$ depends on $\mu$ and on the renormalization scheme $X$ used to define the strong coupling constant: $\alpha_X(\mu)$. 

Therefore, we define
\be
S_{T}(Q)
\equiv S_{(X;N;\mu)}(Q)\equiv\sum_{n=0}^{N}p^{(X)}_n(\frac{\mu}{Q}) \alpha_X^{n+1}(\mu)
\,.
\ee
After truncating, $S_T(Q)$ depends on small variations of $N$ around $N^*$, on $\mu$, and on the scheme $X$ (in this paper we will consider perturbative expansions either in the lattice or in the $\MS$ scheme but the expressions are valid for general renormalization schemes). Overall, we generically label all the summation scheme dependence by $T$. The ambiguity (freedom) of the truncated perturbative series is ``of-the-order'' of the power correction. By this we mean that small variations in $N$ around $N^*$ are of ${\cal O}(e^{-d\frac{2\pi}{\beta_0\alpha_X(\mu)}})$. We also assume that the truncated sum $S_T$ for $N \sim N^*$ to be asymptotic to the full result in the following way
\be
\label{Difference1}
{\rm Observable}(\frac{Q}{\lQ})-S_{T}(Q)={\cal O}(e^{-d\frac{2\pi}{\beta_0\alpha_X(\mu)}})
\,,
\ee
where $d$ is the dimension of the leading NP term of the OPE. 

As the observable is summation scheme independent, the $T$-scheme dependence of $S_T(Q)$ should cancel with the scheme dependence of the NP power corrections. We would like to determine \eq{Difference1} with higher precision. As we have mentioned, one can get right the dimension $d$ of the NP power correction by approaching $N$ to $N^*$. It is more complicated to fix the overall coefficient (and its structure in powers of $\al$ and $\ln \al$) that modulates the NP power correction. This will heavily depend on the freedom in truncating the perturbative series. It also needs some extra information in the relation between 
the perturbative series and the observable. 

In order to quantify this difference, we first search for generalized summation schemes of the perturbative sum that are $T$-scheme independent, i.e. that they are independent of $\mu$, $X$ and $N$. The Borel integral of the Borel transform is a natural candidate. In our case the inverse of the Borel transform needs regularization, as it has singularities in the real axis at positive values of the integration variable. Here we take the principal value (PV) prescription of the perturbative expansion:
\be
S_{\rm PV}(Q) \equiv \int_{0,\rm PV}^{\infty}dt e^{-t/\alpha_X(\mu)} B[S](t)
\,,
\ee
where one takes the arithmetic average of the integral above and below the real axis and 
\be
B[S](t)=\sum_{n=0}^{\infty}\frac{p^{(X)}_n(\frac{\mu}{Q})}{n!}t^n
\,.
\ee
For values of $t$ larger than the radius of convergence of this series, we take the analytic continuation of this function. For instance (this function will be useful later on), 
\bea
\label{Ieq}
I (db)&\equiv& \int_{0,\rm PV}^{\infty} dt e^{-t/\al} \frac{1}{(1-2u/d)^{1+db-\gamma}}=
\al   D_{db-\gamma}(-(2\pi d)/(\beta_0 \alpha))
\\
&\sim& 
\sum_{n=0}^{\infty}\frac{\Gamma(1+db-\gamma+n)}{\Gamma(1+db-\gamma)}\left(\frac{\beta_0}{2\pi d}\right)^n\alpha^{n+1}(\mu)
\,,
\eea
where $u\equiv \beta_0 t/(4\pi)$. For this and related equations we collect a useful set of equalities in Appendix \ref{Sec:Db}.
 
Now, our first task is to show that $S_{\rm PV}$ is indeed $T$-scheme independent. 

In the large $\beta_0$ approximation the PV Borel integral can be shown to be factorization-scale and scheme independent (actually in the large $\beta_0$ approximation both things are the same) \cite{Chyla:1990na}. Beyond the large $\beta_0$ approximation things are more complicated. Nevertheless, we can still show the factorization and scheme independence of $S_{\rm PV}$ under some assumptions. We first consider the renormalization scale dependence. We restrict the discussion to the inclusion of $\beta_1$ to the running of $\al$. Then, renormalization scale independence gives the following relation between coefficients \cite{Chyla:1990na}:
\be
\mu\frac{d}{d\mu}p_0=0;\qquad
\mu\frac{d}{d\mu}p_1=\frac{\beta_0}{2\pi}p_0;
\qquad
\mu\frac{d}{d\mu} p_k
=\frac{\beta_0}{2\pi}kp_{k-1}+\frac{\beta_1}{8\pi^2}(k-1)p_{k-2};\qquad k\geq2
\,.
\ee
Using these relations we can deduce that 
\be
\label{totalderivative}
\mu\frac{d }{d\mu} S_{\rm PV}=
-\alpha
\frac{\beta_1}{8\pi^2}\int_{0,\rm PV}^{\infty}\,du\frac{d}{du}\sum_{j=0}^{\infty}\left(\frac{4\pi}{\beta_0}\right)^{j+2}\frac{1}{j+2}\frac{1}{j!}p_{j}(\tau) e^{-4\pi u/(\beta_0\alpha_X(\mu))}u^{j+2}
\,.
\ee
This is a total derivative and vanishes. It is possible to include $\beta_2$ to the running of $\alpha$. Renormalization scale independence of the perturbative series now gives the following relation between the coefficients of the perturbative expansion
\be
\mu\frac{d }{d\mu}p_k=\frac{\beta_0}{2\pi}kp_{k-1}+\frac{\beta_1}{8\pi^2}(k-1)p_{k-2}+\frac{\beta_2}{32\pi^3}(k-2)p_{k-3}\,,
\qquad k\geq3
\,.
\ee
Though much lengthier expressions show up, it is still possible to deduce that
\be
\mu\frac{d }{d\mu}S_{\rm PV} 
\propto A_1\beta_1 \int_{0,\rm PV}^{\infty}\,du\frac{d}{du}g_1(u)+A_2\beta_2 \int_{0,\rm PV}^{\infty}\,du\frac{d}{du}g_2(u)
\,.
\ee
These are total derivatives. 
The behavior of $g_i(u)$ for small $u$ is $g_i(u) \sim u^{\alpha}$ 
with $ \alpha >0$. For large $u$, $g_i(u) \sim e^{-u/\alpha}h(u)$, where $h(u)$ does not grow exponentially. Therefore, $g(0)=0$ and $g(\infty)=0$, proving the renormalization scale independence of $S_{\rm PV}$. The inclusion of higher order terms seems to produce also total derivatives that vanish. Note that our conclusion disagrees with \cite{Chyla:1990na}. 

We now turn to the scheme dependence. Given the perturbative series in a given scheme:
\be
\label{Series1}
\sum_{k=0}^{\infty}p_k\alpha_X^{k+1}
\,,
\ee
we consider a general change of scheme (but regular enough, such that, for instance, do not introduce spurious singularities in the Borel plane):
\be
\label{ChangeScheme}
\alpha_X=\alpha_{X'}+d_1\alpha_{X'}^2+d_2\alpha_{X'}^3+d_3\alpha_{X'}^4+\cdots
\,.
\ee
The independence on the coefficients $d_i$ of \eq{Series1} 
produces the following relation
\be
\sum_{k=0}^{\infty}\bigg[\left(\frac{d}{dd_i}p_k\right)\alpha_{X}^{k+1}+p_k(k+1)\alpha_{X}^{k}\frac{d}{dd_i}\alpha_{X}\bigg]=0
\,.
\ee
Note also that 
\be 
\frac{d}{dd_i}\alpha_{X}=\alpha_{X}^{i+1}\left(1+{\cal O}(\alpha_{X})\right)
\,.
\ee
Overall we get
\be
\frac{d}{dd_1}p_k=-k p_{k-1}+2d_1(k-1)p_{k-2}-(5d_1^2-2d_2)(k-2)p_{k-3}+\cdots
\ee
\be
\frac{d}{dd_2}p_k=-(k-1)p_{k-2}+3d_1(k-2)p_{k-3}-(9d_1^2-3d_2)(k-3)p_{k-4}+\cdots
\,.
\ee
For simplification one could work in schemes that make higher order terms (the dots) to vanish. Then, making a similar computation to the one we did to get the scale dependence, we get that the PV integral does not change under these variations, as we get total derivatives, which vanish:
\be
\frac{d}{dd_1} S_{\rm PV}=0 \qquad \frac{d}{dd_2} S_{\rm PV}=0
\,.
 \ee
Therefore, $S_{\rm PV}$ is $T$-scheme independent. 

We do not enter in this paper into global definitions of the Borel integral of the observable itself, which may not exist \cite{tHooft:1977xjm}. For the purposes of this paper, it is enough that we can define the Borel transform of the perturbative series and its Borel sum (with the PV prescription). We then assume that difference between the Borel sum regulated using the PV prescription and the complete NP result obtained from full QCD can be absorbed in the NP terms of the OPE. An analytic proof (of disproof) of that is tantamount to given a NP proof of the OPE in QCD, which is, at present, beyond reach. Since we assume that such generalized resummation scheme preserves the structure of the NP OPE, the difference with the observable has to exactly scale as the NP corrections of the OPE:
\be
{\rm Observable}(\frac{Q}{\lQ})
=
S_{\rm PV}(\al(Q))+K_X^{\rm (PV)}\al_X^{\gamma}(Q)\frac{\Lambda_X^{d}}{Q^d}\left(1+{\cal O}(\al_X(Q))\right)
+{\cal O}(\frac{\Lambda_X^{d'}}{Q^{d'}})
\,,
\label{Obtruncated}
\ee
where the last term refers to higher order terms in the OPE ($d' > d$). $K_X^{\rm (PV)}$ is independent of $\mu$ and $Q$. We also demand $K_X^{\rm (PV)}$ to transform as $\Lambda_X^{-d}$ under changes of scheme of the strong coupling, $\al_X$, i.e. the combination $K_X^{\rm PV}\Lambda_X^{d}$ is scheme independent. Indeed, since the structure of the NP OPE should be preserved, alternative generalized summation schemes should be different from the $S_{\rm PV}$ by a term exactly proportional to the $\mu$ and scheme independent quantity
\be
\propto K_X^{\rm (PV)}\al_X^{\gamma}(Q)\frac{\Lambda_X^{d}}{Q^d}\left(1+{\cal O}(\al_X(Q))\right)
\,.
\ee 
Note also that the exponent $\gamma$ and the ${\cal O}(\al_X(Q))$ can be determined by RG analyses. In some cases RG analysis says that there is no ${\cal O}(\al_X(Q))$ corrections (the Wilson coefficient is identically 1). This indeed would be the case of the B-meson mass. 

$S_{\rm PV}$ has the handicap, though, that it needs the full analytic structure of the Borel transform in the Borel plane, i.e. it requires the knowledge of the perturbative series to all orders. This can make them unpractical\footnote{In the resolution of one-dimensional differential equations this cannot be much of a problem, since it is possible to compute perturbation series to very high orders, and one has good analytic control on the NP corrections, as they can be evaluated via instantons. Nevertheless, this is much of an issue for us where in realistic scenarios we will only have approximated evaluations of the leading singularity in the Borel plane.}. Remarkably enough, however, this problem can be bypassed by relating $S_{\rm PV}$ with truncated versions of the perturbative series. 
This is the strategy we follow: devising truncated sums that we can relate with the PV result. This allows us to control the scheme dependence and error of using $S_T$. Quite remarkably, this approach also allows us to quantify the error of using approximate expressions for $S_{\rm PV}$, since we do not know the complete perturbative series. 

For fixed $\mu$, the $N \rightarrow \infty$ limit of $S_T(Q)$ diverges, since the perturbative series is divergent. Therefore, if we want to keep $\mu$ finite, we have to keep $N$ finite as well. Alternatively, if we want to take $N \rightarrow \infty$, then we should as well send $\mu \rightarrow \infty$. Therefore, we explore two possibilities. One is to take $\mu \sim Q$ in $N \sim N^*$, the other is to take $\mu \rightarrow \infty$ (correlated with $N \sim N^* \rightarrow \infty$):
\begin{enumerate}
\item[{1)}]
$N$ and $\mu \sim Q$ large but finite:
\be
\label{eq:NP}
N=N_P(\alpha)\equiv d\frac{ 2\pi}{\beta_0\alpha_X(\mu)}\big(1-c\,\alpha_X(\mu)\big)
\,,
\ee
\item[{2)}]
$N \rightarrow \infty$ and $\mu \rightarrow \infty$ in a correlated way. We consider two options:
\be
\label{eq:muinfty}
{\rm A)} \quad N+1=N_S(\alpha)\equiv d\frac{ 2\pi}{\beta_0\alpha_X(\mu)}
 \,; \quad 
{\rm B)} \quad N=N_A(\alpha) \equiv d\frac{ 2\pi}{\beta_0\alpha_X(\mu)}\big(1-c'\alpha_X(Q)\big)
,
\ee
\end{enumerate}
where $c'>0$ but $c$ is arbitrary otherwise. Note that in case 1), $c$ can partially simulate changes on the scale and scheme 
of $\alpha_X$.

We will study case 1) and 2) in the following two subsections.

\subsection{$N$ large and $\mu \sim Q \gg \lQ$. \eq{eq:NP}. Case 1)}
 
We first study option 1). Now the truncated sum reads ($N_P$ is defined in \eq{eq:NP})
\be
S_{P}(Q)
\equiv\sum_{n=0}^{N_P}p^{(X)}_n(\frac{\mu}{Q}) \alpha_X^{n+1}(\mu)
\,.
\ee
We want to estimate what is the leading contribution to the difference between the PV sum and its truncated sum. This difference is dominated by the leading renormalon. Therefore, we focus on the contribution associated with it:
\bea
\delta S_{\rm PV}&=& Z^X_{O_d}\frac{\mu^d}{Q^d} [I(db)+b_1 I(db-1) +\cdots]
\\
\nn
&=& Z^X_{O_d}\frac{\mu^d}{Q^d}\sum_{n=0}^{N}\frac{\Gamma(1+db-\gamma+n)}{\Gamma(1+db-\gamma)}
\bigg[
1+b_1\frac{db-\gamma}{db-\gamma+n}
\\
\nn
&&
+b_2\frac{(db-\gamma)^2}{(n+db-\gamma)(n+db-\gamma-1)}+\cdots\bigg]
\left(\frac{\beta_0}{2\pi d}\right)^n\alpha_X^{n+1}(\mu)
+\Omega
\,,
\eea
where $I$ is defined in \eq{Ieq}. The finite sum stands for the contribution to $S_{P}(Q)$ associated with the leading renormalon. $\Omega$ is the terminant \cite{Dingle} of the asymptotic series when we truncate at $\alpha^{N+1}$:
\be
\Omega=\Delta\Omega(db)+b_1\Delta\Omega(db-1)+w_2\Delta\Omega(db-2)+\cdots
\ee
where
\be
w_2=\frac{b_2(db-\gamma)}{db-\gamma-1}
\ee
and $\Delta \Omega$ admits the following integral (but not a Borel integral) representation
\be
\label{DeltaOmega}
\Delta\Omega(db)\equiv Z^X_{O_d}
\frac{\mu^d}{Q^d}\frac{1}{\Gamma(1+db-\gamma)}\left(\frac{\beta_0}{2\pi d}\right)^{N+1}\alpha_X^{N+2}(\mu)\int_{0,\rm PV}^{\infty}dx\frac{x^{db-\gamma+N+1}e^{-x}}{1-x\frac{\beta_0\alpha_X(\mu)}{2\pi d}}
\,.
\ee
With these definitions $\Omega$ has the desired asymptotic expansion:
\bea
\Omega&\sim& Z^X_{O_d}\frac{\mu^d}{Q^d}\sum_{n=N+1}^{\infty}\frac{\Gamma(1+db-\gamma+n)}{\Gamma(1+db-\gamma)}
\bigg[
	1+b_1\frac{db-\gamma}{db-\gamma+n}
\\
\nn
&&
	+b_2\frac{(db-\gamma)^2}{(n+db-\gamma)(n+db-\gamma-1)}+\cdots\bigg]
\left(\frac{\beta_0}{2\pi d}\right)^n\alpha_X^{n+1}(\mu)
\,.
\eea
Even if \eq{DeltaOmega} is not in a Borel integral form, this integral is amenable for a saddle approximation analysis (still, note also that we can evaluate it numerically exactly). 
We consider the integral
\be
H=\int_{0,\rm PV}^{\infty}dx\frac{x^{db-\gamma+N+1}e^{-x}}{1-x\frac{\beta_0\alpha_X(\mu)}{2\pi d}}
=
\Gamma(db-\gamma+N+1)D_{db-\gamma+N+1}\left(\frac{2\pi d}{\beta_0 \al_X(\mu)}\right)
\,,
\ee
where $D_b(x)$ is defined in Appendix \ref{Sec:Db}. 
Setting (to avoid considering non-integer values of $N$, for a given value of $\mu$ we will restrict to values of $c$ that ensures that $N_P$ is integer)
\be
N=N_P=\frac{2\pi d}{\beta_0\alpha_X(\mu)}-\frac{2\pi dc}{\beta_0}
\,,
\ee
 the integral $H$ has the following expansion (this result is obtained by explicit computation and checked with an alternative computation using the recursion relations one can find in \cite{Dingle})
\bea
\nn
	H&=&-\left(\frac{2\pi d}{\beta_0\alpha_X(\mu)}\right)^{2+bd-\gamma+\frac{2\pi d}{\beta_0\alpha_X(\mu)}-\frac{2\pi dc}{\beta_0}}e^{\frac{-2\pi d}{\beta_0\alpha_X(\mu)}}\alpha_X^{1/2}(\mu)\bigg\{\frac{\beta_0^{1/2}}{d^{1/2}}\bigg[-\eta_c+\frac{1}{3}\bigg]
\\
\nn
&&
	+\alpha_X(\mu)\frac{\beta_0^{3/2}}{\pi d^{3/2}}\bigg[-\frac{1}{12}\eta_c^3+\frac{1}{24} \eta_c-\frac{1}{1080}\bigg]+\alpha_X^2(\mu)\frac{\beta_0^{5/2}}{\pi^2d^{5/2}}\bigg[-\frac{1}{160}\eta_c^5-\frac{1}{96}\eta_c^4
\\
\nn
&&
+\frac{1}{144}\eta_c^3+\frac{1}{96}\eta_c^2-\frac{1}{640}\eta_c-\frac{25}{24192}\bigg]+\mathcal{O}\left(\alpha_X^3(\mu)\right)\bigg\}
\,,
\eea
where $\eta_c\equiv-bd+\frac{2\pi d}{\beta_0}c+\gamma-1$. Thus
\bea
\nn
\Delta\Omega(bd)&=&-\frac{Z^X_{O_d}\mu^d}{\Gamma(1+bd-\gamma)Q^d}\bigg(\frac{2\pi d}{\beta_0}\bigg)^{bd-\gamma+1}e^{\frac{-2\pi d}{\beta_0\alpha_X(\mu)}}\alpha_X^{1/2-bd+\gamma}(\mu)\bigg\{\frac{\beta_0^{1/2}}{d^{1/2}}\bigg[-\eta_c+\frac{1}{3}\bigg]
\\
\nn
&&
+\alpha_X(\mu)\frac{\beta_0^{3/2}}{\pi d^{3/2}}\bigg[-\frac{1}{12}\eta_c^3+\frac{1}{24} \eta_c-\frac{1}{1080}\bigg]+\alpha_X^2(\mu)\frac{\beta_0^{5/2}}{\pi^2d^{5/2}}\bigg[-\frac{1}{160}\eta_c^5-\frac{1}{96}\eta_c^4
\\
\nn
&&
+\frac{1}{144}\eta_c^3+\frac{1}{96}\eta_c^2-\frac{1}{640}\eta_c-\frac{25}{24192}\bigg]+\mathcal{O}\left(\alpha_X^3(\mu)\right)\bigg\}
\,,
\eea
and $\Omega$ reads
\begin{equation}
\label{Omegaexp}
\Omega=\sqrt{\al_X(\mu)}K_X^{(P)}\frac{\mu^d}{Q^d}e^{-\frac{d2\pi}{\beta_0 \alpha_X(\mu)}}
\left(\frac{\beta_0\alpha_X(\mu)}{4\pi}\right)^{-db}
\alpha_X^{\gamma}(\mu)\bigg(1+\bar K_{X,1}^{(P)}\al_X(\mu)+\bar K_{X,2}^{(P)}\al_X^2(\mu)+O(\alpha_X^3(\mu))\bigg)
\,,
\end{equation}
or in terms of $\lQ$,
\be
\label{OmegaLambda}
\Omega=\sqrt{\al_X(\mu)}K_X^{(P)}\frac{\Lambda_X^d}{Q^d}\alpha_X^{\gamma}(\mu)\bigg(1+K_{X,1}^{(P)}\al_X(\mu)+K_{X,2}^{(P)}\al_X^2(\mu)+O(\alpha_X^3(\mu))\bigg)
\,,
\ee
where
\bea
K_X^{(P)}&=&\frac{-Z^X_{O_d}}{\Gamma(1+bd-\gamma)}\left(\frac{2\pi d}{\beta_0}\right)^{bd-\gamma+1}\left(\frac{\beta_0}{4\pi}\right)^{bd}\left(\frac{\beta_0}{d}\right)^{1/2}\bigg[-\eta_c+\frac{1}{3}\bigg]
\\
%\ee
%\bea
\bar K_{X,1}^{(P)}&=&\frac{\beta_0/(\pi d)}{-\eta_c+\frac{1}{3}}\bigg[-b_1 \left(bd-\gamma \right) \left(\frac{1}{2}\eta_c+\frac{1}{3}\right)
-\frac{1}{12}\eta_c^3+\frac{1}{24}\eta_c-\frac{1}{1080}\bigg]
\\
%\eea
%\be
K_{X,1}^{(P)}&=&\bar K_{X,1}^{(P)}-\frac{b \beta_0 d s_1}{2\pi}
\\
%\ee
%\bea
\bar K_{X,2}^{(P)}&=&\frac{\beta_0^2/(\pi d)^2}{-\eta_c+\frac{1}{3}}
\bigg[-w_2 (b d-\gamma -1) (b d-\gamma ) \left(\frac{1}{4}\eta_c+\frac{5}{12}\right)
\\
\nn
&&
+b_1(bd-\gamma)\left(-\frac{1}{24}\eta_c^3-\frac{1}{8}\eta_c^2
-\frac{5}{48}\eta_c-\frac{23}{1080}\right)-\frac{1}{160}\eta_c^5
\\
\nn
&&
-\frac{1}{96}\eta_c^4+\frac{1}{144}\eta_c^3
+\frac{1}{96}\eta_c^2-\frac{1}{640}\eta_c-\frac{25}{24192}\bigg]
\\
%\eea
%\be
K_{X,2}^{(P)}&=&\frac{1}{8\pi^2}\big(8\pi^2\bar K_{X,2}^{(P)}-4bd\pi s_1\beta_0\bar K_{X,1}^{(P)}+b^2d^2s_1^2\beta_0^2+2b^2ds_2\beta_0^2\big)
\,.
\eea

Let us note that \eq{OmegaLambda} also has a factor $\al^{\gamma}(\mu)$ besides the prefactor $\sqrt{\al(\mu)}$. In this paper we will only consider situations where $\gamma=0$.  To properly account for this factor one has to perform a resummation of $\ln (\mu/Q)$ terms that effectively transform $\al^{\gamma}(\mu)$ into $\al^{\gamma}(Q)$ in \eq{OmegaLambda}. For one example of a case with $\gamma\not=0$ where this is done, see, for instance, \cite{Campanario:2005np}.

In the large $\beta_0$ it is possible to write $\Omega$ in a Borel integral form. It reads 
\be
\Omega=Z^X_{O_d}\frac{\mu^d}{Q^d}\frac{1}{\Gamma(1-\gamma)}\bigg(\frac{4\pi}{\beta_0}\bigg)^{-\gamma+1}\alpha_X^{\gamma}(\mu)\bigg(\frac{2}{d}\bigg)^{N+1}\int_{0,\rm PV}^{\infty}du\,e^{\frac{-4\pi u}{\beta_0\alpha_X(\mu)}}\frac{u^{-\gamma+N+1}}{1-\frac{2u}{d}}
\,.
\ee
After integration we obtain ($\eta_c^{(\beta_0)}\equiv \frac{2\pi d}{\beta_0}c+\gamma-1$)
\bea
\nn
\Omega&=&-\frac{Z^X_{O_d}\Lambda_X^d}{\Gamma(1-\gamma)Q^d}\bigg(\frac{2\pi d}{\beta_0}\bigg)^{-\gamma+1}\alpha_X^{1/2+\gamma}(\mu)\bigg\{\frac{\beta_0^{1/2}}{d^{1/2}}\bigg[-\eta_c^{(\beta_0)}+\frac{1}{3}\bigg]
\\
\nn
&&
+\alpha_X(\mu)\frac{\beta_0^{3/2}}{\pi d^{3/2}}\bigg[-\frac{1}{12}\eta_c^{(\beta_0)3}+\frac{1}{24} \eta_c^{(\beta_0)}-\frac{1}{1080}\bigg]+\alpha_X^2(\mu)\frac{\beta_0^{5/2}}{\pi^2d^{5/2}}\bigg[-\frac{1}{160}\eta_c^{(\beta_0)5}
\\
\nn
&&
-\frac{1}{96}\eta_c^{(\beta_0)4}+\frac{1}{144}\eta_c^{(\beta_0)3}+\frac{1}{96}\eta_c^{(\beta_0)2}
-\frac{1}{640}\eta_c^{(\beta_0)}-\frac{25}{24192}\bigg]+\mathcal{O}\left(\alpha_X^3(\mu)\right)\bigg\}
\,.
\eea
Obviously this result coincides with the full result when setting $b_1=\beta_1=\cdots=0$. 

Subleading NP renormalons give subleading power corrections. A function with a finite radius of convergence in the $\alpha$ plane yields a Borel transform that is an analytic function in the whole complex $u$ plane. Such function generates corrections smaller than any NP correction (i.e. of order $(K/N)^N \sim e^{-K \frac{2\pi}{\beta_0\alpha_X(\mu)}\ln(\frac{2\pi}{\beta_0\alpha_X(\mu)})}$)

Since $\Omega$ gives the leading NP correction to $S_{\rm PV}$ we can write
\be
S_{\rm PV}=S_P+\Omega+ \cdots
\,.
\ee
Overall we obtain
\begin{subequations}
\label{PVP1}
\begin{empheq}[box=\fbox]{align*}
S_{\rm PV}(Q) & =S_{P}(Q;\mu)+\sqrt{\al_X(\mu)}K_X^{(P)}\frac{\Lambda_X^d}{Q^d}\alpha_X^{\gamma}(\mu)
\\
&\times\bigg(1+K_{X,1}^{(P)}\al_X(\mu)+K_{X,2}^{(P)}\al_X^2(\mu)+O(\alpha_X^3(\mu))\bigg)+{\rm NP \; power \; corrections}
\tag{46}
\end{empheq}
\end{subequations}
or
\begin{subequations}
\label{PVP2}
\begin{empheq}[box=\fbox]{align*}
S_{\rm PV}(Q) & =S_{P}(Q;\mu)+\sqrt{\al_X(\mu)}K_X^{(P)}\frac{\mu^d}{Q^d}e^{-\frac{d2\pi}{\beta_0 \alpha_X(\mu)}}
\left(\frac{\beta_0\alpha_X(\mu)}{4\pi}\right)^{-db}
\alpha_X^{\gamma}(\mu)
\\
&\times\bigg(1+\bar K_{X,1}^{(P)}\al_X(\mu)+\bar K_{X,2}^{(P)}\al_X^2(\mu)+O(\alpha_X^3(\mu))\bigg)+{\rm NP \; power \; corrections}
\tag{47}
\end{empheq}
\end{subequations}
Note that with this method we do not expect a bad behavior when we take $c \rightarrow 0$: The result is smooth, unlike what will happen with method 2B). Remarkable enough, this 
result quantifies the error of determinations of NP corrections obtained by truncating the sum at (or around) the minimal term, which is of ${\cal O}(\sqrt{\al(\mu)}\lQ^d)$ irrespective of the scale and scheme (in particular this applies to the analysis in \cite{Bali:2014sja}). We now can do better, as we now can compute these subleading terms that before went into the error. Therefore, we can increase the precision with which the genuine NP term can be determined. 

If the precision of the computation is high enough one may consider going beyond the leading power accuracy and include the first correction to the above equations in the hyperasymptotic expansion. It would read
\be
\label{SPV2final}
S_{\rm PV}(Q)=S_P(Q;\mu)+\Omega(\mu)+\sum_{n=N_P+1}^{N'_P}(p_n-p_n^{(as)})\al_X^{n+1}(\mu)+\Omega'(\mu)+\cdots
\,,\ee
where $N'_P$ stands for the power in $\al$ where the perturbative series will mix with the subleading renormalon and $\Omega'$ can be easily deduced from \eq{Omegaexp} adapting dimension and anomalous dimension to the next renormalon .  

The truncated sum $S_P(Q;\mu)$ depends on $\mu$ but not $S_{\rm PV}$. This has the important consequence that we can determine the $\mu$ dependence of $S_P(Q;\mu)$ with 
$\lQ$ power accuracy, and also to control the scheme dependence. We obtain
\bea
\nn
\mu \frac{d}{d \mu} S_P(Q;\mu)&=&-K_X^{(P)}\frac{\Lambda_X^d}{Q^d}\alpha_X^{\frac{3}{2}+\gamma}(\mu)\bigg(-\frac{\beta_0}{4\pi}(1+2\gamma)+\alpha_X(\mu)\frac{1}{16\pi^2}\big(-12\pi\beta_0K_{X,1}^{(P)}
\\
\nn
&&
-8\pi\beta_0\gamma K_{X,1}^{(P)}-\beta_1-2\beta_1\gamma\big)+\alpha_X^2(\mu)\frac{1}{64\pi^3}\big(-2 \beta_2 \gamma -\beta_2-8 \pi  \beta_1 \gamma  K_{X,1}^{(P)}
\\
\nn
&&
-12 \pi  \beta_1 K_{X,1}^{(P)}-32 \pi ^2 \beta_0 \gamma  K_{X,2}^{(P)}-80 \pi ^2 \beta_0 K_{X,2}^{(P)}\big)+\mathcal{O}\big(\alpha_X^3(\mu)\big)\bigg)
\\
&&
-\mu \frac{d}{d \mu}\sum_{n=N_P+1}^{N'_P}(p_n-p_n^{(as)})\al_X^{n+1}(\mu)+\cdots
\,.
\eea

We will typically take $\mu=k Q$, where $k$ is a constant of order 1 to avoid large factors. 
Note also that the Taylor expansion in powers of $\al$ of the last term in \eq{SPV2final} starts at $n=N_P+1$. This effectively transform this term in a NP power correction. Moreover, the fact that the leading renormalon is subtracted from the perturbative series expansions further suppress this contribution. A naive estimate can be obtained by saturating the coefficients by the next renormalon. For the case of the static potential, the next renormalon is located at $u=3/2$. This produces that the series roughly scales as 
\be
\label{estimate}
\sim \left(\frac{1}{3}\right)^{\frac{2\pi}{\beta_0 \al_X(\mu)}}
e^{-\frac{2\pi}{\beta_0 \al_X(\mu)}}=e^{-\frac{2\pi}{\beta_0 \al_X(\mu)}\left(1+\ln(3)\right)}\,,
\ee
which is obviously subleading, but still more important than the next NP correction. We will visualize the size of the different terms of the hyperasymptotic expansion in more detail in Sec. \ref{Sec:V1} for the case of the static potential in the large 
$\beta_0$ approximation.

The correction associated with an analytic function in the whole complex Borel plane (of order 
$\al^N \sim e^{-\#N\ln(N)}$) is smaller than any NP correction (of order 
$ e^{-\#'N}$, where $\#'$ is finite and bigger the further away the renormalon singularity is from the origin). 
Still, one can also worry about the role played by the logs generated in the perturbative computation: $\ln(\mu /Q)$ . Assuming they are large, the leading contribution to the order $\alpha^N$ is of ${\cal O}( \alpha^N \ln^N(\mu/Q))$.
Since it is still $1/N!$ suppressed compared with the renormalon contributions, it can
 be written as $e^{-\#N\ln(N/(\mu/Q))}$. Obviously if $k$ is made 
parametrically big it could jeopardize the hierarchy of the corrections that we have here. Therefore, we will always keep $k$ parametrically of ${\cal O}(1)$. 

\medskip

We now illustrate the above general discussion using the particular case of the heavy quark mass (we neglect ultraviolet renormalons).  We then have $m_{\rm PV}={\bar m}[1+S_P({\bar m};\mu)+\Omega_m(\mu)+\cdots]$, where ${\bar m} \equiv m_{\MS}(m_{\MS})$, we set $d=1$, and also set the Wilson coefficient of the nonperturbative correction to 1 \cite{Beneke:1994rs} in $S_P$ and $\Omega_m$.

We now compare our analysis with existing threshold masses. 
We focus on the RS mass \cite{Pineda:2001zq} and relatives\footnote{Conceptually they are equivalent to the kinetic 
\cite{Bigi:1994em} or PS mass \cite{Beneke:1998rk}, as they have an explicit cut-off as well. These other schemes are different at low orders but they share the same asymptotic behavior.}. The RS mass is defined in the following way:
\be
m_{\RS}(\nu_f)=m_{\rm OS}-\delta m_{\RS}^{(0)}\equiv {\bar m}+\sum_{n=0}^{N}r_n^{\RS}(\mu;\nu_f)\al^{n+1}(\mu) 
\,,
\ee
where
\be
m_{\rm OS}={\bar m}+\sum_{n=0}^{N}r_n(\mu)\al^{n+1}(\mu) 
\,,
\ee
and (in \cite{Pineda:2001zq} $Z^X_m$ was named $N_m$)
\be
\delta m_{\RS}^{(n)}=\sum_{s=n}^N r_s^{(\rm as)}(\nu_f)\als^{s+1}(\nu_f)\,, \qquad r_s^{(\rm as)}(\nu_f)=Z^X_m\,\nu_f\,\left({\beta_0
\over 2\pi}\right )^s \,\sum_{k=0}^\infty
c_k{\Gamma(s+1+b-k) \over \Gamma(1+b-k)}
\,,
\ee
where one typically takes $N=N_{max}\equiv$ the maximal number of coefficients of the perturbative expansion that are known exactly (we assume that $N_{max}$ is not that high that we have to worry about subleading renormalon). 
 In order to lessen the $\nu_f$ scale dependence, the RS'$\equiv$ RS$^{(1)}$ was also defined:
 \be
 m_{\RS'}(\nu_f)=m_{\rm OS}-\delta m_{\RS'}={\bar m}+r_0\al(\mu)+\sum_{n=1}^{N}r_n^{\RS'}(\mu;\nu_f)\al^{n+1}(\mu) 
 .
 \ee 
 It is obvious that one could generalize to RS$^{(n)}$ where the subtraction starts at order $\al^{n+1}$:
 \be
 m_{\RS^{(n)}}(\nu_f)=m_{\rm OS}-\delta m_{\RS^{(n)}}={\bar m}+\sum_{s=0}^nr_s(\mu)\al^{s+1}(\mu)
+\sum_{s=n}^{N}r_s^{\RS^{(n)}}(\mu;\nu_f)\al^{s+1}(\mu) 
.
 \ee 
Nevertheless, we can not increase $n$ arbitrarily, otherwise the renormalon is not canceled. Moreover, the value of $n$ for which there is no cancellation of the renormalon will depend on $\mu$. Therefore, when including higher orders one should do it with care once approaching to the minimal term. Another issue is the $\nu_f$ dependence. To connect with the approach used in this paper we should take $\nu_f=\mu$. Note that then $r_s^{\RS^{(n)}}(\mu)=r_s(\mu)-r_s^{(as)}(\mu)$. In the original applications of the RS schemes this could be a problem, since the natural scale in the pole mass is different from the natural scale in the static potential\footnote{If the scales are widely separated, this problem could be overcome using the resummation of logarithms of $\nu_f$, as first worked out in \cite{Bali:2003jq}.}. To connect with the approach used in this paper, we control the scale dependence by fixing $n=N=N_P(\mu)$. This smoothly connect the RS schemes with the schemes where the series is truncated at the minimal term. One can then add $\Omega_m$ and higher orders terms in the hyperasymptotic expansion of $m_{\rm PV}$.

We now consider the threshold mass named $m_{\rm BR}$, defined in \cite{Lee:2003hh} (see also \cite{Lee:2002sn}). The author directly works with the Borel transform and then regulate the Borel integral using the PV prescription. 
The complete expression of the Borel transform is not known. Therefore, in practice, an approximated expression is used that agrees with the known terms of the pole mass perturbative expansion till $N=N_{max}=2$ (the known coefficients at that time) and incorporates the leading singularity in the Borel plane. The author also makes a conformal mapping of the Borel transform.  The $\mu$ dependence of $m_{\rm BR}$ was usually fixed to $\mu=m$, except in \cite{Lee:2005hf}. To make a quantitative comparison with our analysis, we leave aside the conformal mapping and make explicit the $\mu$ scale dependence in $m_{\rm BR}$.  The key point then is the comparison of $N(=2)$ with $N_P$. If $N< N_P(\mu)$ there is power-like $\mu$ dependence that gets uncancelled with the contribution of $S_P$. In other words
\be
{\bar m}\left[\frac{m_{\rm BR}^{(N)}(\mu)}{{\bar m}}-1-S_P({\bar m};\mu)-\Omega(\mu)\right]=\sum_{N+1}^{N_P}r^{(as)}_n(\mu)\al^{n+1}(\mu)
\,.
\ee
Note that this produces an strong (linear) renormalization scale dependence ($r_n^{(as)} \sim \mu$) that is missed if one sets $\mu={\bar m}$. This problem is potentially more severe in top physics (see for instance \cite{Beneke:2016cbu}), since one includes orders in perturbation theory beyond those presently known if the perturbative expansion is made with $\al(m_t)$.

For $N=N_P$ we exactly have that 
\be
{\bar m}\left[\frac{m_{\rm BR}^{(N)}(\mu)}{{\bar m}}-1-S_P({\bar m};\mu)-\Omega(\mu)\right]=0
\,.
\ee
For $N>N_P$ we have
\be
{\bar m}\left[\frac{m_{\rm BR}^{(N)}(\mu)}{{\bar m}}-1-S_P({\bar m};\mu)-\Omega(\mu)\right]=\sum_{n=N_P+1}^{N}(r_n-r_n^{(as)})\al^{n+1}(\mu)
\,.
\ee
Overall, the only problematic situation would be if $N < N_P$. 
For $N \geq N_P$, $m_{\rm BR}$ and $m_{\rm PV}$ are equal within the approximation used, and our analysis reorganizes the result within a hyperasymptotic expansion. This allows us to quantitatively control the $\mu$ dependence, and to parametrically state the error, of the result (for a given truncation) with NP power accuracy using a hyperasymptotic counting. 

We can also connect our results with $m_{\rm MRS}$, defined in \cite{Brambilla:2017hcq}, in the following way (the expression of ${\cal J}$ can be found in Eq. (2.17) of \cite{Brambilla:2017hcq}).
\be
m_{\rm MRS}=m_{\RS}({\bar m})+{\cal J}({\bar m})={\bar m}+\sum_{n=0}^{N}(r_n({\bar m})-r_n^{(as)}({\bar m}))\al^{n+1}({\bar m}) +{\cal J}({\bar m})
\,.
\ee
In this definition, $\mu$ has been fixed to ${\bar m}$. By doing so we cannot estimate the error associated with the $\mu$ dependence of $m_{\rm MRS}$. Therefore, we introduce it and generalize the definition of $m_{\rm MRS}$ in the following way [we could indeed write a more general definition by putting a different scale for the renormalon term: $m_{\rm MRS}(\nu_f)=m_{\rm RS}(\nu_f)+{\cal J}(\nu_f)$. This would still achieve renormalon cancellation]: 
\be
m_{\rm MRS}(\mu)={\bar m}+\sum_{n=0}^{N}(r_n-r_n^{(as)})\al^{n+1}(\mu) +{\cal J}(\mu)
\,,
\ee
which makes explicit the $\mu$ scale dependence of the definition. In principle one could think that, since it is related with RS mass, this would make a linear dependence in $\mu$ appear. Remarkably enough this is not the case. We can relate this expression with the quantities defined above. Indeed the difference between $m^{(N)}_{\rm MRS}(\mu)$ and $m^{(N)}_{\rm BR}(\mu)$ is proportional to $\lQ$:
\be
\label{BRMRS}
m^{(N)}_{\rm BR}(\mu)- m^{(N)}_{\rm MRS}(\mu)=-\cos(\pi b)\frac{4\pi \Gamma(-b)}{2^{1+b}\beta_0}Z_m^X\Lambda_X
\,.
\ee
This quantity diverges in the large $\beta_0$ limit, which makes it not possible to take the large $\beta_0$ limit of $m^{(N)}_{\rm MRS}(\mu)$ (alternative definitions were then proposed in \cite{Brambilla:2017hcq}).  The possibility to subtract this term from the PV regulated Borel integral was also considered in \cite{Lee:2002sn}, though with a different (but related) motivation. In this respect, we note that subtracting this quantity from the PV result has been criticized in \cite{Caprini:2003tr}, on the basis of analytic properties of the observable. Nevertheless, this discussion is not directly relevant for us\footnote{It would be if we were able to relate the PV Borel integral with a NP definition of the observable.}, as adding or subtracting this term would just be equivalent to a change of resummation scheme that can be absorbed in the genuine NP power correction. Note though that this difference is parametrically bigger than $\bar m \Omega_m$, since the latter scales like ${\cal O}(\sqrt{\al}\lQ)$. In any case, since the difference with the PV result is a scale/scheme independent quantity proportional to $\lQ$, the comparison with our analysis runs in complete parallel to the previous discussion of $m_{\rm BR}^{(N)}$ with respect to $N$. Again, problems will show up if $N<N_P$, but for $N \geq N_P$, $m_{\rm MRS}$ is equal to $m_{\rm PV}$ within the accuracy of the computation, except for \eq{BRMRS}. Therefore, it can be  written in terms of a modified version of the hyperasymptotic expansion discussed in this section. 

A more extensive discussion and a quantitative analysis for the case of the top, bottom and charm quark masses will be carried out in \cite{HyperII}.

\subsection{$(N,\mu) \rightarrow \infty$. \eq{eq:muinfty}. Case 2A)}

As promising as method 1) is, it is worth it to explore alternatives that yield results that are explicitly $N$ (and therefore $\mu$) independent. They may also lead to a better analytic understanding of the observable. This can be achieved by taking $\mu$ and $N$ going to infinity in a correlated way. The simplest possibility one may consider is taking the limit as in 2A) in \eq{eq:muinfty}.
 
The case 2A) was studied in the large $\beta_0$ limit in  \cite{Sumino:2003yp,Sumino:2005cq} for the case of the static potential 
(a more general case, including subleading corrections to the running of $\al$, was also considered in \cite{Sumino:2005cq}). It was observed that $S_T$ was logarithmically divergent in $N$ and the proportionality coefficient found. Nevertheless, it was not possible to get a direct connection of this coefficient with the normalization of the leading renormalon in the Borel plane. This problem has been solved in \cite{Mishima:2016vna}, where it has been shown how to relate the coefficient of the $\ln N$ term with the normalization of the renormalon. 
This analysis has also been done for the Adler function. Unfortunately, the validity of these findings is restricted to the large $\beta_0$ approximation. 

Beyond the large $\beta_0$ approximation only the static potential has been studied \cite{Sumino:2003yp,Sumino:2005cq}. Remarkably enough the $\ln N$ (and an associated $\ln (\ln r \lQ)$) behavior survives, albeit with different coefficients. This  may point to a certain universality (beyond large $\beta_0$) of this result. Unfortunately, it is not known now how to relate such coefficient with the normalization of the renormalon. This would be very useful for analyses beyond the large $\beta_0$. 

We will discuss all this in more detail in Sec. \ref{Sec:V2A} where we study the static potential in the large $\beta_0$ approximation in this limit. 

\subsection{$(N,\mu) \rightarrow \infty$. \eq{eq:muinfty}. Case 2B)}
\label{Sec:2B}

We have seen that $S_T$ was logarithmic divergent in $N$ when taking the limit 2A). It was also not possible to connect 
$S_T$ with its Borel sum. We now consider the limit 2B). In this case one truncates before reaching the minimum, i.e. for $N<N^*=\frac{d 2 \pi}{\beta_0 \al_X(\mu)}$. This will yield a finite result. The other point we address is the relation of $S_T$ in the limit 2B) with its Borel sum.

For some specific models of sign alternating perturbative series, it was soon realized that the $N \rightarrow \infty$ limit of their associated truncated sums could be related with a modified version of the Borel integral \cite{Stevenson:1982qw,Maxwell:1983fm}, if such $N \rightarrow \infty$ limit is performed in an specific way. For instance, it was shown that 
\be
\lim_{N \rightarrow \infty} \sum_{k=0}^{N-1} p_k(\tau(N)) \alpha^{k+1}(\tau(N))=
\int_0^{\frac{4\pi}{\beta_0\chi_0}} \frac{e^{-t/\alpha}}{\pi+t} dt
\,,
\ee   
where 
\be
\tau \equiv \frac{\beta_0}{2} \ln (\mu/ \Lambda)= \frac{\pi}{\alpha} \quad 
 \tau(N)=\frac{\beta_0}{4}\chi_0N+{\cal O}(\ln N)
 \ee
 with
 \be
 \chi_0=\frac{4}{\beta_0}0.278 \qquad p_k(\tau_0)=(-1)^k\frac{1}{\pi^{k+1}} k!
 \,.
\ee

Later work generalized this result to more general series expansions, even to some that show a non-sign alternating series (but assuming that their Borel transform has a finite radius of convergence), and for arbitrary $\chi$ (as far as it satisfies some conditions). Their results can be summarized in the following equation:
\be
\lim_{N \rightarrow \infty} \sum_{k=0}^{N-1} p_k(\tau(N)) \alpha^{k+1}(\tau(N))=
\int_0^{\frac{4\pi}{\beta_0\chi}} e^{-t/\alpha(\tau_0)} \sum_{j=0}^{\infty}\frac{p_j(\tau^0)}{j!}t^jdt
\,,
\ee   
where
\be
 \frac{\alpha(\tau)}{\pi}=1/(\tau(N)+\tau_0) \quad 
 \tau(N)=\frac{\beta_0}{4}\chi N
\,,
\ee
and we require $\chi$ to be such that $\sum_{j=0}^{\infty}\frac{p_j(\tau^0)}{j!}t^j$ is analytic for $|t| < \frac{4\pi}{\beta_0\chi}$. Therefore, we can indeed sum the Borel series unambiguously inside the disc. 

This was originally proven in \cite{Chyla:1990ki,Chyla:1990na} by brute force computation. It was also proven using a different method (integration in the complex plane) in \cite{VanAcoleyen:2003gc} (in this last reference the ${\cal O}(1/\sqrt{N})$ corrections were also computed). In both cases the running of the strong coupling is restricted to follow the large $\beta_0$ approximation. 

Whereas the above result applies to arbitrary perturbative series (with the qualifications mentioned above), the running of $\alpha$ is constrained to follow the large $\beta_0$ approximation. This is an important constraint if we want to consider the case of QCD, where the perturbative expansion of the beta function is not a monomial but has more terms. One can bypass this constraint if Eq. (2) in \cite{VanAcoleyen:2003gc} is understood as a change of scheme instead of a change of a renormalization scale. It is also possible to generalize the derivation of \cite{VanAcoleyen:2003gc} for a strong coupling with a general beta function. In this generalization new $1/N$ terms are generated. Alternatively, one can slightly modify how the $\mu \rightarrow \infty$ is taken in \eq{eq:muinfty}. Instead of case 2B) one can take
\be
N_A'(\al) \equiv  d\frac{ 2\pi}{\beta_0\alpha_X(\mu)}\big(1-c'\alpha_X(Q)\big)- d\frac{ 2\pi}{\beta_0\alpha_X(Q)}\big(1-c'\alpha_X(Q)\big)
\,.
\ee 
The difference with $N_A$ vanishes when $\mu \rightarrow \infty$. With this modified scaling it is possible to show that Eq (5) in \cite{VanAcoleyen:2003gc} holds taking $k=N\chi$ with $\chi=d/(1-c'\al(Q))$. The derivation is then analogous to the derivation in \cite{VanAcoleyen:2003gc}. Overall, we are then able to obtain (taking the $\mu \rightarrow \infty$ limit according to 2B) of \eq{eq:muinfty})
\be
\label{SAST}
\lim_{\mu \rightarrow \infty; 2B)} S_T(Q) \equiv S_A(Q)\equiv \int_0^{\frac{4\pi}{\beta_0\chi}}dt e^{-t/\alpha_X(Q)} B[S](t)
\ee
beyond the large $\beta_0$ approximation, where $\frac{1}{\chi}<\frac{d}{2}$. In particular, we will take $1/\chi$ close to $d/2$, and parameterise it in the following way:
\be
\label{chi2}
\frac{1}{\chi}=\frac{d}{2}-\frac{d}{2}c'\alpha(Q)
\,,\ee
where $c' >0$ (this is the reason we took $c'>0$ in \eq{eq:muinfty}). The reason for the sign of $c'$ is that we have to approach to the closest singularity to the origin in the Borel plane from the left. Indeed, in \cite{Chyla:1990ki,Chyla:1990na}, in the context of the large $\beta_0$ approximation, it was shown that in order the integral to be well defined one needed $\frac{1}{\chi}<\frac{d}{2}$. It was also noticed that by taking the limit $\frac{1}{\chi} \rightarrow \frac{d}{2}$ the correct exponent (of the NP power correction) is obtained, i.e. the difference is of the order of the leading NP term of the OPE. Nevertheless, one does not get the right prefactor. This was quantified in \cite{VanAcoleyen:2003gc}, where it was first shown that using \eq{chi2}, and expanding in 
$\al$, the ambiguity is of the order of the higher order condensate with the right $\alpha$ dependence of the prefactor. 
 
 The leading renormalon (the singularity in the Borel plane closest to the origin) gives the main contribution to the difference between $S_{PV}$ and $S_A$:
\be
S_{\rm PV}-S_A=\int_{\frac{4\pi}{\beta_0\chi},\rm PV}^{\infty}dt\,e^{\frac{-t}{\alpha_X(Q)}}Z_X\frac{1}{(1-\frac{\beta_0}{2\pi d}t)^{1+db-\gamma}}+\cdots
\,.
\ee
This yields
\begin{subequations}
\label{PVA}
\begin{empheq}[box=\fbox]{align*}
S_{\rm PV}&=S_A+K^{(A)}_X\frac{\Lambda_X^d}{Q^d}\alpha_X^{\gamma}(Q)(1+\mathcal{O}(\alpha_X))
\\
&=S_A+K^{(A)}_X e^{-\frac{2\pi d}{\beta_0 \alpha_X(Q)}}
\left(\frac{\beta_0\alpha_X(Q)}{4\pi}\right)^{-db}
\alpha_X^{\gamma}(Q)(1+\mathcal{O}(\alpha_X))
\tag{71}
\end{empheq}
\end{subequations}
where
\be
K^{(A)}_X=\frac{2\pi d}{\beta_0}Z_X\bigg(\frac{\beta_0}{4\pi}\bigg)^{bd}
\int_{-c',\rm PV}^{\infty}dx\,e^{\frac{-2\pi dx}{\beta_0}}\frac{1}{(-x)^{1+db-\gamma}}
\,.
\ee

Subleading corrections to the leading renormalon are of the form ($1+n > 0$)
\be 
\int_{\frac{4\pi}{\beta_0\chi},\rm PV}^{\infty}dt\,e^{\frac{-t}{\alpha_X(Q)}}\frac{1}{(1-\frac{\beta_0}{2\pi d}t)^{db-\gamma-n}}
\sim 
e^{-\frac{d2\pi}{\beta_0 \alpha_X(Q)}}
\left(\frac{\beta_0\alpha_X(Q)}{4\pi}\right)^{-db}
\alpha_X^{\gamma+1+n}(Q)
\,.
\ee
This gives ${\cal O}(\al^{1+n})$ corrections. 

All subleading renormalons potentially contribute to the same order:
\be
\propto e^{-\frac{d2\pi}{\beta_0 \alpha_X(Q)}}
\alpha_X(Q)
\,.\ee 
This contribution is ${\cal O}(\al^{1+db-\gamma})$ suppressed with respect the leading term. This is a problem if one wants to obtain subleading corrections to the leading NP term, as one would need to know the normalization coefficient of all subleading renormalons. 
 
An issue observed in \cite{VanAcoleyen:2003gc}, in the context of the large $\beta_0$ approximation, was that when $1/\chi \rightarrow d/2$, i.e. when the integrand approaches the singularity of the Borel transform, the truncated PV integral diverges, and it is not a good approximation of the PV integral (for instance see Figs. 2 and 3 in \cite{VanAcoleyen:2003gc}). Therefore, it is better not to make the combination $c' \alpha(Q)$ very small. We study this problem in the example we will consider in the following section.

This observation also makes that we can not use the results obtained in this section to the case 2A) obtained in the previous section, as it means setting $\chi=2/d$, i.e. exactly at the singularity in the Borel plane. (yet it would be very interesting a dedicated study to see if the analysis of this section can be generalized to the case $\chi=2/d$). 

\subsection{Strategy}

In summary, we have two alternative expressions (\eqs{PVP1}{PVA}) to determine 
$S_{\rm PV}(Q)$ with $\lQ$ power-like precision. Remarkably enough, we can achieve such precision even though we do not know the complete perturbative series expansion. The reason is that we can relate $S_{\rm PV}(Q)$ with the truncated sum of the perturbative series for both methods.  We also obtain an analytic expression for the leading power correction that accounts for the difference between the truncated sum and the PV result. One important feature of this result is that, in both cases, the leading power correction can be determined if the strength and structure of the leading singularity in the Borel plane is known. This result is also true beyond the large $\beta_0$ approximation. Such results are scheme independent. 
 
There are important differences between both methods beyond the above general properties. The first one is that the method 2B) (the ``$\mu \rightarrow \infty$ method") yields a finite NP correction in the limit $Q \rightarrow \infty$. This is not so for the method 1) (the ``$\mu =Q$ method"). For the latter, the leading NP correction gets multiplied by the small factor $\sqrt{\al(Q)}$, which vanishes (albeit weakly) in the $Q \rightarrow \infty$ limit. In principle, this makes the second method better. Nevertheless, one should also keep in mind that, in order to profit from this property, one needs to have physical data for as large as possible $Q$.  Since in both cases the leading corrections are known analytically this could not make a practical difference. A numerical analysis can check which one is better. A more serious problem with the ``$\mu \rightarrow \infty$ method" is that, in order to take the $\mu \rightarrow \infty$ limit, one needs the running of $\alpha$ with higher and higher precision. In the large $\beta_0$ limit, the running of $\alpha$ is known exactly, so this is not a problem, but it will be once we move beyond this approximation. One also needs higher and higher order coefficients of the perturbative expansion as one takes the $\mu \rightarrow \infty$ limit. Again in the large 
$\beta_0$ limit the coefficients can be generated to any arbitrary finite order\footnote{For the static potential this is indeed so, but even for the pole mass this is numerically demanding.} but not beyond the large $\beta_0$ limit. In the real case, the most we will have is the asymptotic behavior of the high order coefficients.

Another important issue is that with the ``$\mu =Q$ method" we are potentially capable of computing corrections to the leading NP effect. The ${\cal O}(\Lambda^d\alpha)$ corrections are still related with the leading renormalon and can be computed. The effect of subleading renormalons give power suppressed corrections. For the ``$\mu  
\rightarrow \infty$ method" the ${\cal O}(\Lambda^d\alpha)$ corrections receive corrections from all subleading renormalons. In practice, this makes it impossible to compute these corrections in a controlled way. 

In general it is impossible to obtain closed results for the PV regulated perturbative sum on which to test the above results. This is only possible in the large $\beta_0$ approximation for a few cases. Here, we use one of them as a laboratory to check the methods we will apply to physical cases. 
The question here is to quantify the difference between the PV result (which we take as a ``fake" NP data), and the truncated perturbative expansions (for large values of $N$). 
Obviously such comparison is made in the short distance limit where the OPE should apply. In \Sec{Sec:Vs}, we check our formulas (in the large 
$\beta_0$ approximation) for the case of the static potential. This example will allow us to quantify (in practice) when the complete result is well approximated by \eqs{PVP1}{PVA}. In particular, we try to answer the following questions: How large $Q$ has to be in both cases\footnote{This is expected to be dependent on $n_f$. The bigger $n_f$ the smaller the renormalon effect. Therefore, any discussion with $n_f=0$ should be understood as an upper bound of the importance of renormalons.}, how large $\mu$ has to be for \eq{PVA} to hold. We also study the dependence of the answer to the scale/scheme used for the strong coupling (we use lattice and $\MS$ scheme). 

The method that leads to \eq{PVA} requires $\mu \rightarrow \infty$. Formally, this means that we need all the coefficients $p_n$. As in realistic cases we do not have this information, we 
check the dependence on approximating the exact perturbative coefficients (starting at different orders)  by their asymptotic expansions in the large $\beta_0$ approximation. In this case we will be able to see the error introduced by considering different orders from which one approximates the coefficients by the asymptotic behavior.  
What we will not be able to test in the large $\beta_0$ approximation is the dependence on the higher order coefficients of the beta function, which are needed for \eq{PVA} (since we need to run $\al(\mu)$ to $\mu=\infty$). This is relegated to subsequent work.  

Note that all the scheme dependence (in the broad sense: $T$=\{$N$, $X$, $\mu$\}) has disappeared up to terms beyond the accuracy we achieve. We also obtain expressions for the difference between different truncation schemes. 

Overall, we express the observable in the following two alternative ways
\begin{subequations}
\label{ObservableA}
\begin{empheq}[box=\fbox]{align*}
{\rm Observable}(\frac{Q}{\lQ})
&=
S_{P}(Q;\mu)+K_X^{\rm (PV)}\al_X^{\gamma}(Q)\frac{\Lambda_X^{d}}{Q^d}\left(1+{\cal O}(\al_X(Q))\right)+\Omega(\mu)
\\
&+\sum_{n=N_P+1}^{N'_P}(p_n-p_n^{(as)})\al^{n+1}(\mu)+\dots
\tag{75}
\end{empheq}
\end{subequations}
\be
\label{Observable2B}
\setlength{\fboxsep}{2mm}
\fbox{
$\displaystyle{{\rm Observable}(\frac{Q}{\lQ})
=
S_{A}(Q;\chi)+(K_X^{\rm (PV)}+K_X^{\rm (A)})\al_X^{\gamma}(Q)\frac{\Lambda_X^{d}}{Q^d}\left(1+{\cal O}(\al_X(Q))\right)+\cdots
}$}
\ee
up to exponentially suppressed terms. Note that $\Omega$ scales like ${\cal O}(\sqrt{\al_X(Q)}\frac{\Lambda_X^d}{Q^d})$. Both methods have $\lQ$ power accuracy but with the method 1) we have enough theoretical precision to determine the subleading ${\cal O}(\al_X)$ corrections or even subleading terms in the OPE (hyperasymptotic) expansion (provided the ``experimental'' data is precise enough).

\section{The static potential in the large $\beta_0$ approximation}
\label{Sec:Vs}

The large $\beta_0$ approximation cannot be obtained from a well defined limit of the parameters of QCD. 
Still, it is useful to test techniques that can be used beyond the large $\beta_0$ 
approximation in a place where we know the exact solution. In this respect the static potential is an ideal object, since we have a lot of analytic control for it.

\subsection{$V_{\rm PV}(r)$}

The QCD static potential is written in terms of its Fourier transform as 
\be
V(r)=-\frac{2C_F}{\pi}\int_0^\infty dq \frac{\sin qr}{qr}\alpha_v(q)\ .
\label{Vr}
\ee
This equation defines $\alpha_v(q)$ in the V-scheme. In the large-$\beta_0$ approximation, we know the behavior of $\alpha_v(q)$ as a series in powers of $\alpha_X\equiv\alpha_X(\mu)$
\be
\alpha_v(q)=\alpha_X\sum_{n=0}^\infty L^n=\alpha_X\frac{1}{1-L}\ ,
\label{alsV}
\ee
where $L=\frac{\beta_0\alpha_X}{2\pi}\ln(\frac{\mu e^{-c_X/2}}{q})$. If $X=\MS$ then $c_{\MS}=-5/3$ (in the large $\beta_0$ approximation). If $X=V$ then $c_V=0$.  If $X={\rm latt}$, we take the $n_f=0$ number for a Wilson action: $c_{\rm latt}=-8.38807$ \cite{Hasenfratz:1980kn}, as we will only use this scheme for checking the consistency between the results obtained with different schemes. We also define $\tilde \Lambda 
= \Lambda_X e^{-c_X/2}$ and $\rho=\tilde \Lambda r$. 
Note that $\tilde \Lambda$ is scheme independent. 
 
\eq{Vr} is ill defined but not its Borel transform. It reads \cite{Aglietti:1995tg}
\be
B[V](t(u))=B(t(u))=\frac{-C_F}{\pi^{1/2}}\frac{1}{r} e^{-c_Xu}\bigg(\frac{\mu^2r^2}{4}\bigg)^u\frac{\Gamma(1/2-u)}{\Gamma(1+u)}
\,,
\ee
which is a meromorphic function in the $u$ complex plane. 

We then define (where the single poles of the Borel transform are regulated using the PV prescription)
\be
V_{\rm PV}(r)=\int_{0,\rm PV}^{\infty} dt e^{-t/\alpha(\mu)}B[V](t(u))
\,.
\ee
We can also regulate \eq{Vr} via 
\be
V_{\rm PV}(r)=-\frac{2C_F}{\pi}\int_{0,\rm PV}^\infty dq \frac{\sin qr}{qr}\alpha_v(q)\ .
\ee
We have checked that the numerical determinations of both definitions give the same. We can then use this PV prescription as a NP definition of the observable, to which to test our methods and approximations. Note that this definition is indeed scheme independent. On the other hand the result is an oscillating function of $r$, which violates general properties of the static potential (energy) of two static sources in the fundamental representation \cite{Bachas:1985xs}. These state that the potential should be concave (we should also keep in mind that we are working in the large $\beta_0$ limit, which is not a well-defined limit of QCD). 

We now consider the short distance limit ($r \rightarrow 0$) of $V_{\rm PV}(r)$. In other words, we analyze its OPE. 
First, we study how well we can approximate $V_{\rm PV}(r)$ by its perturbative expansion at weak coupling. Thus, we 
approximate the potential by the truncated perturbative sum:
\be
\label{VNb0}
V_N\equiv \sum_{n=0}^N V_n \al^{n+1}
\,.
\ee 
For fixed $\mu$, the $N \rightarrow \infty$ limit of $V_N$ diverges since the perturbative expansion is asymptotic. 
Therefore, we have to be careful in the definition used for the truncated sum. For such object, we use the two definitions discussed in Sec. \ref{Sec:Scheme} (with $Q=1/r$). For both of them we will need the normalization of the leading renormalon. In the large $\beta_0$ it reads
\be
Z_V=-2\frac{C_F}{\pi}e^{-\frac{c_X}{2}}
\,.
\ee
It agrees with the result from the pole mass \cite{Beneke:1994sw} after using that the renormalon of the pole mass cancels with the renormalon of the static potential \cite{Pineda:1998id}. 

We will perform computations with $n_f=0$ and $n_f=3$. In the first case we will work in lattice units (aiming to compare with quenched lattice simulations) and use $\Lambda_{\MS}(n_f=0)=0.602 r_0^{-1} \approx 238$ MeV \cite{Capitani:1998mq}. 
In the large $\beta_0$ approximation (with $n_f=0$), this yields $\al(M_{\tau}) \approx 0.29$.  
In the second case we take $\Lambda_{\MS}(n_f=3)=174$ MeV. This last number we fix such that it gives a reasonable value at the $\tau$ mass in the large $\beta_0$ approximation: $\al(M_{\tau}) \approx 0.3$ (see for instance \cite{Boito:2018yvl}). 

We then confront $V_{\rm PV}$ with the results obtained with these methods. 

\subsection{$N$ large and $\mu \sim 1/r \gg \lQ$. \eq{eq:NP}. Case 1)}
\label{Sec:V1}

We truncate at $N=N_P$ ($N_P$ is defined in \eq{eq:NP}) in Eq. (\ref{VNb0}).
\be
V_P\equiv \sum_{n=0}^{N_P} V_n \al^{n+1}
\,.
\ee
Applying \eq{SPV2final} to the static potential in the large $\beta_0$ approximation, the relation between $V_{\rm PV}$ and $V_{P}$ reads
\be
\label{Vb0PV1}
V_{\rm PV}=V_{P}+\frac{1}{r}\Omega_V+\sum_{n=N_P+1}^{3N_P} (V_n-V_n^{(\rm as)}) \al^{n+1}
+\frac{1}{r}\Omega_V'+o(\lQ^3 r^2)
\,,
\ee
where $\Omega_V$ reads for this case
\be
\label{eq:OmegaV}
\Omega_V=
\sqrt{\al_X(\mu)}K_X^{(P)} r\ \Lambda_X\left(1+K_{X,1}^{(P)}\al_X(\mu)+{\cal O}(\alpha_X^2)\right)
\,,
\ee
with
\be
K_X^{(P)}=\frac{4C_Fe^{-c_X/2}(-\frac{6\pi c}{\beta_0}+4)}{3\beta_0^{1/2}}
\,,
\qquad
%\ee
%\be
K_{X,1}^{(P)}=\frac{\beta_0(-\frac{2\pi c}{\beta_0}+1)^3+\frac{\beta_0}{2} (\frac{2\pi c}{\beta_0}-1)-\frac{\beta_0}{90}}{4\pi(-\frac{6\pi c}{\beta_0}+4)}
\,,
\ee
and so on. Note that in the large $\beta_0$ we identically have $\Lambda_X=\mu e^{-2\pi/(\beta_0 \alpha_X(\mu))}$. This makes that $K_{X,i}^{(P)}=\bar K_{X,i}^{(P)}$. A similar expression applies to $\Omega'_V \sim \sqrt{\al_X(\mu)}(r\lQ)^3$.

By incorporating the last two terms in \eq{Vb0PV1} we are sensitive to the next renormalon. Note that subleading renormalons give $\lQ$ power-suppressed corrections. The further away the singularity in the Borel plane, the more suppressed the correction is. For the next-to-leading singularity we have
\be
\delta V \sim  \int_{0,\rm PV}^{\infty} du e^{-\frac{4\pi}{\beta_0 \alpha(1/r)}}
\frac{(\frac{2}{3}u)^N}{1-\frac{2}{3}u} 
\sim
Z^{(3/2,X)}_V \Lambda_X e^{-4\pi/(\beta_0 \alpha_X(1/r))}
\,.
\ee

\begin{center}
\begin{figure}
\includegraphics[width=0.765\textwidth]{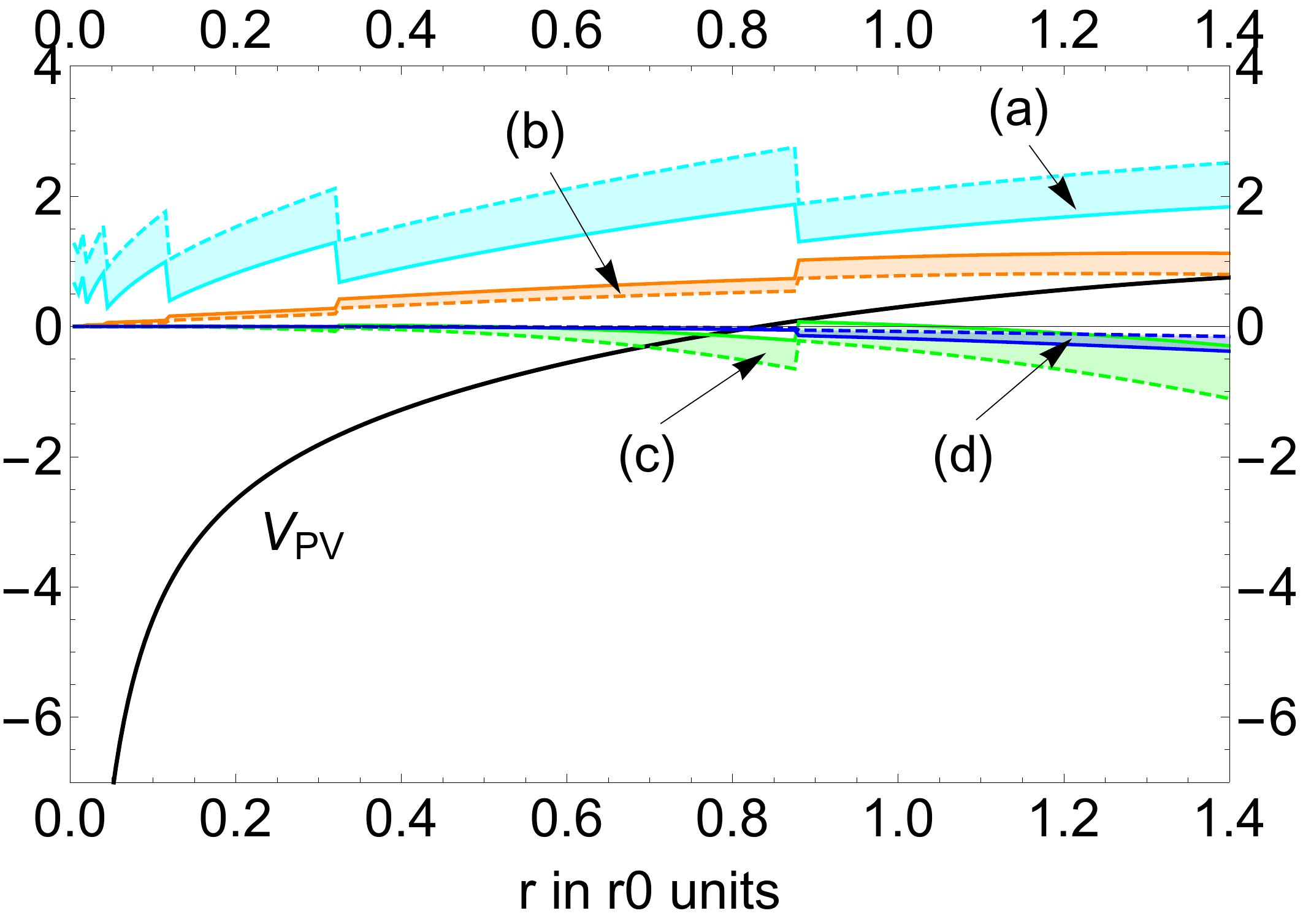}
%\includegraphics[width=0.76\textwidth]{PlottingSmallestPositivecAndNegativeHyper.png}
%\put(0,189){{\large (a)}}
%\put(0,139){{\large (d)}}
%\put(0,125){{\large (c)}}
%\put(0,165){{\large (b)}}
\vspace{0.1in}
\includegraphics[width=0.822\textwidth]{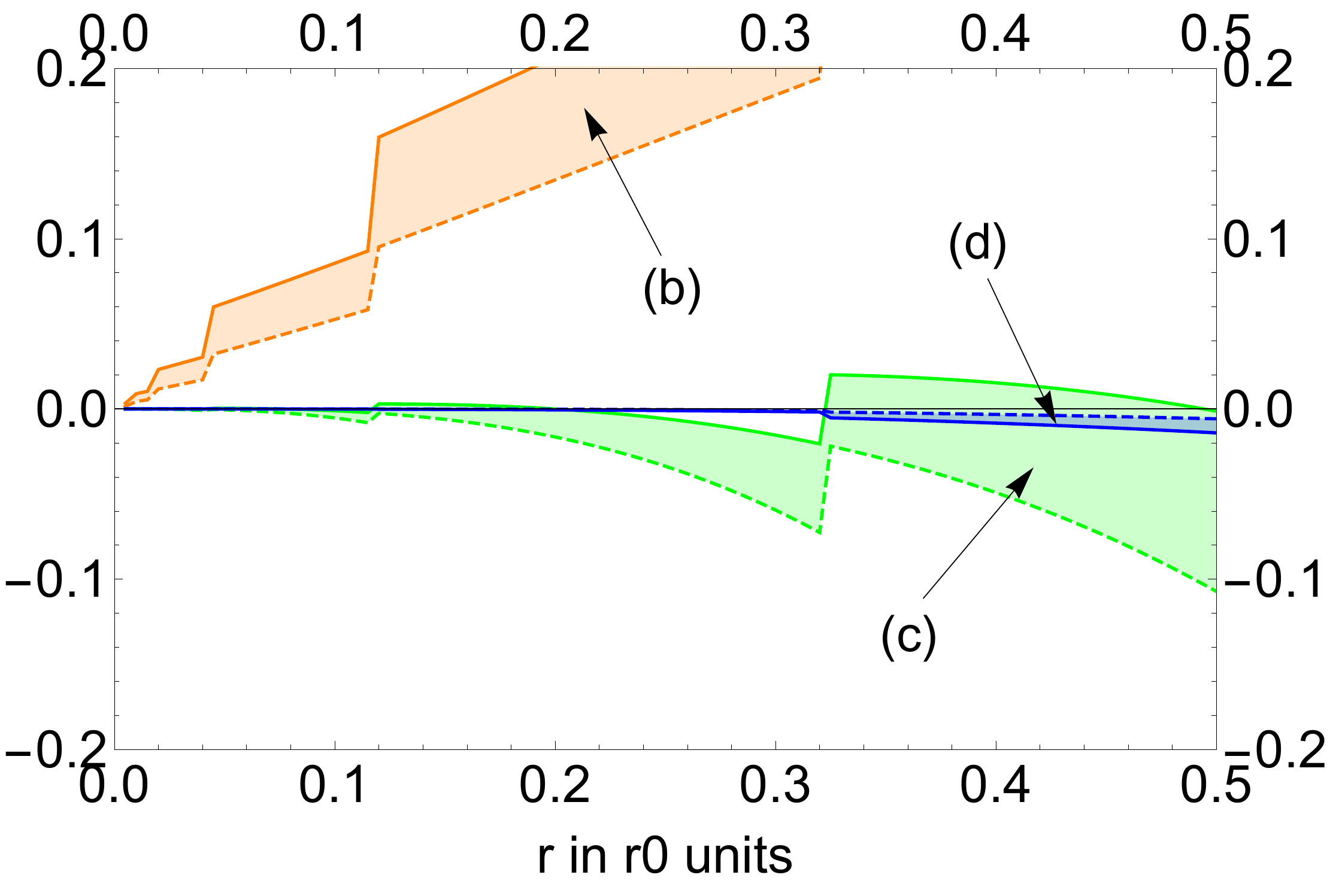}
%\includegraphics[width=0.76\textwidth]{PlottingSmallestPositivecAndNegativeHyperZomed.png}
%\put(-140,189){{\large (b)}}
%\put(0,105){{\large (d)}}
%\put(0,75){{\large (c)}}
\caption{{\bf Upper panel}: We plot $V_{\rm PV}$ (black line) and the differences: (a) $V_{\rm PV}-V_P$ (cyan), (b) $V_{\rm PV}-V_P-\frac{1}{r}\Omega_V$ (orange),
 (c) $V_{\rm PV}-V_P-\frac{1}{r}\Omega_V-\sum_{n=N_P+1}^{3N_P} (V_n-V_n^{(\rm as)}) \al^{n+1}$ (green),
and  (d) $V_{\rm PV}-V_P-\frac{1}{r}\Omega_V-\sum_{n=N_P+1}^{3N_P} (V_n-V_n^{(\rm as)}) \al^{n+1}-\frac{1}{r}\Omega'_V$ (blue)
in the lattice scheme with $n_f=0$ light flavours. For each difference, 
the bands are generated by the difference of the prediction produced by the smallest positive or negative possible values of $c$ that yields integer values for $N_P$. {\bf Lower panel}: As in the upper panel but in a smaller range. $r_0^{-1} \approx 400$ MeV. \label{Fig:VPVb0nf0latt}}
\end{figure}
\end{center}
\begin{center}
\begin{figure}
\includegraphics[width=0.765\textwidth]{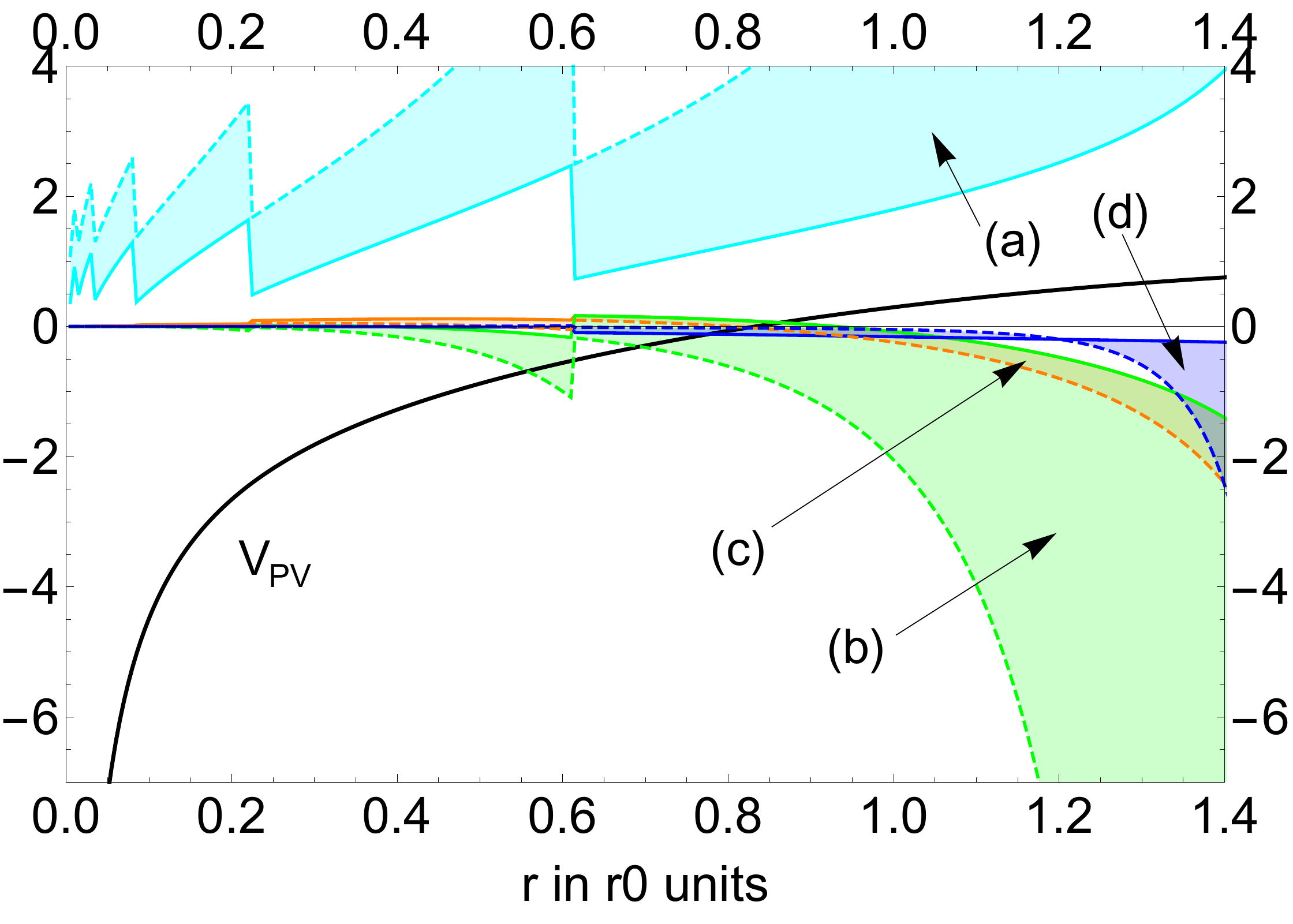}
%\includegraphics[width=0.76\textwidth]{PlottingSmallestPositivecAndNegativeHyperMS.png}
%\put(0,220){{\large (a)}}
%\put(0,129){{\large (d)}}
%\put(0,105){{\large (c)}}
%\put(0,65){{\large (b)}}
\vspace{0.1in}
\includegraphics[width=0.82\textwidth]{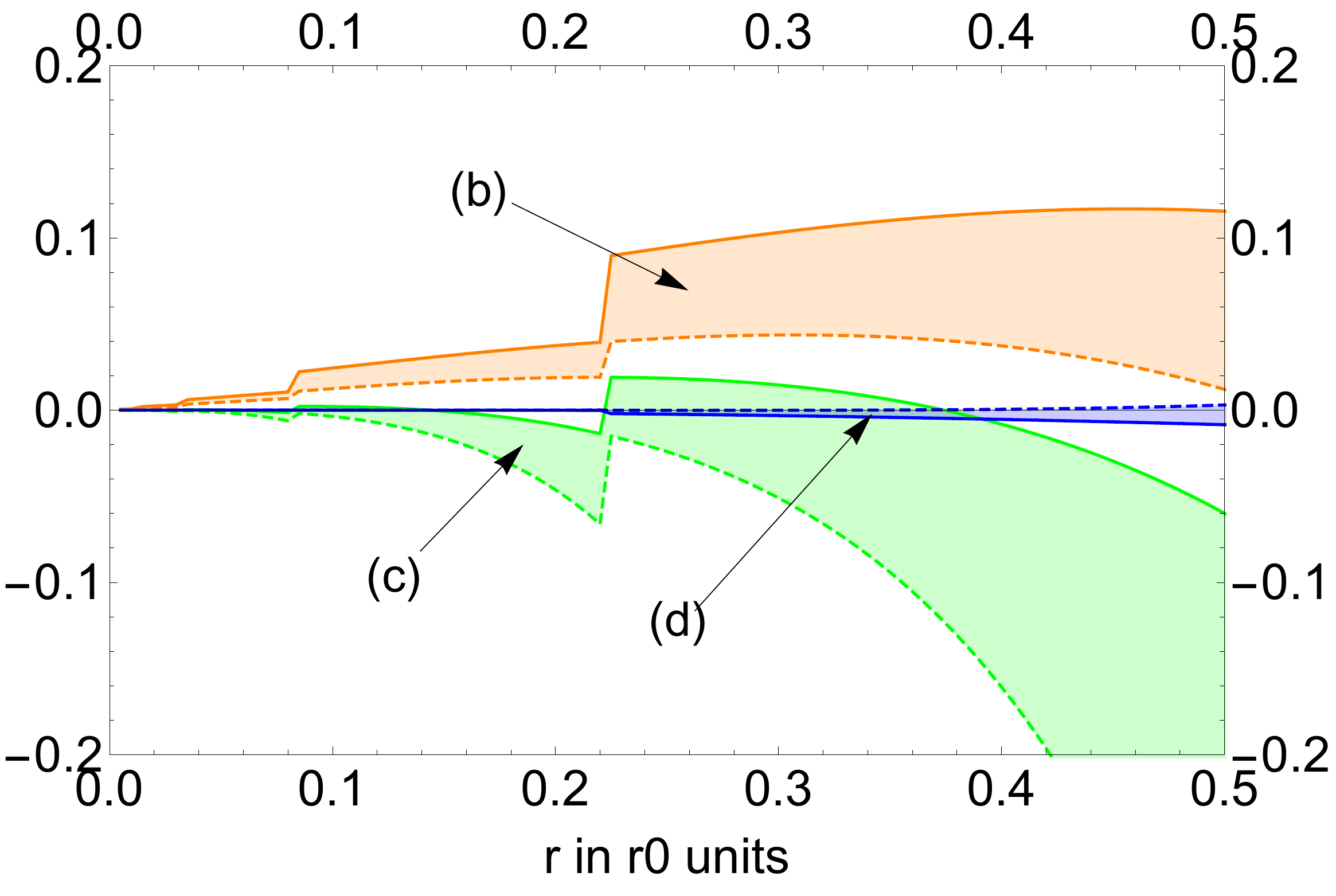}
%\includegraphics[width=0.76\textwidth]{PlottingSmallestPositivecAndNegativeHyperZoomed.png}
%\put(0,145){{\large (b)}}
%\put(0,110){{\large (d)}}
%\put(0,55){{\large (c)}}
\caption{As in Fig. \ref{Fig:VPVb0nf0latt} but in the $\MS$ scheme. \label{Fig:VPVb0nf0MS}}
\end{figure}
\end{center}
\begin{center}
\begin{figure}
\includegraphics[width=0.765\textwidth]{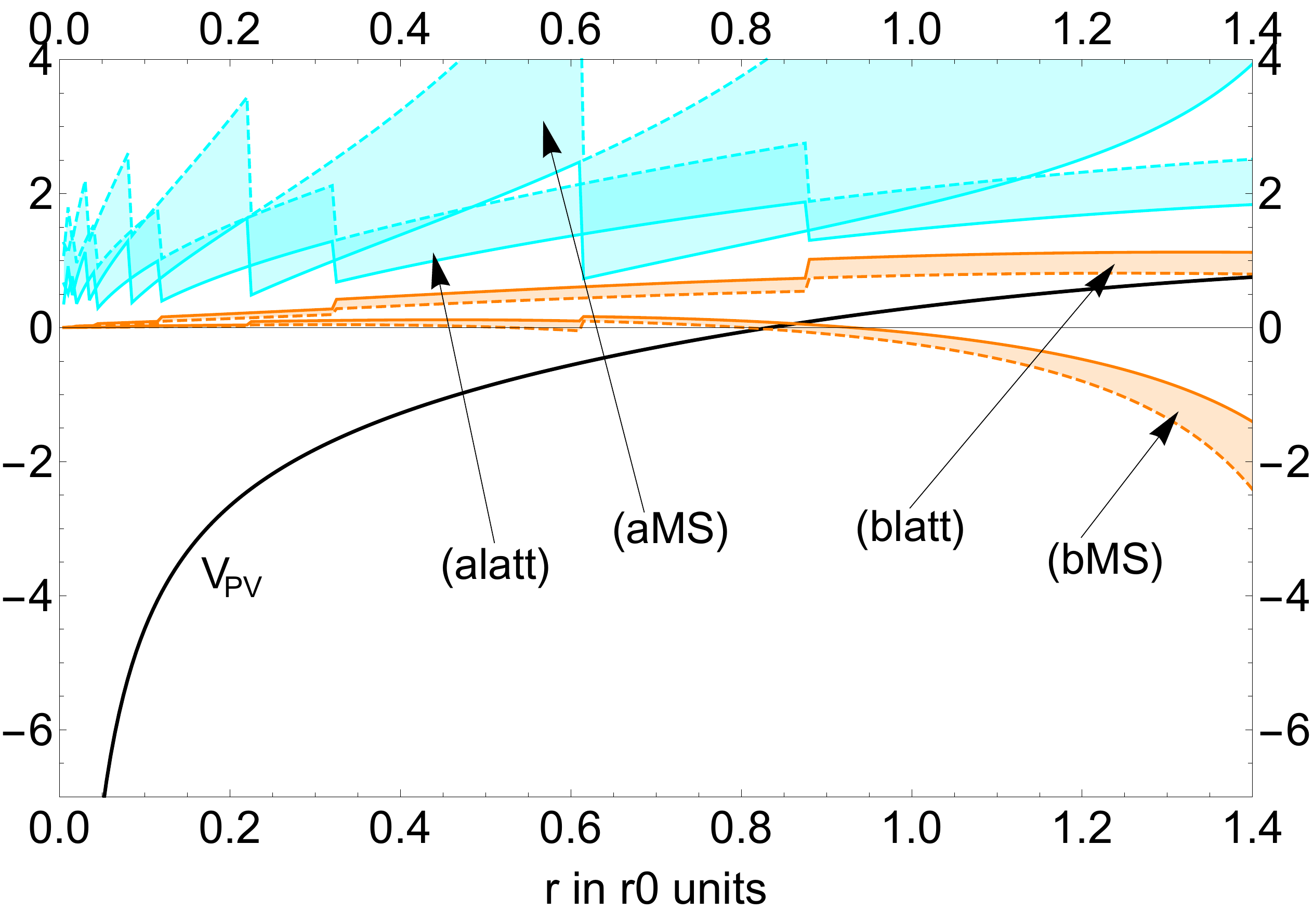}
%\includegraphics[width=0.76\textwidth]{PlottingSmallestPositivecAndNegativeWithoutDeltaVCombined.png}
%\put(0,220){{\large (aMS)}}
%\put(0,190){{\large (alatt)}}
%\put(0,159){{\large (blatt)}}
%\put(0,105){{\large (bMS)}}
\vspace{0.1in}
\includegraphics[width=0.82\textwidth]{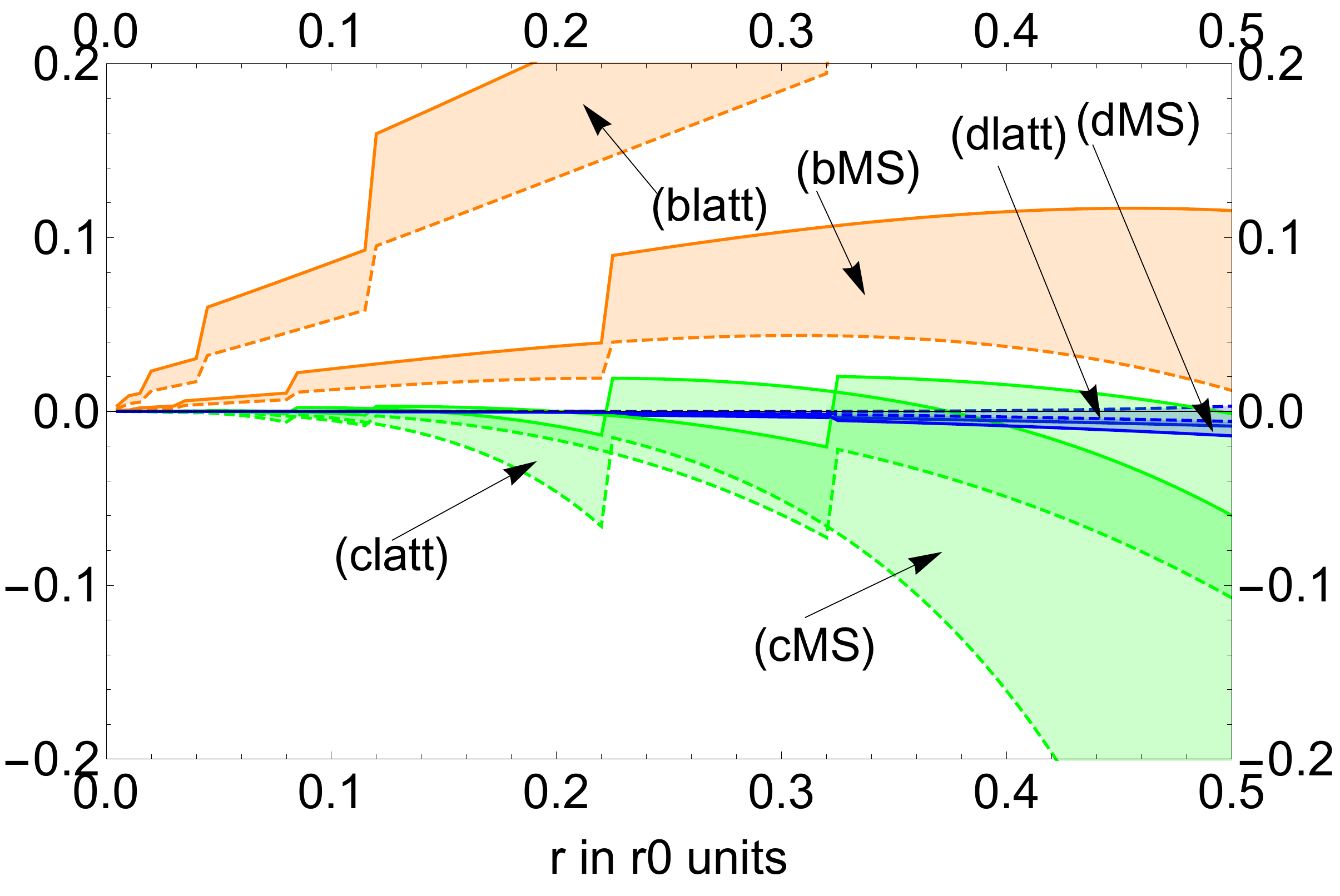}
%\includegraphics[width=0.76\textwidth]{PlottingSmallestPositivecAndNegativeHyperZoomedCombined.png}
%\put(-160,200){{\large (blatt)}}
%\put(0,145){{\large (bMS)}}
%\put(0,70){{\large (clatt)}}
%\put(0,25){{\large (cMS)}}
%\put(0,103){{\large (dlatt)}}
%\put(0,112){{\large (dMS)}}
\caption{\label{Fig:VPVb0nf0lattvsMS} Comparison of lattice and $\MS$ scheme results for $n_f=0$. {\bf Upper panel}: We plot $V_{\rm PV}$ and the differences: (a) $V_{\rm PV}-V_P$, and (b) $V_{\rm PV}-V_P-\frac{1}{r}\Omega_V$ 
in the lattice and $\MS$ scheme with $n_f=0$ light flavours. {\bf Lower panel}: Figs. \ref{Fig:VPVb0nf0latt} and \ref{Fig:VPVb0nf0MS} combined.}
\end{figure}
\end{center}

Unlike in the limit 2) (see expressions in Sec. \ref{Sec:VPVmuinfty}), in the limit case 1), \eq{eq:NP}, we do not have direct analytic control in the relation between $V_{\rm PV}$ and $V_{P}$ (unlike what will happen in Sec. \ref{Sec:VPVmuinfty} when using the limit case 2), \eq{eq:muinfty}). Nevertheless, we can numerically compute both and check that their difference complies with the theoretical expectations. We can study (even if in the large $\beta_0$ approximation) up to which values of $r$ the OPE is a good approximation of $V_{\rm PV}$. Remarkably enough we can actually check more than one term of the OPE (hyperasymptotic) expansion. We also explore the scheme dependence by performing the computation in the lattice and the $\MS$ scheme (actually in the large $\beta_0$ approximation this is equivalent to a change of scale). We will do these analyses for the cases with $n_f=0$ and $n_f=3$. The first in view of comparing with quenched lattice simulations, the second to simulate a more physical scenario, for which we can draw some conclusions that could be applied beyond the large-$\beta_0$ limit. In Figs. \ref{Fig:VPVb0nf0latt}, \ref{Fig:VPVb0nf0MS}, and \ref{Fig:VPVb0nf0lattvsMS} we plot $V_{\rm PV}$, $V_{\rm PV}-V_P$, 
$V_{\rm PV}-V_P-\frac{1}{r}\Omega_V$, 
$V_{\rm PV}-V_P-\frac{1}{r}\Omega_V-\sum_{n=N_P+1}^{3N_P} (V_n-V_n^{(\rm as)}) \al^{n+1}$,
and  $V_{\rm PV}-V_P-\frac{1}{r}\Omega_V-\sum_{n=N_P+1}^{3N_P} (V_n-V_n^{(\rm as)}) \al^{n+1}-\frac{1}{r}\Omega'_V$ with $n_f=0$ light flavours. We do such computation in the lattice (Fig. \ref{Fig:VPVb0nf0latt}) and the $\MS$ (Fig. \ref{Fig:VPVb0nf0MS}). In Fig. \ref{Fig:VPVb0nf0lattvsMS} we compare the results in the lattice and $\MS$ scheme. 
We observe a very nice convergent patter in all cases down to surprisingly small scales. To visualize the dependence on $c$ for each case, we show the band generated by the smallest positive and negative possible values of $c$ that yields integer values for $N_P$. The size of the band generated by the different values of $c$ (the $c$ dependence) decreases as we introduce more terms in the hyperasymptotic expansion. This is particularly so when including $\Omega_V$ ($\Omega_V'$) to its associated sum.

Let us discuss the results in more detail. We first observe that the $r$ dependence of $V_{\rm PV}$ is basically eliminated in $V_{\rm PV}-V_P$, as expected. This happens both in the lattice and $\MS$ scheme. The latter shows an stronger $c$ dependence. This is to be expected, as in the $\MS$, we truncate at smaller orders in $N$. This makes the truncation error bigger. Note that the lattice scheme can be understood (in the large $\beta_0$ approximation) as the $\MS$ scheme with a larger factorization scale. As we can see in the upper panel of Fig. \ref{Fig:VPVb0nf0lattvsMS}, both schemes yield consistent predictions for $V_{\rm PV}-V_P$. We can draw some interesting observations out of this analysis. For $V_{\rm PV}-V_P$ it is better to choose a larger factorization scale, if we have enough coefficients of the perturbative expansion. This is particularly so at large distances: We can still get sound results up to very large distances in the lattice scheme. 

We now turn to $V_{\rm PV}-V_P-\frac{1}{r}\Omega_V$. Adding the new correction produces a better agreement with expectations (which we recall is to get zero).   
After the introduction of $\frac{1}{r}\Omega_V$, the $\MS$ scheme yields more accurate results than the lattice scheme. This can already be seen in the upper panel of Fig. \ref{Fig:VPVb0nf0lattvsMS}, and in greater detail in the lower panel of Fig. \ref{Fig:VPVb0nf0lattvsMS}. 

$V_{\rm PV}-V_P-\frac{1}{r}\Omega_V$ shows some dependence on $1/r$, which is more pronounced in the lattice than in the $\MS$ scheme. As in the large $\beta_0$ the difference between both schemes is equivalent to a change of scale, this results points to that $\mu=1/r$ in $\MS$ is close to the natural scale and minimize higher order corrections. Note that the lattice scheme computation is equivalent to the $\MS$ scheme choosing $\mu_{latt}= \mu_{\MS}e^{-\frac{c_{latt}}{2}}e^{\frac{c_{\MS}}{2}}$. This gives around a factor 30!!. 
Once $\sum_{n=N_P+1}^{3N_P} (V_n-V_n^{(\rm as)}) \al^{n+1}$ is incorporated in the prediction most of the difference disappears and the lattice scheme is marginally better. Nevertheless, after introducing $\Omega_V'$, the $\MS$ becomes marginally better again. In any case, the difference between schemes gets smaller and smaller as we go to higher orders in the hyperasymptotic expansion, in particular at short distances.

We also want to stress that this analysis opens the window to apply perturbation theory at rather large distances.  Note that in the upper panel plots in Figs. \ref{Fig:VPVb0nf0latt}, \ref{Fig:VPVb0nf0MS}, and \ref{Fig:VPVb0nf0lattvsMS}, we have gone to very large distances. 

As some concluding remarks let us emphasize the following points. The truncated sum is more or less constant with relatively large uncertainties. This is to be expected, as the next correction in magnitude is $\Omega_V$ which is approximately constant (mildly modulated by $\sqrt{\al(\mu)}$). After introducing this term the error is much smaller and we can see more structure. In particular we are sensitive to $\sum_{n=N_P+1}^{3N_P} (V_n-V_n^{(\rm as)}) \al^{n+1}$. Here we find (at the level of precision we have now) a sizable difference between lattice and $\MS$. This can be expected: $\sum_{n=N_P+1}^{3N_P} (V_n-V_n^{(\rm as)}) \al^{n+1}$ is the object we expect to be more sensitive to the scale.  

In the lattice and $\MS$ scheme, we observe a very nice convergence pattern up to (surprisingly) rather large scales. The agreement with the theoretical prediction (which is zero) is perfect at short distances. The estimated error is also expected to be small. It would be interesting to see if this also happens beyond the large $\beta_0$. 

Another interesting observation is that truncated sums behave better in the lattice scheme than in the $\MS$ scheme. Nevertheless, this could be missleading. The sums are truncated at the minimal term. Therefore, one needs more terms in the lattice scheme. If the number of terms is not an issue (which could be the case with dedicated numerical stochastic perturbation theory (NSPT) \cite{DiRenzo:1994sy,DiRenzo:2004hhl} computations in the lattice scheme) then the lattice scheme looks better. But as soon as $\Omega_V$ is introduced in the computation $\MS$ behaves better (at least in the large $\beta_0$ approximation).

\begin{center}
\begin{figure}
\includegraphics[width=0.765\textwidth]{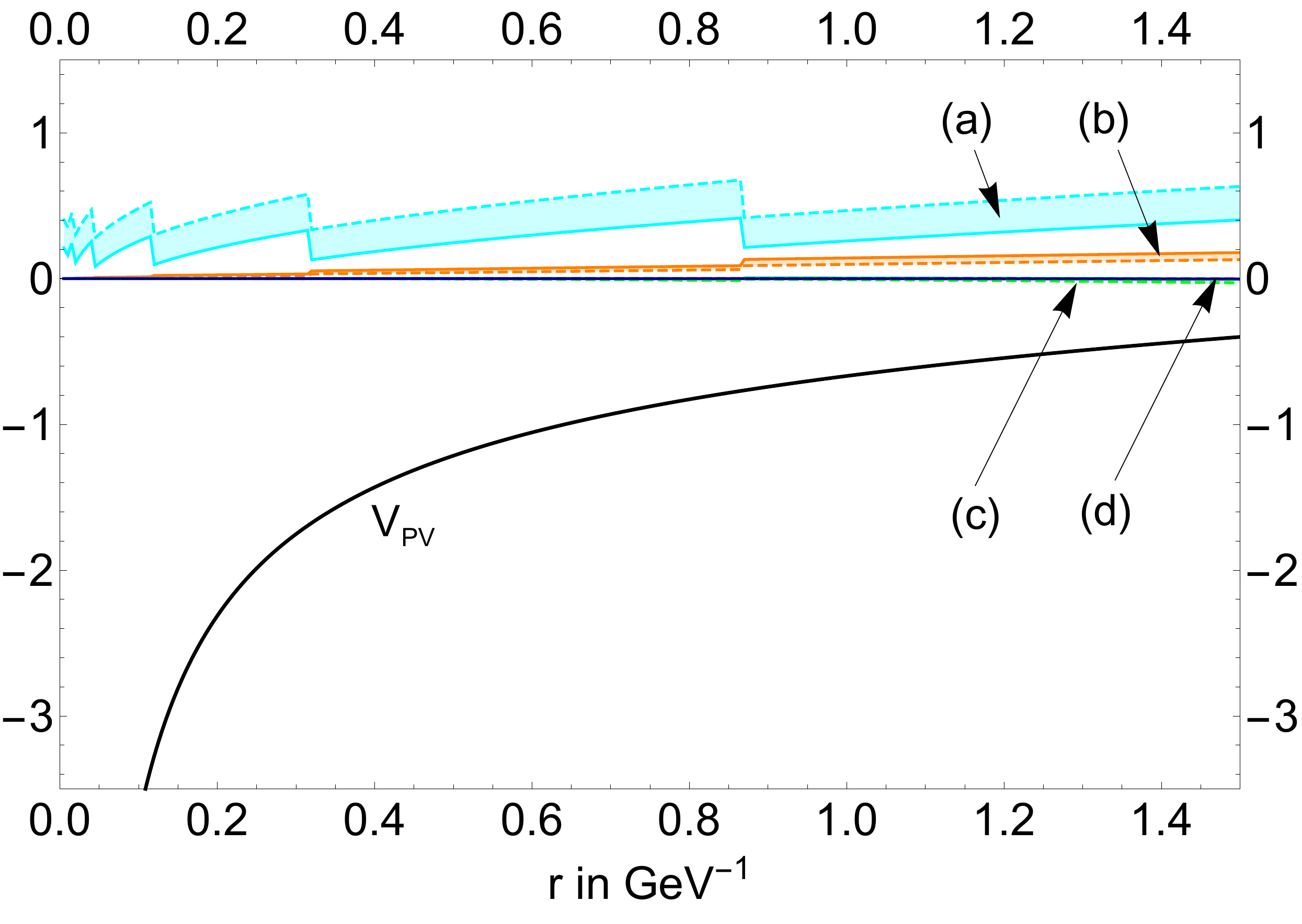}
%\includegraphics[width=0.75\textwidth]{PlottingSmallestPositivecAndNegativeHypernf3.png}
%\put(0,179){{\large (a)}}
%\put(0,155){{\large (c,d)}}
%\put(0,165){{\large (b)}}
\vspace{0.1in}
\includegraphics[width=0.852\textwidth]{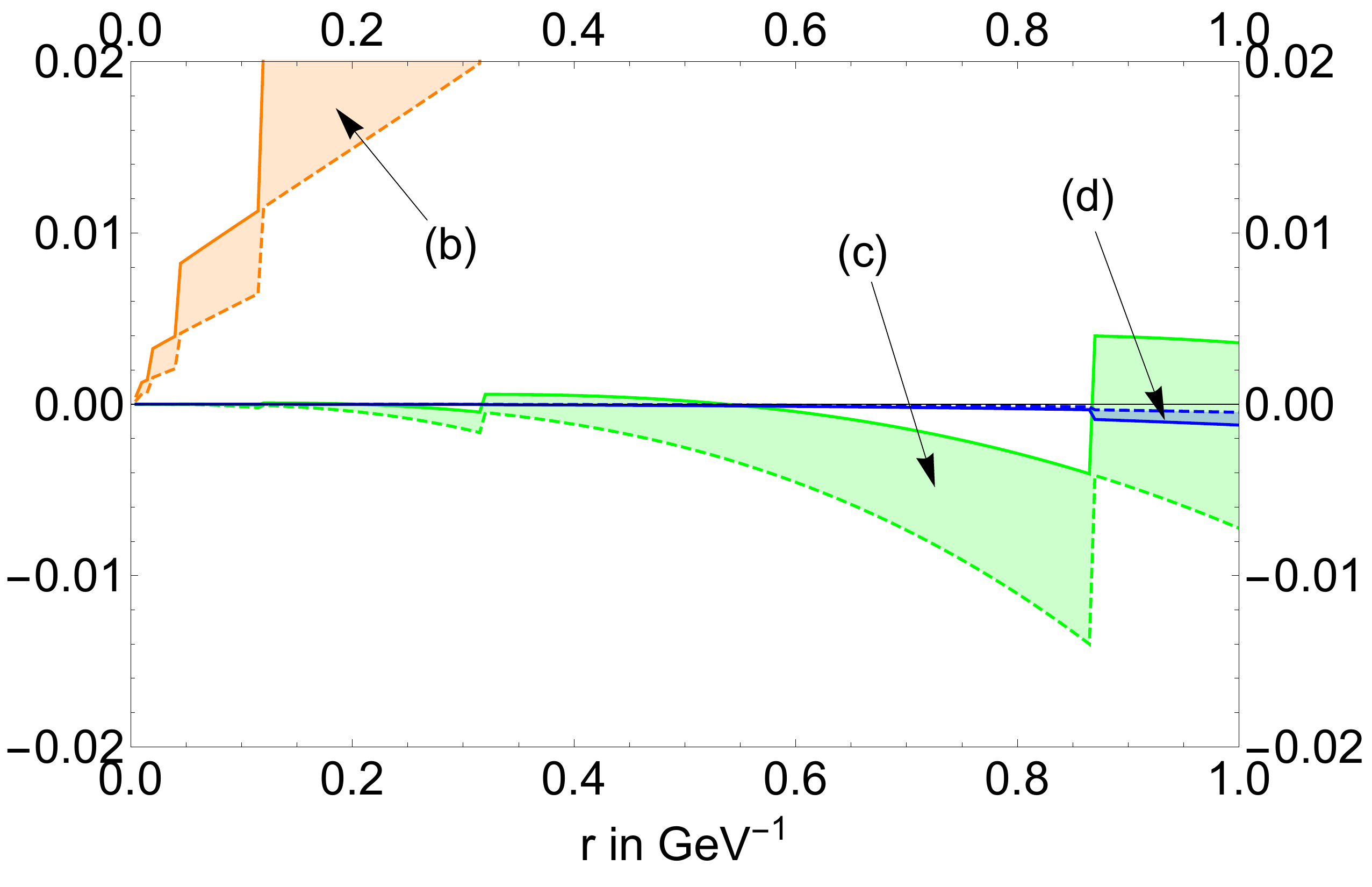}
%\includegraphics[width=0.75\textwidth]{PlottingSmallestPositivecAndNegativeHyperZoomednf3.png}
%\put(-220,189){{\large (b)}}
%\put(0,105){{\large (d)}}
%\put(0,75){{\large (c)}}
\caption{As in Fig. \ref{Fig:VPVb0nf0latt} but with $n_f=3$ light flavours.  \label{Fig:VPVb0nf3latt}}
\end{figure}
\end{center}

\begin{center}
\begin{figure}
\includegraphics[width=0.765\textwidth]{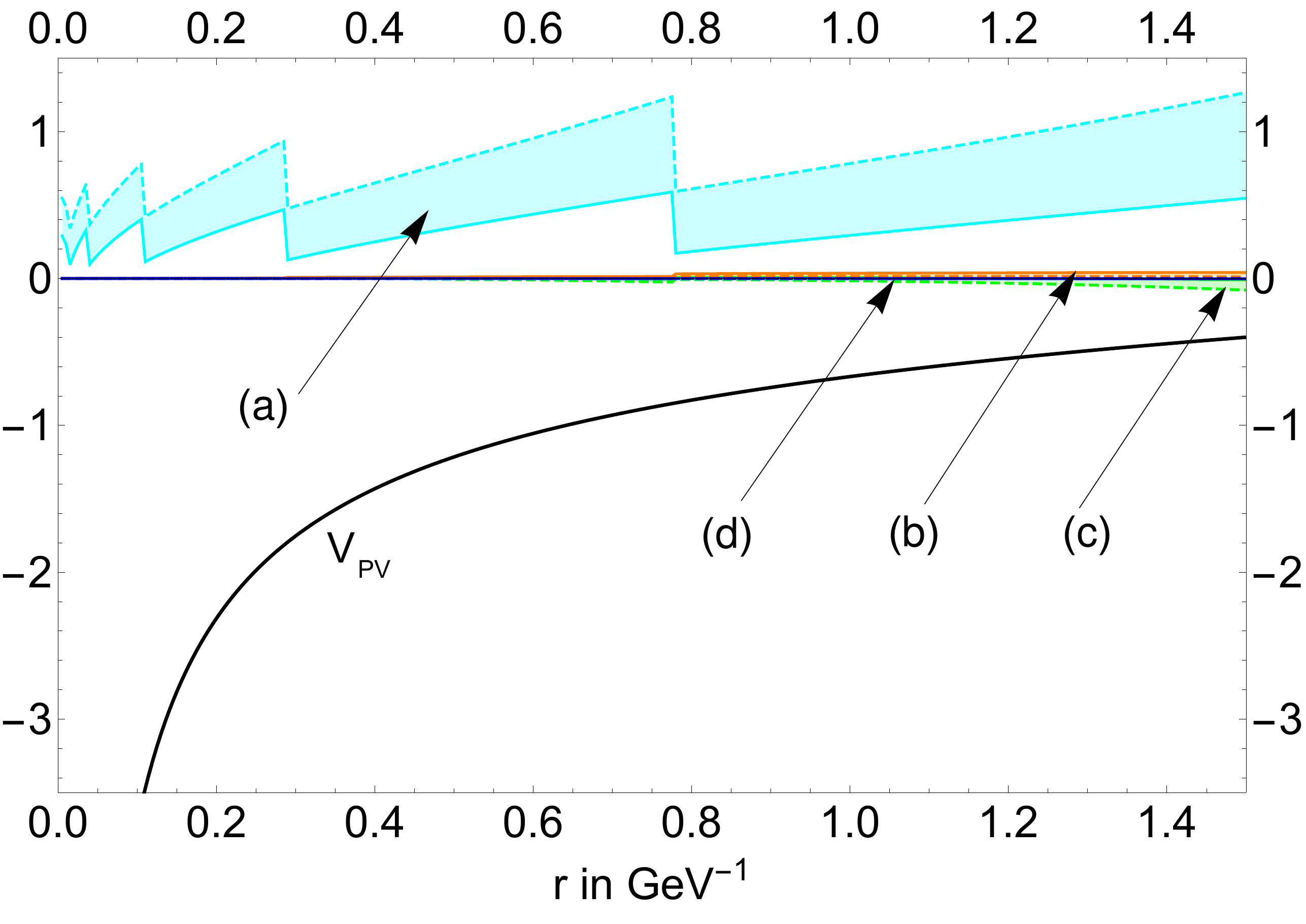}
%\includegraphics[width=0.75\textwidth]{PlottingSmallestPositivecAndNegativeHyperMSnf3.png}
%\put(0,199){{\large (a)}}
%\put(0,155){{\large (b,c,d)}}
\vspace{0.1in}
\includegraphics[width=0.852\textwidth]{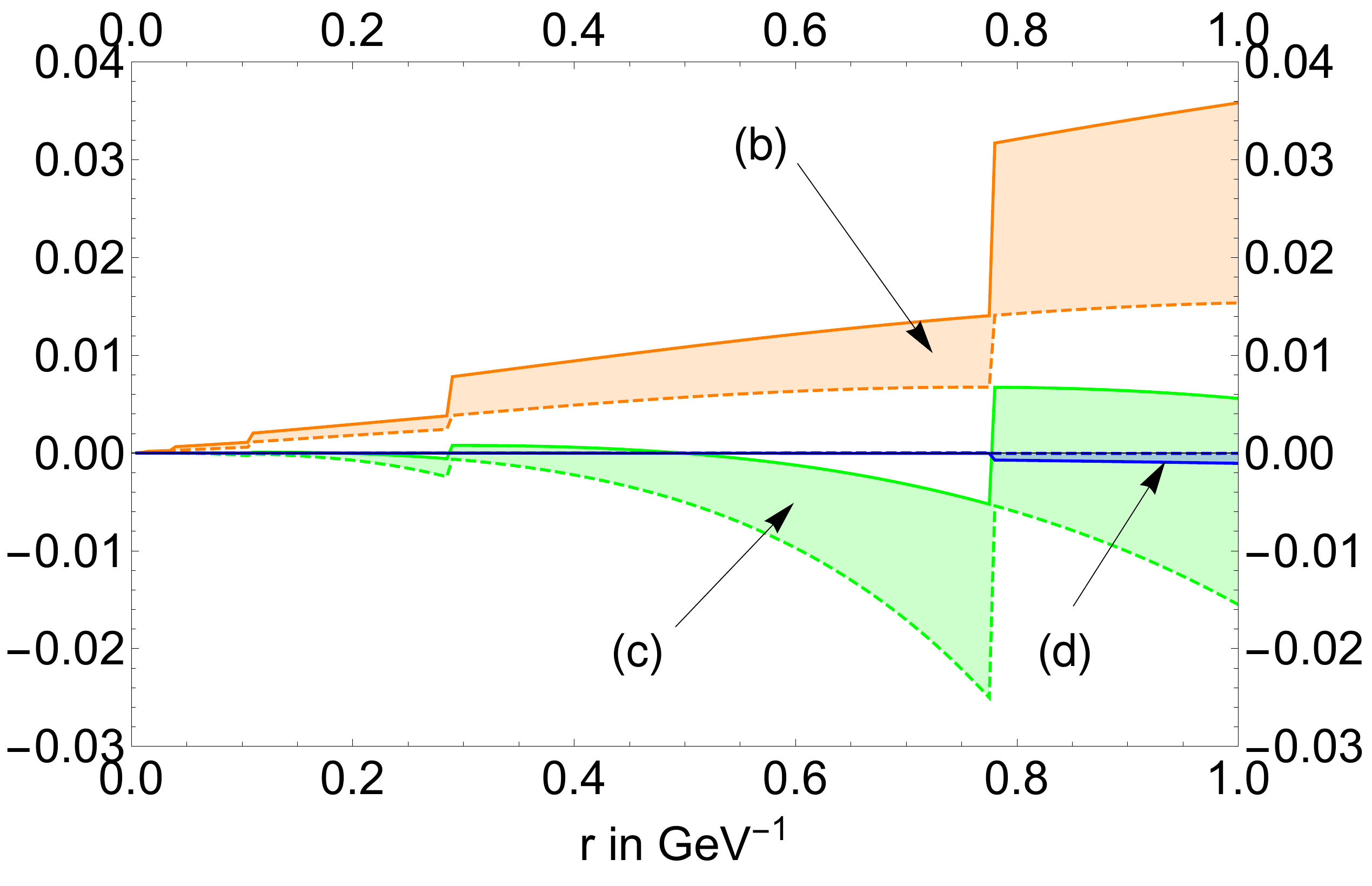}
%\includegraphics[width=0.75\textwidth]{PlottingSmallestPositivecAndNegativeHyperZoomedMSnf3.png}
%\put(0,199){{\large (b)}}
%\put(0,105){{\large (d)}}
%\put(0,75){{\large (c)}}
\caption{As in Fig. \ref{Fig:VPVb0nf0latt} but with $n_f=3$ light flavours and in the $\MS$ scheme. \label{Fig:VPVb0nf3MS}}
\end{figure}
\end{center}

\begin{center}
\begin{figure}
\includegraphics[width=0.76\textwidth]{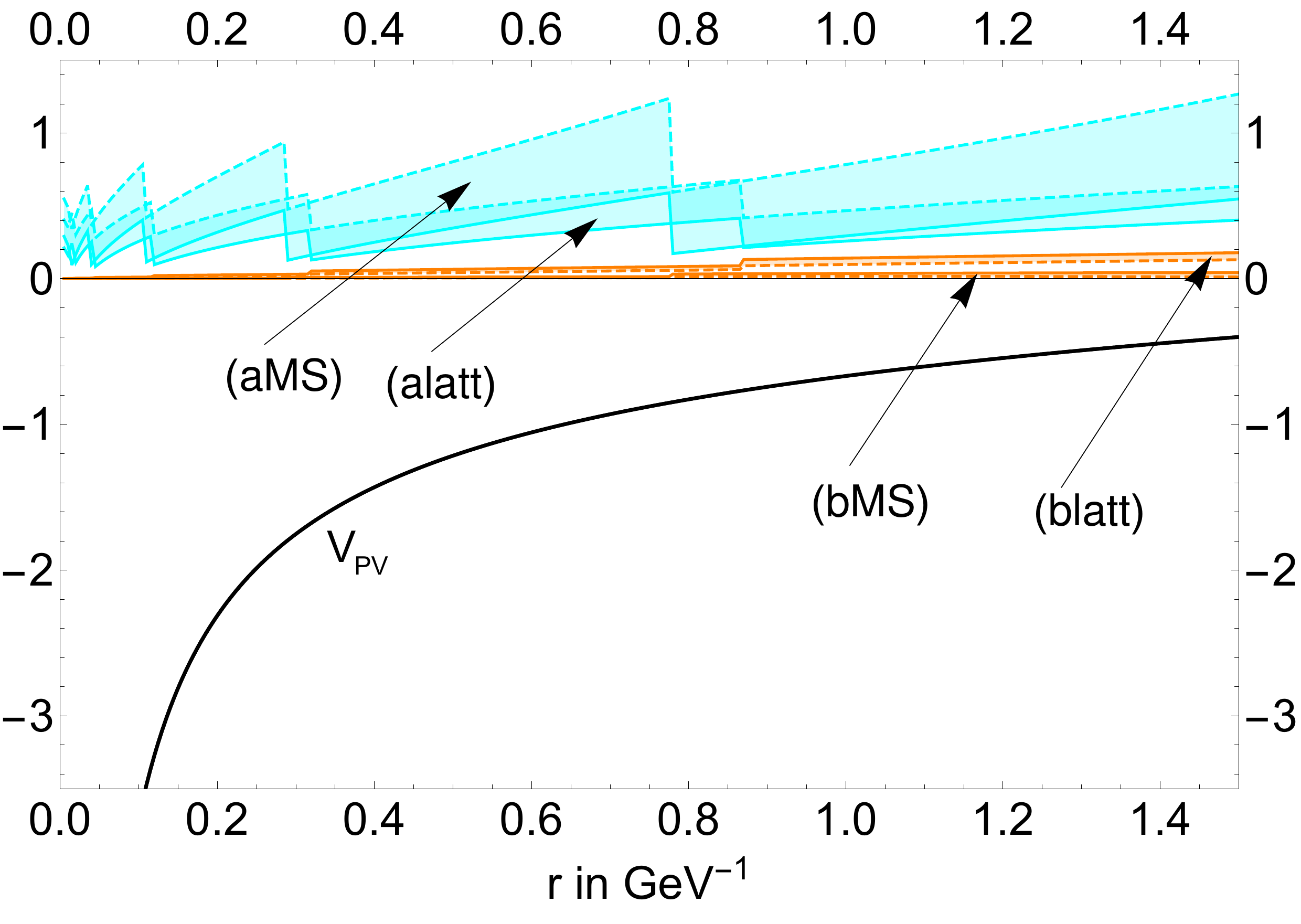}
%\includegraphics[width=0.75\textwidth]{PlottingSmallestPositivecAndNegativeExcludingDeltaVCombined.png}
%\put(0,205){{\large (aMS)}}
%\put(0,182){{\large (alatt)}}
%\put(0,163){{\large (blatt)}}
%\put(0,153){{\large (bMS)}}
\vspace{0.1in}
\includegraphics[width=0.84\textwidth]{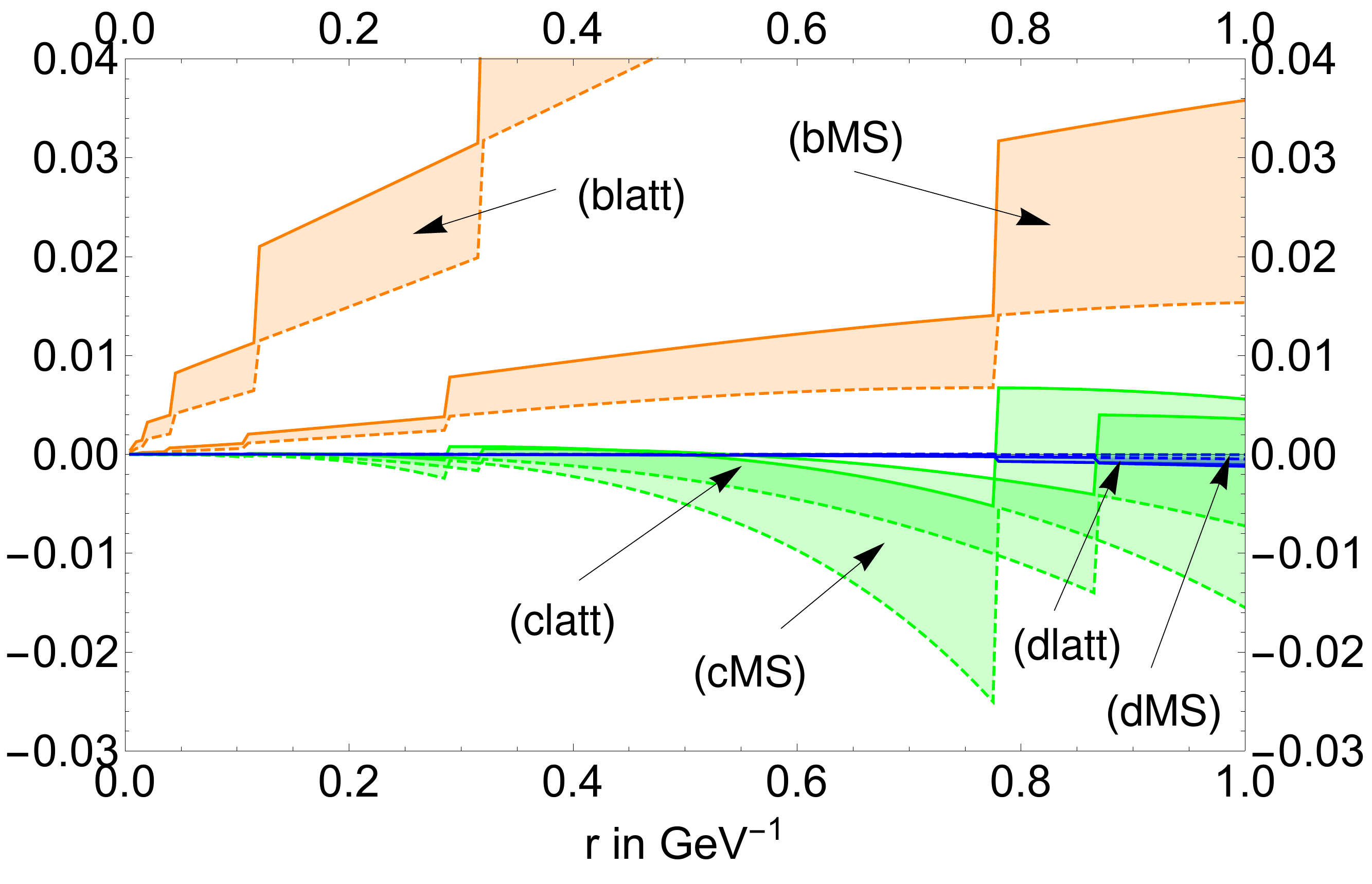}
%\includegraphics[width=0.75\textwidth]{PlottingSmallestPositivecAndNegativeHyperZoomedCombinednf3.png}
%\put(-230,200){{\large (blatt)}}
%\put(0,195){{\large (bMS)}}
%\put(0,85){{\large (clatt)}}
%\put(0,45){{\large (cMS)}}
%\put(0,103){{\large (dlatt)}}
%\put(0,112){{\large (dMS)}}
\caption{\label{Fig:VPVb0nf3lattvsMS} 
Comparison of lattice and $\MS$ scheme results for $n_f=3$. {\bf Upper panel}: We plot $V_{\rm PV}$ and the differences: (a) $V_{\rm PV}-V_P$, and (b) $V_{\rm PV}-V_P-\frac{1}{r}\Omega_V$ 
in the lattice and $\MS$ scheme with $n_f=3$ light flavours. {\bf Lower panel}: Fig. \ref{Fig:VPVb0nf3latt} and Fig. \ref{Fig:VPVb0nf3MS} combined.}
\end{figure}
\end{center}

We now turn to the $n_f=3$ case. 
We note that $\lQ$ for the physical case ($n_f=3$) is smaller than for the $n_f=0$ case (if one sets the physical scale according to $r_0^{-1} \approx 400$ MeV). On top of that the running is less important. All this points to that the convergence should be even better than in the $n_f=0$ case (and it was quite good already there). We show our results in Figs. 
\ref{Fig:VPVb0nf3latt}, \ref{Fig:VPVb0nf3MS} and \ref{Fig:VPVb0nf3lattvsMS} (these are the analogous of Figs. \ref{Fig:VPVb0nf0latt}, \ref{Fig:VPVb0nf0MS} and \ref{Fig:VPVb0nf0lattvsMS} but with $n_f=3$). These plots confirm our expectations. Down to scales as low as 667 MeV we see no sign of breakdown of the hyperasymptotic expansion. This is so in both the lattice and the $\MS$ schemes. Note that the precision we get is extremely high as we go to small scales: Using truncation 
(c): $V_P+\frac{1}{r}\Omega_V+\sum_{n=N_P+1}^{3N_P} (V_n-V_n^{(\rm as)}) \al^{n+1}$,
 one gets $V_{\rm PV}$ in both schemes with a precision well below 1 MeV at scales of the order of the mass of the bottom. Using truncation (d): $V_P+\frac{1}{r}\Omega_V+\sum_{n=N_P+1}^{3N_P} (V_n-V_n^{(\rm as)}) \al^{n+1}+\frac{1}{r}\Omega'_V$, the error is astonishingly small (see Fig. \ref{Fig:Hyper} for an extra zoom in this region). The rest of the discussion follows parallel the one for $n_f=0$. 
In the above numerics, we have used the exact expression for $\Omega_V$ and $\Omega'_V$. In full QCD, we will not know the exact expression. 
Therefore, it makes sense to study how well the exact result is reproduced by the semiclassical expansion obtained in \eq{Omegaexp}. 
We compare in Table \ref{Table:OmegaVnf0latt} and \ref{Table:OmegaVnf3latt} for an illustrative set of values the exact result and the truncated semiclassical expansion. 
We observe that the exact result is very well saturated by the first terms of the expansion computed in \eq{Omegaexp}. Truncating the expansion produces differences much smaller than the typical precision of the different terms of the hyperasymptotic expansion. 
As expected $n_f=3$ is better than $n_f=0$.  Note that in the large $\beta_0$ approximation we exactly have $\Lambda=\mu e^{-2\pi/(\beta_0\al(\mu))}$.
 
%\begin{center}
%\begin{figure}
%\includegraphics[width=0.75\textwidth]{TableOfTheOmegasnf0MS.png}
%\includegraphics[width=0.75\textwidth]{TableOfTheOmegasnf0latt.png}
%\caption{\label{Table:OmegaVnf0latt}
%$1/r\Omega_V$; $n_f=0$. Upper panel MS. Lower panel Latt. GeV units. Lattice seems to be even better. But both are very good. }
%\end{figure}
%\end{center}
\renewcommand{\arraystretch}{1.5}
\begin{table}
\begin{tabular}{|c|c|c|c|c|c|}
\hline
\multicolumn{6}{|c|}{$\overline{\text{MS}}$-Scheme ($n_f=0$)} \\
\hline
 $r$ in $r_0$ & $c$ & $\frac{1}{r}\Omega$Exact & $\left| \frac{\Omega\text{LO}}{\Omega\text{Exact}}-1\right|\times 10^2$ & $\left| \frac{\Omega \text{NLO}}{\Omega\text{Exact}}-1\right|\times 10^3$ & $\left| \frac{\Omega \text{NNLO}}{\Omega\text{Exact}}-1\right|\times 10^4$ \\ \hline
 1.5 & 0.178629 & 8.3643 & 22.4162 & 47.5334 & 1969.22 \\
 %1.4 & 0.299415 & 5.3378 & 11.359 & 6.51733 & 969.082 \\
 %1.3 & 0.429156 & 3.8821 & 4.68914 & 22.121 & 504.236 \\
 1.2 & 0.569288 & 2.9883 & 0.40329 & 24.9029 & 253.624 \\
 %1.1 & 0.721619 & 2.3599 & 2.34987 & 22.2889 & 109.027 \\
 1.0 & 0.88848 & 1.8767 & 3.9895 & 17.132 & 24.530 \\
 %0.9 & 1.07293 & 1.4795 & 4.72302 & 10.7416 & 22.2529 \\
 0.8 & 1.27914 & 1.1346 & 4.61687 & 3.82012 & 43.2013 \\
 %0.7 & 1.51291 & 0.82027 & 3.57888 & 3.20768 & 44.7456 \\
 0.6 & 0.032079 & 2.3128 & 5.14476 & 0.979725 & 9.61123 \\
 %0.5 & 0.35127 & 1.7649 & 1.39576 & 1.87651 & 8.08095 \\
 %0.45 & 0.535725 & 1.5127 & 0.0444862 & 2.52049 & 5.32011 \\
 0.4 & 0.741928 & 1.2686 & 1.15011 & 2.54579 & 1.8752 \\
 %0.35 & 0.975702 & 1.0278 & 1.85671 & 1.89987 & 1.6066 \\
 %0.3 & 1.24557 & 0.78420 & 2.04338 & 0.556517 & 4.26306 \\
 %0.25 & 1.56477 & 0.53013 & 1.42695 & 1.47196 & 4.82749 \\
 0.2 & 0.20472 & 1.4294 & 1.53909 & 0.352876 & 1.59471 \\
 %0.15 & 0.708366 & 1.0018 & 0.64194 & 0.922709 & 0.386691 \\
 0.1 & 1.41822 & 0.51943 & 1.16526 & 0.275048 & 1.36791 \\
 %0.05 & 0.881007 & 0.74107 & 0.749328 & 0.411744 & 0.0952108 \\
 0.01 & 0.197248 & 0.91480 & 0.654315 & 0.073017 & 0.120936 \\
\hline \hline
\multicolumn{6}{|c|}{Lattice-Scheme ($n_f=0$)} \\
\hline
 $r$ in $r_0$ & $c$ & $\frac{1}{r}\Omega$Exact & $\left| \frac{\Omega\text{LO}}{\Omega\text{Exact}}-1\right|\times 10^3$ & $\left| \frac{\Omega \text{NLO}}{\Omega\text{Exact}}-1\right|\times 10^4$ & $\left| \frac{\Omega \text{NNLO}}{\Omega\text{Exact}}-1\right|\times 10^{5}$ \\ \hline
 1.5 & 0.810107 & 0.78253 & 6.49313 & 4.43451 & 0.0787894 \\
% 1.4 & 0.930894 & 0.71233 & 8.07844 & 3.82254 & 1.55701 \\
% 1.3 & 1.06063 & 0.63912 & 9.14257 & 2.79432 & 3.03694 \\
 1.2 & 1.20077 & 0.56237 & 9.54184 & 1.29876 & 4.36981 \\
% 1.1 & 1.3531 & 0.48138 & 9.05508 & 0.714089 & 5.33188 \\
 1.0 & 1.5200 & 0.39525 & 7.3017 & 3.2941 & 5.5745 \\
% 0.9 & 1.70441 & 0.30278 & 3.49552 & 6.51033 & 4.49394 \\
 0.8 & 0.159911 & 1.0434 & 9.33533 & 0.811786 & 2.31443 \\
% 0.7 & 0.393685 & 0.91023 & 2.60058 & 2.20927 & 1.91711 \\
 0.6 & 0.663557 & 0.76543 & 3.00401 & 2.81903 & 0.726526 \\
% 0.5 & 0.982748 & 0.60443 & 6.75049 & 2.08063 & 1.09887 \\
% 0.45 & 1.1672 & 0.51563 & 7.49691 & 0.993224 & 2.03308 \\
 0.4 & 1.37341 & 0.41946 & 7.02749 & 0.69835 & 2.69681 \\
% 0.35 & 1.60718 & 0.31385 & 4.53507 & 3.09564 & 2.62707 \\
% 0.3 & 0.126349 & 0.95070 & 8.40191 & 0.385756 & 1.21249 \\
% 0.25 & 0.44554 & 0.79315 & 1.0556 & 1.60551 & 0.91614 \\
 0.2 & 0.836198 & 0.61277 & 4.4603 & 1.74206 & 0.162621 \\
% 0.15 & 1.33984 & 0.39589 & 5.94439 & 0.298258 & 1.49101 \\
 0.1 & 0.29899 & 0.79056 & 3.42696 & 0.82486 & 0.671662 \\
% 0.05 & 1.51249 & 0.30040 & 4.08796 & 1.07969 & 0.984137 \\
 0.01 & 0.828727 & 0.49592 & 2.87157 & 0.729908 & 0.0478304 \\
\hline
\end{tabular}
\caption{\label{Table:OmegaVnf0latt}
$1/r\Omega_V$ for $n_f=0$ in $r_0^{-1}$ units compared with \eq{eq:OmegaV} truncated at different powers of $\al$. Upper panel computed in the $\MS$ scheme. Lower panel in the lattice scheme. Lattice seems to be better but both schemes yield very good results.}
\end{table}

\renewcommand{\arraystretch}{1.5}
\begin{table}
\begin{tabular}{|c|c|c|c|c|c|}
\hline
\multicolumn{6}{|c|}{$\overline{\text{MS}}$-Scheme ($n_f=3$)} \\
\hline
r in GeV$^{-1}$ & $c$ & $\frac{1}{r}\Omega$Exact & $\left| \frac{\Omega\text{LO}}{\Omega\text{Exact}}-1\right|\times 10^2$ & $\left| \frac{\Omega \text{NLO}}{\Omega\text{Exact}}-1\right|\times 10^3$ & $\left| \frac{\Omega \text{NNLO}}{\Omega\text{Exact}}-1\right|\times 10^4$ \\ \hline
 1.5 & 0.491648 & 0.50579 & 0.447496 & 2.60006 & 4.23428 \\
 %1.4 & 0.590473 & 0.45611 & 1.06283 & 2.5711 & 2.20842 \\
 %1.3 & 0.696625 & 0.40687 & 1.54825 & 2.31872 & 0.14396 \\
 1.2 & 0.811277 & 0.35770 & 1.88563 & 1.83577 & 1.80928 \\
 %1.1 & 0.935912 & 0.30817 & 2.04814 & 1.11882 & 3.47202 \\
 1.0 & 1.0724 & 0.25779 & 1.9932 & 0.16927 & 4.6228 \\
 %0.9 & 1.22335 & 0.20594 & 1.64439 & 1.00689 & 4.9702 \\
 0.8 & 1.39206 & 0.15183 & 0.834303 & 2.4094 & 4.06597 \\
 %0.7 & 0.150938 & 0.51181 & 1.66306 & 0.300555 & 1.60672 \\
 0.6 & 0.371743 & 0.42591 & 0.250112 & 0.811721 & 1.16451 \\
 %0.5 & 0.632899 & 0.33656 & 0.807976 & 0.896668 & 0.166134 \\
 %0.45 & 0.783817 & 0.28961 & 1.14644 & 0.715352 & 0.44167 \\
 0.4 & 0.952529 & 0.24035 & 1.30166 & 0.350684 & 1.00395 \\
 %0.35 & 1.1438 & 0.18789 & 1.19016 & 0.219646 & 1.35618 \\
 %0.3 & 1.3646 & 0.13094 & 0.618902 & 1.01637 & 1.19862 \\
 %0.25 & 0.193365 & 0.40623 & 0.91725 & 0.222216 & 0.512232 \\
 0.2 & 0.512995 & 0.31554 & 0.2984 & 0.466805 & 0.21132 \\
 %0.15 & 0.925069 & 0.21213 & 0.944059 & 0.20938 & 0.374642 \\
 0.1 & 0.0734605 & 0.38329 & 1.171 & 0.02592 & 0.228532 \\
 %0.05 & 1.06632 & 0.15962 & 0.735045 & 0.0072883 & 0.253138 \\
 0.01 & 0.506882 & 0.23072 & 0.15953 & 0.132861 & 0.0290686 \\
\hline\hline
\multicolumn{6}{|c|}{Lattice-Scheme ($n_f=3$)} \\
\hline
  r in GeV$^{-1}$ & $c$ & $\frac{1}{r}\Omega$Exact & $\left| \frac{\Omega\text{LO}}{\Omega\text{Exact}}-1\right|\times 10^3$ & $\left| \frac{\Omega \text{NLO}}{\Omega\text{Exact}}-1\right|\times 10^4$ & $\left| \frac{\Omega \text{NNLO}}{\Omega\text{Exact}}-1\right|\times 10^{5}$ \\ \hline
 1.5 & 0.645661 & 0.22392 & 3.99452 & 1.80993 & 0.0121182 \\
% 1.4 & 0.744486 & 0.20490 & 5.08078 & 1.59025 & 0.399743 \\
% 1.3 & 0.850638 & 0.18486 & 5.84582 & 1.1929 & 0.803435 \\
 1.2 & 0.965291 & 0.16363 & 6.1928 & 0.589165 & 1.1845 \\
% 1.1 & 1.08993 & 0.14099 & 5.9731 & 0.25109 & 1.48041 \\
 1.0 & 1.2264 & 0.11668 & 4.9332 & 1.3597 & 1.5880 \\
% 0.9 & 1.37737 & 0.090317 & 2.5594 & 2.77683 & 1.32197 \\
 0.8 & 0.113682 & 0.30451 & 6.74004 & 0.327704 & 0.714606 \\
% 0.7 & 0.304952 & 0.26789 & 2.08274 & 1.00314 & 0.615626 \\
 0.6 & 0.525757 & 0.22732 & 1.87499 & 1.33636 & 0.251716 \\
% 0.5 & 0.786913 & 0.18134 & 4.62225 & 1.03572 & 0.353618 \\
% 0.45 & 0.937831 & 0.15563 & 5.23695 & 0.529505 & 0.683398 \\
 0.4 & 1.10654 & 0.12752 & 5.01592 & 0.28887 & 0.935909 \\
% 0.35 & 1.29781 & 0.096370 & 3.39336 & 1.48334 & 0.945347 \\
% 0.3 & 0.0862223 & 0.28680 & 6.44642 & 0.159698 & 0.451612 \\
% 0.25 & 0.347379 & 0.24130 & 1.02653 & 0.828621 & 0.359602 \\
 0.2 & 0.667008 & 0.18826 & 3.14974 & 0.950153 & 0.0488073 \\
% 0.15 & 1.07908 & 0.12333 & 4.4577 & 0.114258 & 0.601042 \\
 0.1 & 0.227474 & 0.24525 & 2.88066 & 0.447528 & 0.294791 \\
% 0.05 & 1.22033 & 0.095908 & 3.27279 & 0.593678 & 0.455968 \\
 0.01 & 0.660895 & 0.15997 & 2.22517 & 0.485566 & 0.0175813 \\
\hline
\end{tabular}
\caption{\label{Table:OmegaVnf3latt}
$1/r\Omega_V$ for $n_f=3$ in GeV units compared with \eq{eq:OmegaV} truncated at different powers of $\al$. Upper panel computed in the $\MS$ scheme. Lower panel in the lattice scheme. Lattice seems to be better but both schemes yield very good results.}
\end{table}

\begin{center}
\begin{figure}
\includegraphics[width=0.95\textwidth]{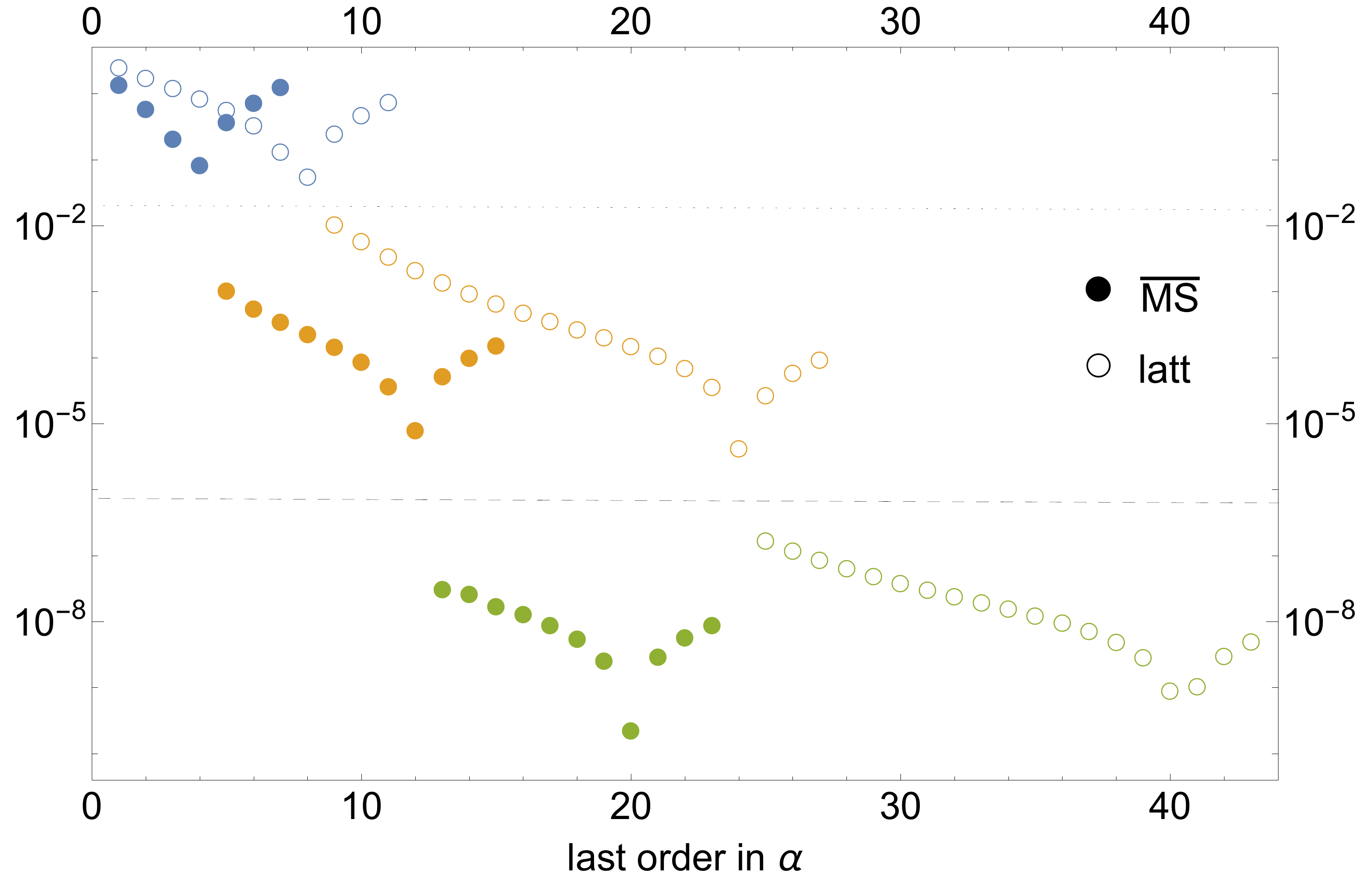}
\caption{\label{Fig:Boyd}
$|V_{\rm PV}-V_{\rm PV}^{\rm Hyperasymptotic}|$ for $r=0.1$ GeV$^{-1}$. Points above the horizontal dotted line are |$V_{\rm PV}-V_N$|. Points between the horizontal dotted and horizontal dashed lines are $|V_{\rm PV}-V_P-\frac{1}{r}\Omega_V-\sum_{n=N_P+1}^{N} (V_n-V_n^{(\rm as)}) \al^{n+1}|$ with $c=0.073$ 
%0.0734605
and 
%0.227474
$c=0.227$ (the smallest positive values that yield integer $N_P$) in the $\MS$ and lattice scheme respectively. Points below the horizontal dashed lines are $|V_{\rm PV}-V_P-\frac{1}{r}\Omega_V-\sum_{n=N_P+1}^{3N_P} (V_n-V_n^{(\rm as)}) \al^{n+1}-\frac{1}{r}\Omega'_V-\sum_{n=3N_P+1}^{N} (V_n-V_n^{(\rm as)}) \al^{n+1}|$, where in the last sum the two first renormalons are subtracted. Jumps correspond to the inclusion of $\Omega_V$ and  $\Omega_V'$. Full points have been computed in the $\MS$ scheme and empty points in the lattice scheme. We work with $n_f=3$.}
\end{figure}
\end{center}

An alternative, very effective, presentation of the above results can be done by plotting the relative accuracy of the prediction at each order in $\al$ and at each order of the superasymptotic expansion. We note that we have one observable for each value of $r$. Therefore, for illustration, we make the comparison with the observable for $r=0.1$ GeV$^{-1}$, and for the theoretical prediction we take the 
smallest positive value of $c$ corresponding to lattice or $\MS$. We show the results in Fig. \ref{Fig:Boyd}. We stress that several terms of the hyperasymptotic expansion are included. We can also see gaps each time the NP exponential terms are included. Indeed to reach the precision where $\Omega_V'$ is relevant, we used the exact (numerical) expression of $\Omega_V$, since the NNLO truncated expression is not precise enough.  We also nicely see that, once reached the minimum, both schemes yield similar precision, but in the lattice scheme (bigger factorization scale $\mu$) more terms of the perturbative expansions are needed to reach the same precision. One important lesson one may extrapolate from this exercise is that, for a fixed order computation, the smaller the renormalization scale $\mu$, the better. One can obtain much better precision for an equal number of perturbative coefficients. Another observation is that the minimal term determined numerically need not to coincide with the minimal term computed using $n=N_P$ (though it should not be much different). The difference reflects how much the exact coefficient is saturated by the asymptotic expression. 

\subsection{$(N,\mu) \rightarrow \infty$. \eq{eq:muinfty}. Case 2)}
\label{Sec:VPVmuinfty}

The potential advantage of this method is that we can obtain analytic results that are $\mu$ independent. We profit from earlier analyses in \cite{Sumino:2003yp,Sumino:2005cq} adapted to our case. In all cases the $q$ integrals will be done in the complex plane along similar lines as the computation done in those references. 

We first truncate the sum of the $\alpha_v$ coupling: 
\be
\alpha_{N}(q)\equiv \alpha\sum_{n=0}^{N} L^n=\alpha\frac{1-L^{N+1}}{1-L}\ .
\label{alsN}
\ee
Following \cite{Sumino:2003yp,Sumino:2005cq} we can isolate the $N$-dependence from the leading contribution to the potential at short distances:
\be
V_N(r)=-\frac{2C_F\alpha}{\pi}\int_0^\infty dq \frac{\sin qr}{qr}\frac{1-L^{N+1}}{1-L} 
\equiv\frac{4C_F}{\beta_0}\tilde \Lambda \left[v_1(\tilde{\Lambda} r)+v_2(\tilde{\Lambda} r,N+1) \right]\, ,
\label{VrN}
\ee
where
\be
v_1=\frac{1}{r\tilde{\Lambda}}\int_0^{\infty}dx\,e^{-x}\arctan(\frac{\pi}{2\ln(\frac{r\tilde{\Lambda}}{x})})
\,,
\ee
$\arctan(x)$ is defined in the branch $[0,\pi)$, and
\begin{equation}
v_2=-\frac{\pi}{\rho}\cos \rho-\int_{0,\rm PV}^{\infty}dk\,\frac{\sin k\rho}{k\rho}\frac{1}{\ln1/k}\bigg[1+\frac{1}{N+1}\ln\frac{1}{k}\bigg]^{N+1}
\,.
\end{equation}
We then have that
\be
V_{\rm PV}-V_{N}=\frac{4C_F\tilde{\Lambda}}{\beta_0}\bigg(\frac{-\pi}{\rho}\cos\rho-v_2\bigg)
\,.
\ee
Note that this equality allows us to write $V_{\rm PV}$ in the following way ($v_C=v_1(\rho)-\frac{\pi}{\rho}$ with the notation of  \cite{Sumino:2003yp}):
\be
\label{VPVOPES}
V_{\rm PV}=\frac{4C_F\tilde{\Lambda}}{\beta_0}\bigg(v_C-\frac{\pi}{\rho}(\cos\rho-1)\bigg)
\,.
\ee
In this explicit representation of $V_{\rm PV}$  each term scales differently in powers of $\rho$: ${\cal O}(v_C) \sim \rho^{-1}$, the $\rho^0$ term is set to zero (or incorporated in $v_C$), and each ${\cal O}(\rho^{2n+1})$ term is encoded in $\frac{\pi}{\rho}(\cos\rho-1)$. Still, \eq{VPVOPES} can not be understood as an explicit representation of the OPE, since the NP power corrections scale with odd powers of $\rho$, and indeed there are no ${\cal O}(\rho^{2n})$ terms. However, this splitting naturally leads to define a short distance coupling:
\be
\al_{SD}(1/r)=-r \frac{4}{\beta_0}\tilde \Lambda v_C(r)
\,.
\ee
This definition has nice properties. It is an smooth function $\forall$ $r \in (0,\infty)$, with the right short distance limit:
\be
\al_{SD}(1/r)=\frac{2\pi}{\beta_0}\frac{1}{\ln (\rho r)} \qquad r \rightarrow 0
\,.
\ee
A detailed study of this quantity can be found in \cite{Sumino:2005cq}. Note also that in this definition the whole ${\cal O}(\lQ)$ correction has been included in $\frac{4C_F}{\beta_0}\tilde \Lambda v_C(r)$. The other thing that one could study, since we have the analytic behavior, is the behavior of $\al_{SD}$ beyond the regime where it was originally defined, i.e. at long distances. In this respect, it is interesting to notice that the long distance limit
\be
\al_{SD}(0)=\frac{4\pi}{\beta_0}
\ee
is exactly equal to the value obtained in \cite{Shirkov:1997wi}, within the context of analytic perturbation theory analyses. Nevertheless, one could as well argue that all ${\cal O}(\rho^{2n+1})$ terms are short distances and should be incorporated in $\al_{SD}$. If one does so, $\al_{SD}$ does not have an smooth limit for $\rho \rightarrow 0$ anymore. 
Finally, one could also study the $\beta$ function of $\al_{SD}$. 

\begin{center}
\begin{figure}
\includegraphics[width=0.74\textwidth]{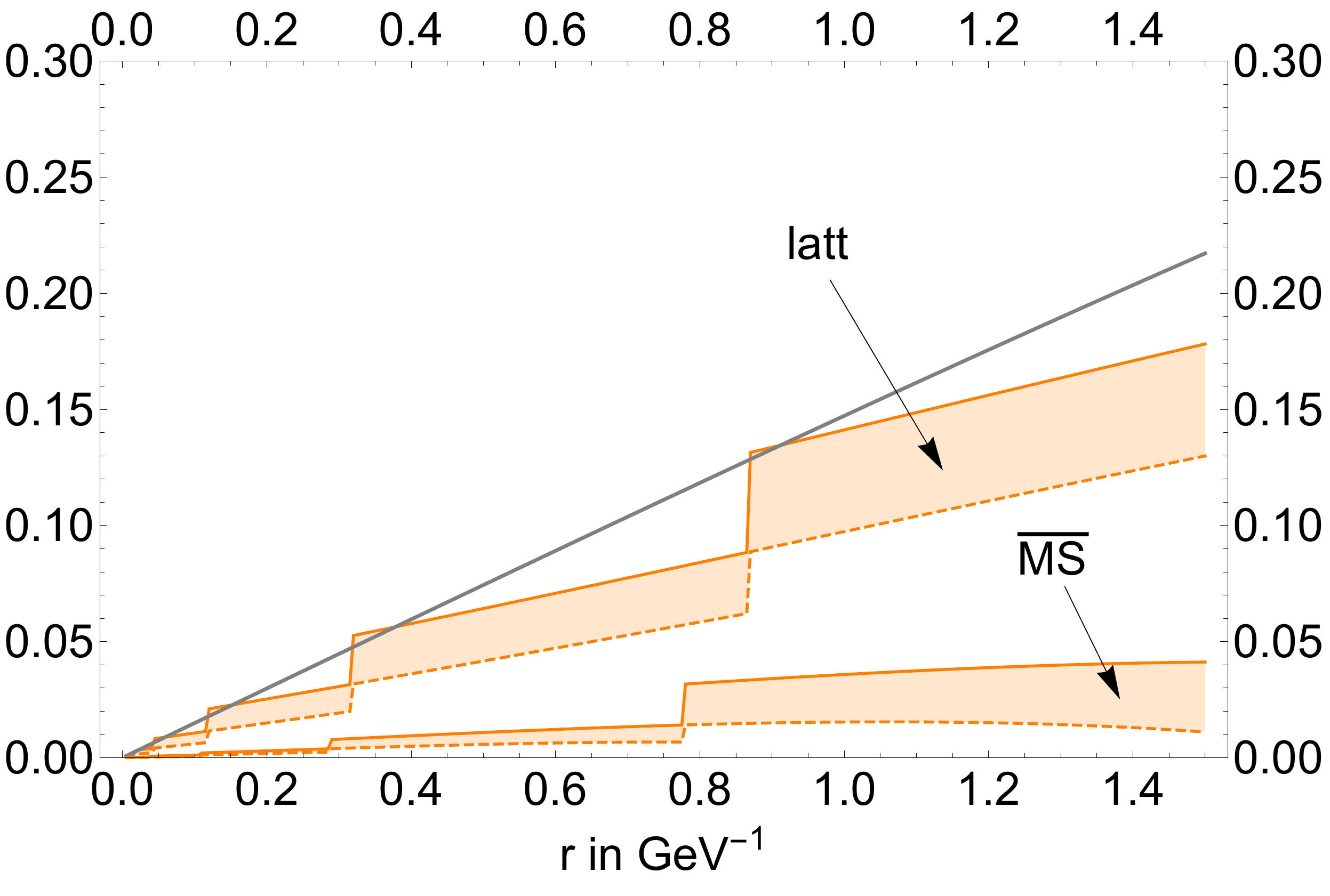}
%\put(0,20){{\large (MS)}}
%\put(0,78){{\large (latt)}}
\vspace{0.06in}
\includegraphics[width=0.772\textwidth]{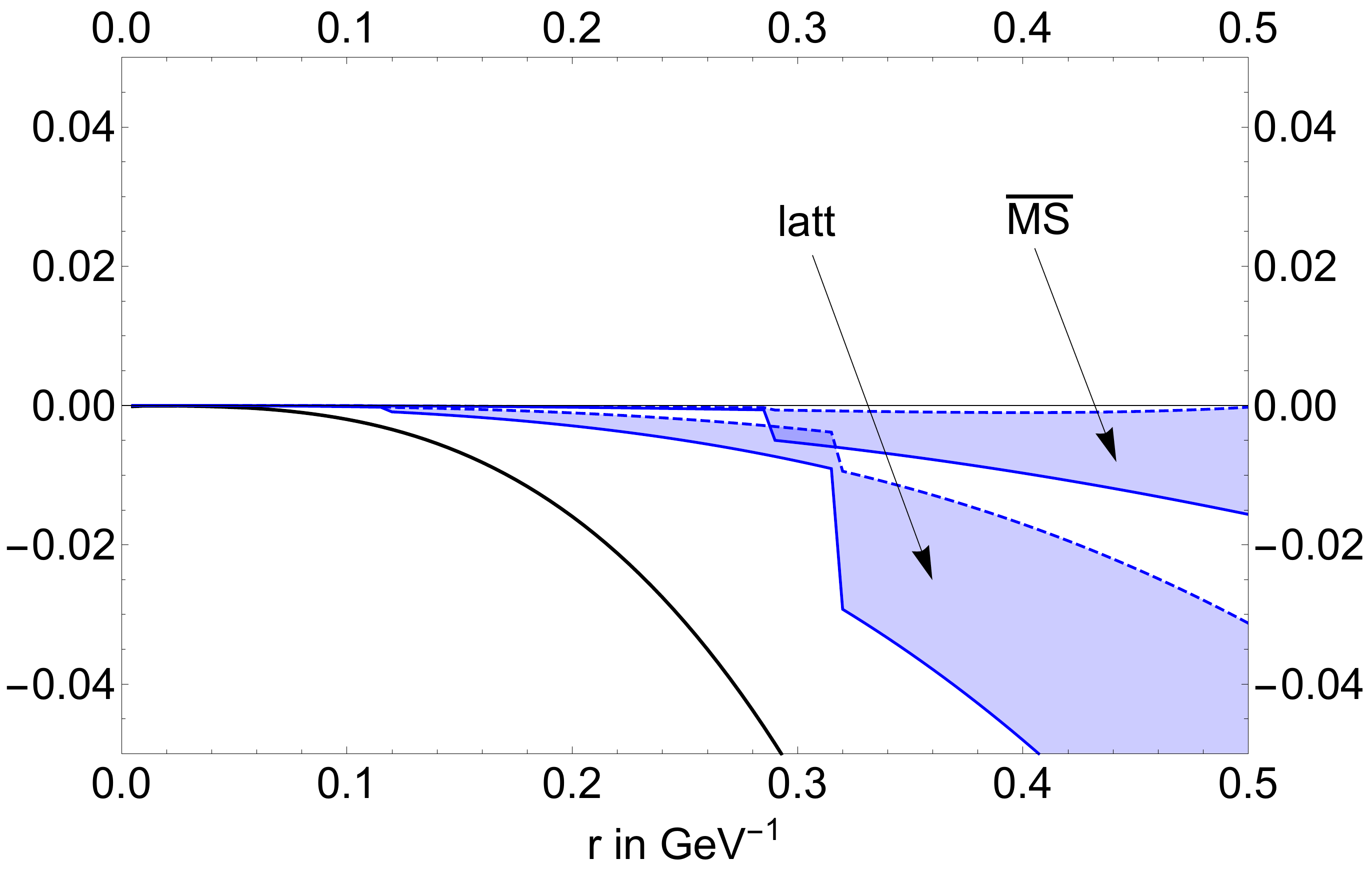}
\put(-385,150){{\large MeV}}
%\put(0,100){{\large (MS)}}
\caption{\label{Fig:Hyper} {\bf Upper panel}: $o(\lQ)$ precision figure. $V_{\rm PV}-\frac{4C_F\tilde{\Lambda}}{\beta_0}v_C$ (black line), and 
$V_{\rm PV}-V_P-\frac{1}{r}\Omega_V$ (orange bands)
in the lattice and $\MS$ scheme with $n_f=3$ (as drawn in Fig. \ref{Fig:VPVb0nf3lattvsMS}). {\bf Lower panel}: $o(\lQ^3r^2)$ precision figure. 
$V_{\rm PV}-\frac{4C_F\tilde{\Lambda}}{\beta_0}(v_C+\pi\rho/2)$ (black line), and 
$V_{\rm PV}-V_P-\frac{1}{r}\Omega_V-\sum_{n=N_P+1}^{3N_P} (V_n-V_n^{(\rm as)}) \al^{n+1}-\frac{1}{r}\Omega'_V$ (blue bands)
in the lattice and $\MS$ scheme with $n_f=3$ (as drawn in Fig. \ref{Fig:VPVb0nf3lattvsMS}). Note that in this last figure the vertical axis is in MeV and the precision is at the level of 10$^{-2}$ MeV!.}
\end{figure}
\end{center}

It has some interest to compare \eq{Vb0PV1}, the hyperasymptotic expansion using method 1), with \eq{VPVOPES}. We can make the comparison at $o(\lQ)$ and at  $o(\lQ^3 r^2)$ in the hyperasymptotic expansion. We show such comparison in Fig. \ref{Fig:Hyper}. At $o(\lQ)$, the leading power correction in \eq{VPVOPES} is of ${\cal O}(\rho)$. We find that \eq{Vb0PV1} is more convergent, which is consistent with the estimated made in \eq{estimate}. Either way, the convergence is extremely good. The precision is much below the MeV. 

In real life we will not have such complete analytic control and must rely on the methods discussed in \Sec{Sec:Scheme}. Therefore, we now apply the limit 2A) and 2B) discussed in \eq{eq:muinfty} to $V_N$.

\subsubsection{Case 2A)}
\label{Sec:V2A}

We now take
\be
\label{NlimitSpot}
N+1=\frac{2 \pi}{\beta_0 \al(\mu)}
\,.
\ee
The large $N$ limit of $v_2$ yields 
\bea
v_2&=&\frac{-\pi}{\rho}+\int_{0}^{\infty}dx\,\frac{e^{-x}-1+x\theta(1-x)}{x^2}\frac{\ln\frac{\rho}{x}}{\ln^2\frac{\rho}{x}+\pi^2/4}
\nn
\\
&&
-\frac{1}{2}(-\gamma_E+\ln2+\ln (N+1))+\frac{1}{2}\ln(\ln^2\rho+\frac{\pi^2}{4})
\eea
up to terms that vanish when $N \rightarrow \infty$. Note that  the 
$N \rightarrow \infty$ limit of $v_2$ (logarithmically) diverges. 
Note also that when $\rho\to 0$ the integral term tends to zero. Thus, the $\rho\sim 0$ limit of $v_2$ is
\begin{equation}
v_2=\frac{-\pi}{\rho}-\frac{1}{2}(-\gamma_E+\ln2+\ln (N+1))+\ln\ln\frac{1}{\rho}\qquad\rho\sim0
\,.
\end{equation}

The difference between the PV and the truncated series can be computed by  complex variable integration following similar lines as in \cite{Sumino:2003yp,Sumino:2005cq}. We find (for large $N$)
\bea
\nn
V_{\rm PV}-V_{N}&=&\frac{4C_F\tilde{\Lambda}}{\beta_0}\bigg(\frac{\pi}{\rho}\big(1-\cos(\rho)\big)-\int_0^{\infty}dx\,\frac{e^{-x}-1+\theta(1-x)x}{x^2}\frac{\ln(\frac{\rho}{x})}{\ln^2(\frac{\rho}{x})+\frac{\pi^2}{4}}
\\
&&
-\frac{1}{2}\ln(\ln^2(\rho)+\frac{\pi^2}{4})+\frac{1}{2}(-\gamma_E+\ln 2+\ln (N+1))\bigg)+o(1/N).
\eea

For large values of $N$ and small values of $r$ (care should be taken when taking the $r \rightarrow 0$ limit) the above expression simplifies to
\be
V_{\rm PV}-V_{N}=\frac{4C_F\tilde{\Lambda}}{\beta_0}\bigg(-\ln\ln(\frac{1}{\rho})+\frac{1}{2}(-\gamma_E+\ln 2+\ln (N+1))\bigg)+o(1/N,r)
\,.
\ee
For completeness, we have also obtained the $\ln(N)$ behavior in a different way. We follow the method recently proposed in \cite{Mishima:2016vna}. There, a summation integral relation was found for a general observable. We applied it to the case of the first IR renormalon of the potential and pole mass. The advantage of this new method is that the $\ln N$ term can be determined if the normalization of the leading renormalon in the Borel plane is known. It would be very interesting to try to generalize this result beyond the large $\beta_0$ approximation, as well as to extend the analysis to the $\ln\ln(\frac{1}{\rho})$ term. 

The fact that we have certain analytic control of the result allows us to address some issues. The first one is to make explicit that truncated sums around the minimal term do not guarantee, per se, that they are finite. In particular, one can see that $V_N$ is divergent in the $N \rightarrow \infty$. Therefore, it would be wrong to assign $V_N$ to the leading term in the hyperasymptotic expansion of $V_{\rm PV}$. On the other hand, we have analytic control on the divergence, which is found to be logarithmic in $N$.\footnote{It is worth mentioning again that this $\ln N$ behavior also appears beyond the large $\beta_0$ approximation in the context of the static potential \cite{Sumino:2005cq}.} In principle, one can subtract this $\ln N$ divergence from $V_N$ (this is completely analogous to subtracting $1/\epsilon$ divergences in perturbative computations using dimensional regularization) to obtain the first term of the hyperasymptotic expansion. Nevertheless, the difference does not still scale like $\lQ$. Instead one has
\be
V_{\rm PV}-[V_{N}+
\frac{4C_F\tilde{\Lambda}}{\beta_0}\ln (N+1)]
=\frac{4C_F\tilde{\Lambda}}{\beta_0}\bigg(-\ln\ln(\frac{1}{\rho})+\frac{1}{2}(-\gamma_E+\ln2)\bigg)+o(1/N,r)
\,,
\ee
which, at short distances, scales as $\lQ\ln\ln(\frac{1}{\rho})$ (this behavior is also seen beyond the large $\beta_0$ approximation in the context of the static potential \cite{Sumino:2005cq}). Therefore, to get  the proper scaling in $\lQ$ of the different terms of the hyperasymptotic expansion requires that the $\lQ\ln\ln(\frac{1}{\rho})$ should be identified and subtracted first from $V_N$. One then has the freedom to subtract ${\cal O}(\lQ)$ finite pieces, which can be absorbed in the next term of the hyperasymptotic expansion. 

We do not do a numerical analysis here, as the method cannot, at present, be generalised beyond the large $\beta_0$ approximation.
 
\subsubsection{Case 2B)}

We now take
\be
\label{NlimitApot}
N+1=\frac{2 \pi}{\beta_0 \al(\mu)}\left(s-1\right)
\qquad {\rm with}\;\; s<2
\,.
\ee
Under these conditions, we can take the $N \rightarrow \infty$ limit (the result does not diverge in this limit). 
Adapting \cite{Sumino:2005cq} derivation to our case we obtain 
\be
\lim_{N\to\infty}v_{2}\equiv v_3=-\frac{\pi}{\rho}-\rho^{s-2}\int_0^{\infty}dx\,\frac{e^{-x}-1}{x^s}\frac{\frac{\pi}{2}\cos(\frac{\pi}{2}[1-s])+\ln\frac{\rho}{x}\sin(\frac{\pi}{2}[1-s])}{\ln^2\frac{\rho}{x}+\frac{\pi^2}{4}}
\,.
\ee
Therefore, we define (using the relation \eq{NlimitApot})\footnote{Since the result we obtain is finite, we could as well taken $N+1 \rightarrow N=N_A$ in \eq{NlimitApot}, and the result does not change. In other words, $V_A^{(\beta_0)}$ does not depend on adding or subtracting an extra term to the sum. This is a pleasant property.} 
\be
V_A \equiv \lim_{N \rightarrow \infty} V_N =v_1+v_3
\,.
\ee
Note that this far, the expressions for $v_1$ and $v_3$ are valid $\forall$ $r$. It is also possible, and most relevant for us, to relate the truncated sum (in the limit $\mu \rightarrow \infty$) with 
 $V_{\rm PV}$. We obtain
\bea
\label{PVNAcs}
&&
V_{\rm PV}-V_{A}=
\\
\nn
&&=\frac{4C_F\tilde{\Lambda}}{\beta_{0}}\bigg(\frac{\pi}{\rho}[1-\cos(\rho)]+(\rho)^{s-2}\int_0^{\infty}dx\,\frac{e^{-x}-1}{x^s}\frac{\pi/2\cos(\frac{\pi}{2}[1-s])+\ln\frac{\rho}{x}\sin(\frac{\pi}{2}[1-s])}{\ln^2\frac{\rho}{x}+\frac{\pi^2}{4}}\bigg)
\,.
\eea
Again this result is valid $\forall$ $r$. We now focus on the 
$\rho \rightarrow 0$ limit. This will allows us to connect with  the limit 2B) of \eq{eq:muinfty}. Nevertheless, this connection has to be done with care. One has to take the limit $r \rightarrow 0$ and $s \rightarrow 2$ in a correlated way,  following the limit 2B) of \eq{eq:muinfty}. 
Therefore, we take
\be
s=2-c'\al(1/r) 
\,.
\ee
Then, the previous expression reads
\bea
\label{PVAc}
&&
V_{\rm PV}-V_{A}
=\frac{4C_F\tilde{\Lambda}}{\beta_0}\bigg(\frac{\pi}{\tilde{\Lambda}r}\big(1-\cos(\tilde{\Lambda}r)\big)
\\
\nn
&&
+(\tilde{\Lambda}r)^{-c'\alpha(1/r)}\int_0^{\infty}dx\,\frac{e^{-x}-1}{x^{2-c\alpha(1/r)}}\frac{\frac{\pi}{2}\cos(\frac{\pi}{2}(-1+c'\alpha(1/r)))+\ln(\frac{\tilde{\Lambda}r}{x})\sin(\frac{\pi}{2}(-1+c'\alpha(1/r)))}{\ln^2(\frac{\tilde{\Lambda}r}{x})+\frac{\pi^2}{4}}\bigg)
\,.
\eea
We can now obtain the $\rho \rightarrow 0$ limit:
\be
\label{PVAcSD}
V_{\rm PV}-V_{A}=\frac{-4C_F\tilde{\Lambda}}{\beta_0}Ei\big(\frac{2\pi c'}{\beta_0}\big)
+o(r)
\,,
\ee
where, for $x\in\mathbb{R}$,
\be
Ei(x)=-\int_{-x,\rm PV}^{\infty}dt\,\frac{e^{-t}}{t}
\,.
\ee
Nicely enough \eq{PVAcSD} agrees with the prediction of \eq{PVA} applied to $V_{\rm PV}$. 

For future reference, we are also interested in the next correction in powers of $\al(1/r)$ of Eq. (\ref{PVAcSD}). We obtain
\be
\label{PVAcSD2}
V_{\rm PV}-V_{A}=\frac{-4C_F\tilde{\Lambda}}{\beta_0}
\left(
Ei\big(\frac{2\pi c'}{\beta_0}\big) 
-
e^{\frac{2\pi c'}{\beta_0}}\frac{\beta_0}{12\pi}(6\gamma_E-1)\al(1/r)
+O(\al^2(1/r))\right)
\,.
\ee
Note though that \eq{PVA} cannot predict the ${\cal O}(\al)$ correction. 

We have already emphasized that obtaining the $\rho \rightarrow 0$ limit was delicate. Let us illustrate this. If we take the $\rho \rightarrow 0$ limit with $s$ fixed (but close to 2), such that  $s<2$ , we obtain 
\bea
v_3&=&-\frac{\pi}{\rho}-\rho^{s-2}\int_0^{\infty}dx\,\frac{e^{-x}-1}{x^s}\frac{\ln\rho\sin(\frac{\pi}{2}[1-s])}{\ln^2\rho}
\\
\nn
&=&-\frac{\pi}{\rho}-\frac{\rho^{s-2}\sin(\frac{\pi}{2}[1-s])}{\ln\rho}\int_0^{\infty}dx\,\frac{e^{-x}-1}{x^s}
\\
\nn
&=&-\frac{\pi}{\rho}-\frac{\rho^{s-2}\sin(\frac{\pi}{2}[1-s])}{\ln\rho}\Gamma(1-s)
\,,
\eea
up to contributions that vanish when $\rho \rightarrow 0$. If we now take 
$s=2-c'\alpha(1/r)$ and take again $\rho\to0$ we obtain
\be
\label{VPVVNA2}
\lim_{s \rightarrow 2}\lim_{\rho \rightarrow 0} \rm{\eq{PVNAcs}}
=\frac{-2C_F\tilde{\Lambda}}{\pi c'}e^{\frac{2c'\pi}{\beta_0}}
\,,
\ee
which is obviously different that \eq{PVAcSD}. In short
\be
\lim_{s \rightarrow 2}\lim_{\rho \rightarrow 0} \rm{\eq{PVNAcs}}
\not=
\lim_{s \rightarrow 2  \& \rho \rightarrow 0(\rm correlated)}  \rm{\eq{PVNAcs}}
\,.
\ee

If we rephrase this discussion in terms of the $c'$ behavior, what we have is that  Eq. (\ref{PVAcSD}) is not obtained by taking the limit $c' \rightarrow 0$ before taking the limit $r \rightarrow 0$ of Eq. (\ref{PVAc}). Indeed, the limit $c' \rightarrow 0$ before taking the limit $r \rightarrow 0$ produces \eq{VPVVNA2}, which does not correspond to the limit 2B) we are following in this paper. As we can see from the explicit computation, both limits yield NP power corrections with the right scaling (pointing out that there is not unique procedure to get/define the NP correction). Nevertheless, the overall coefficient is different, whereas \eq{PVAcSD} diverges logarithmically in $c'$, \eq{VPVVNA2} diverges like $1/c'$ for small $c'$. In this paper we stick to method 2B) as it allows us to go beyond the large $\beta_0$ approximation and to relate the normalization of the power correction with the normalization of the renormalon.  

\begin{center}
\begin{figure}
\includegraphics[width=0.76\textwidth]{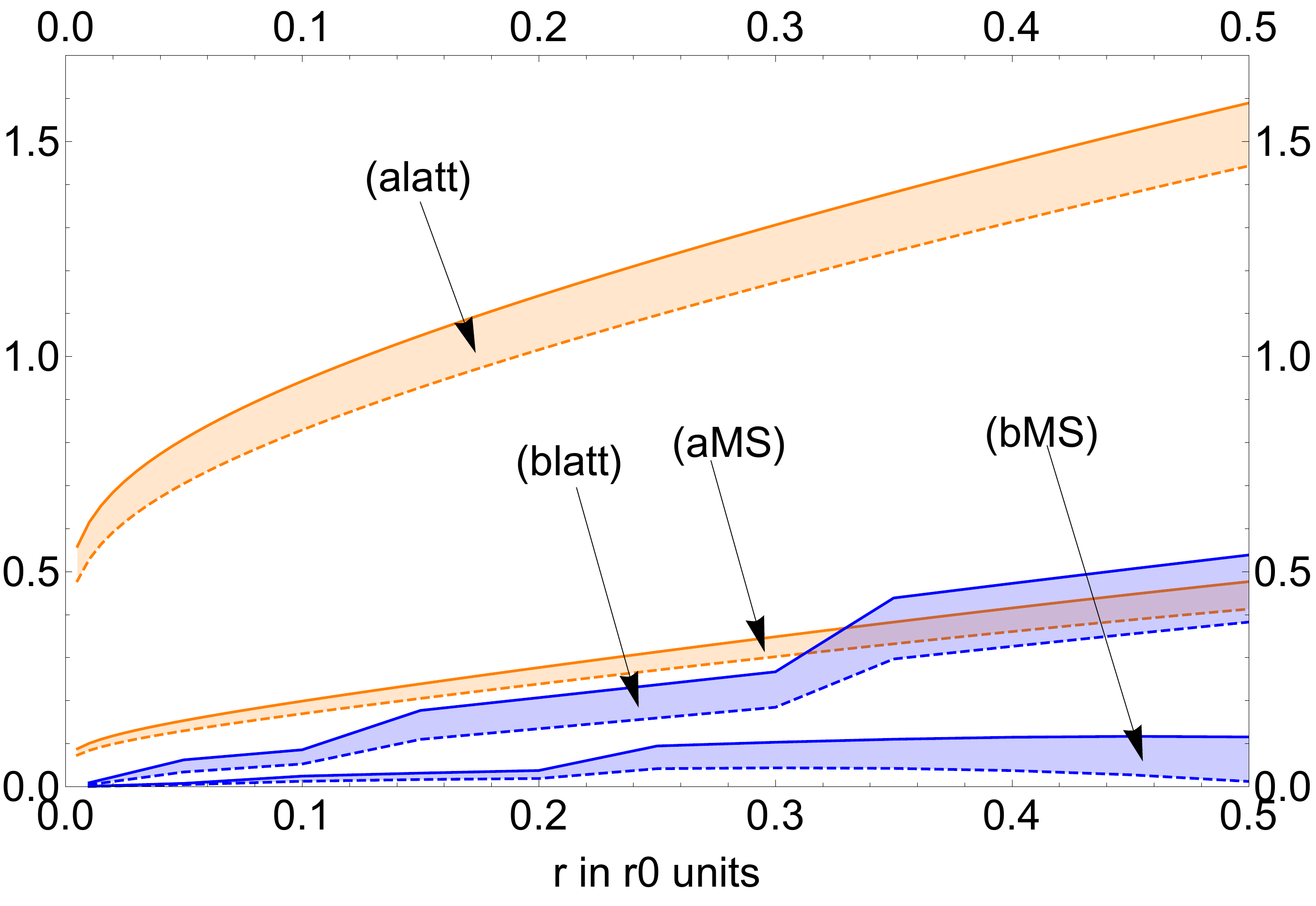}
%\put(0,195){{\large (alatt)}}
%\put(-200,25){{\large (blatt)}}
%\put(0,15){{\large (bMS)}}
%\put(0,63){{\large (aMS)}}
\caption{We plot (a) $V_{\rm PV}-V_A-K_X^{(A)}\Lambda_X$ for $n_f=0$ in the lattice and $\MS$ scheme. For each case, we generate bands by computing $V_A$ with $c'=1$ and $c'=c'_{\rm min}$. We also compare with (b) $V_{\rm PV}-V_P-\frac{1}{r}\Omega_V$ obtained with method 1) with the bands generated for Fig. \ref{Fig:VPVb0nf0lattvsMS}.}
\label{Fig:BothmethodsVPVnf0}
\end{figure}
\end{center}

\begin{center}
\begin{figure}
\includegraphics[width=0.755\textwidth]{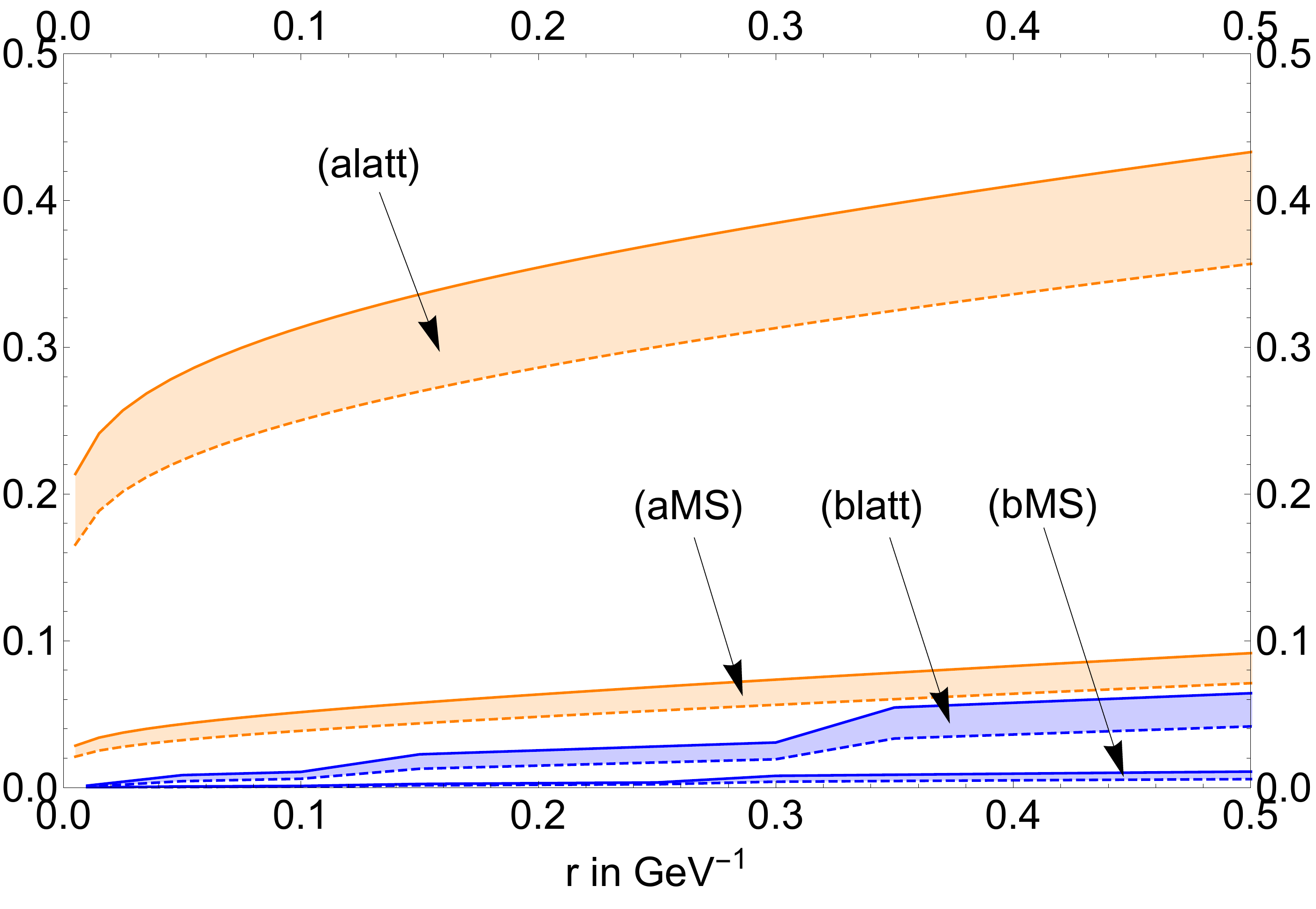}
%\put(0,180){{\large (alatt)}}
%\put(0,30){{\large (blatt)}}
%\put(0,15){{\large (bMS)}}
%\put(0,43){{\large (aMS)}}
\caption{We plot (a) $V_{\rm PV}-V_A-K_X^{(A)}\Lambda_X$ for $n_f=3$ in the lattice and $\MS$ scheme. For each case, we generate bands by computing $V_A$ with $c'=1$ and $c'=c'_{\rm min}$. We also compare with 
(b) $V_{\rm PV}-V_P-\frac{1}{r}\Omega_V$ obtained with method 1) with the bands generated for Fig. 
\ref{Fig:VPVb0nf3lattvsMS}.}
\label{Fig:BothmethodsVPVnf3}
\end{figure}
\end{center}

\begin{center}
\begin{figure}
\includegraphics[width=0.70\textwidth]{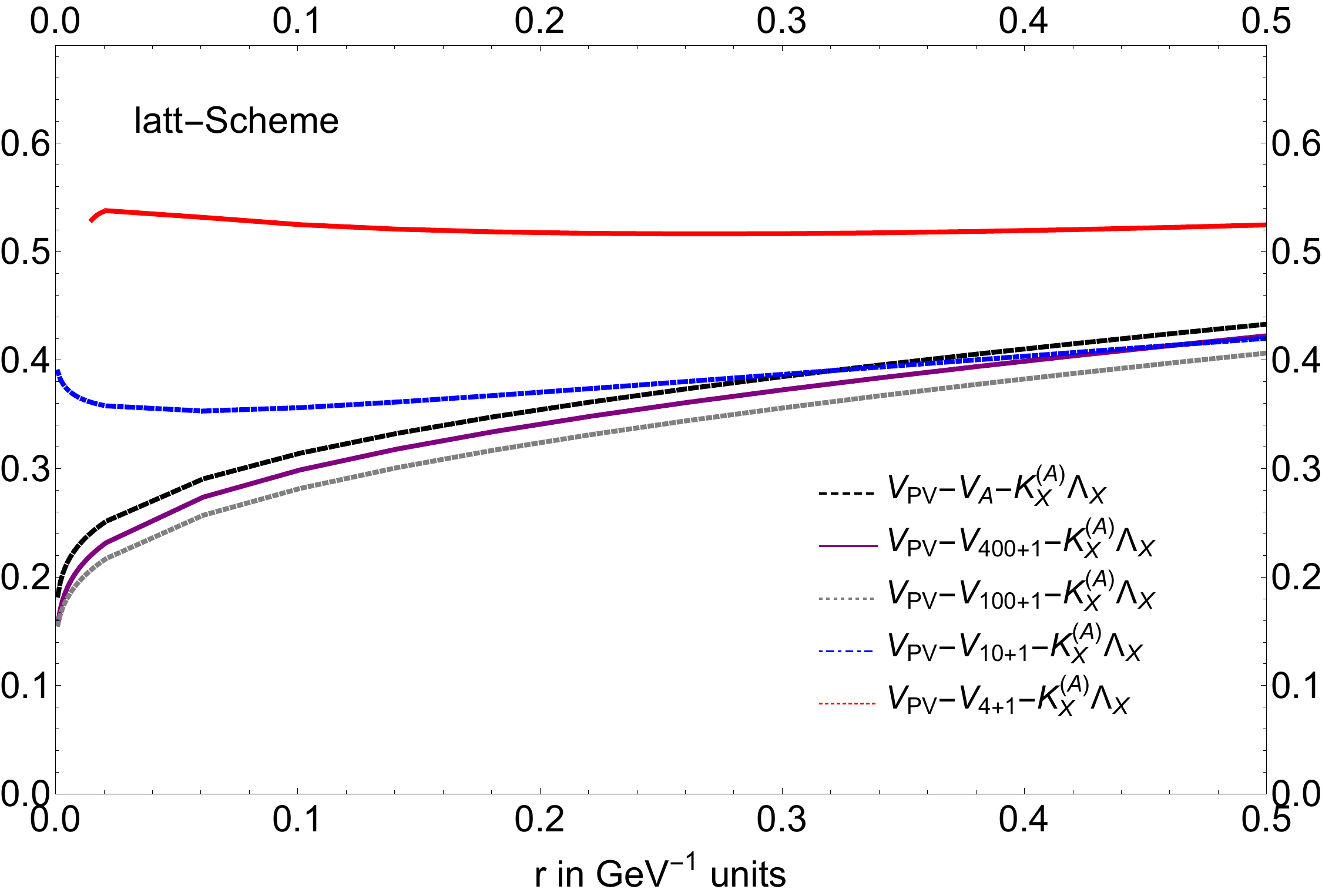}
\includegraphics[width=0.70\textwidth]{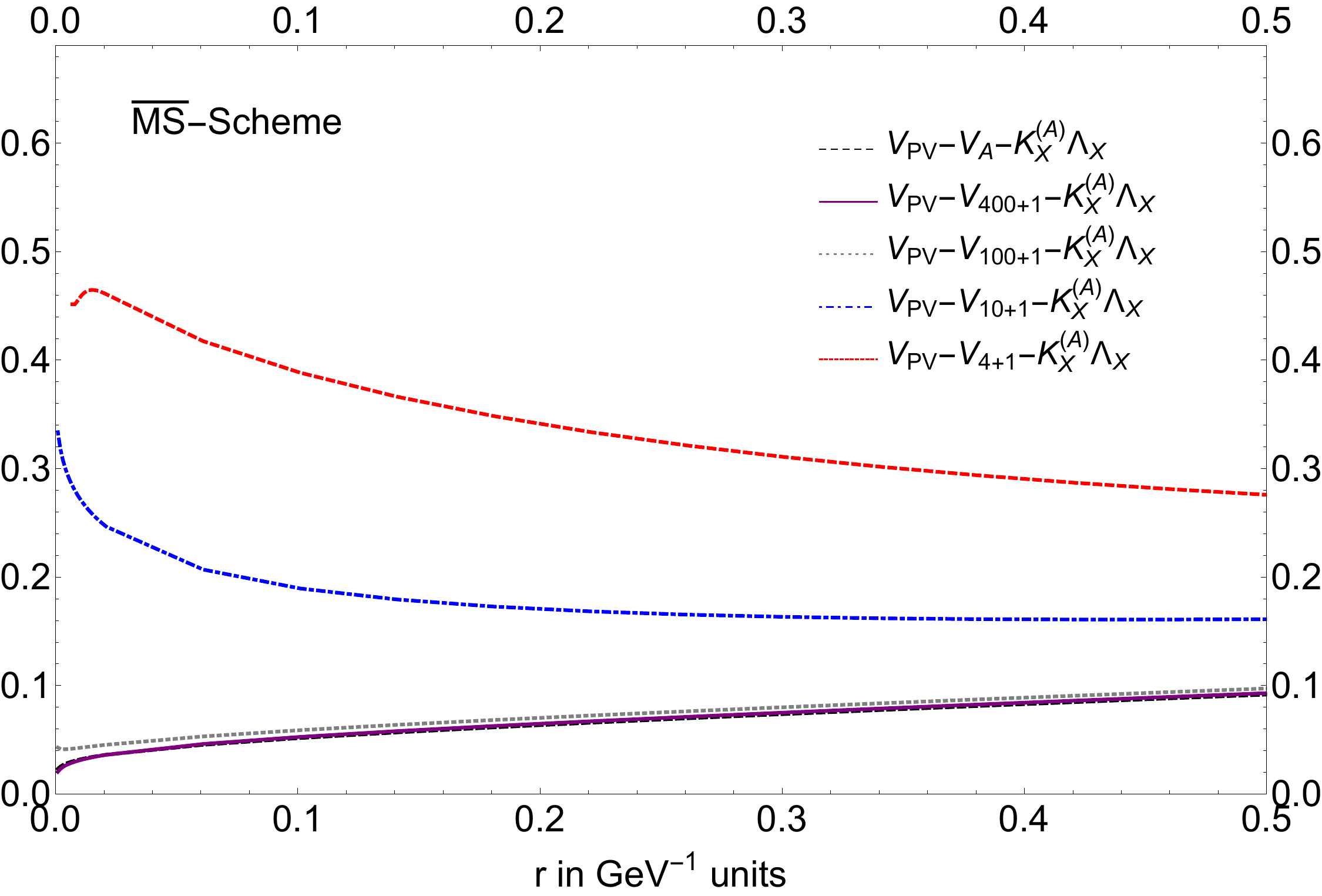}
\caption{{\bf Upper panel}: We plot $V_{\rm PV}-V_A-K_X^{(A)}\Lambda_X$ for $n_f=3$ in the lattice scheme with $c'=1$ versus the truncated sums $V_{\rm PV}-\sum_{n=0}^{N_A}V_n\al^{n+1}(\mu)-K_X^{(A)}\Lambda_X$, where $\mu$ is fixed using $N_A$ defined in \eq{eq:muinfty}. {\bf Lower panel}: As in the upper panel but in the $\MS$ scheme.}
\label{Fig:Convergencec1nf3}
\end{figure}
\end{center}

Finally, note that this method has the pleasant feature that the generated ${\cal O}(\lQ)$ correction complies with the OPE. It also yields results that do not depend on $N$ (and $\mu$) anymore. Still, it has some errors and does not reach the precision of method 1). There is a residual scheme dependence associated with uncomputed terms of ${\cal O}(\al \lQ)$. Part of it can be estimated by the residual dependence in $c'$. In order to estimate it, we compute $V_A$ for different values of $c'$. On the one hand $c'$ cannot be very large, as $c'\al(1/r)$ should be relatively close to zero.  On the other hand we cannot make $c'\al(1/r)$ to get arbitrary close to zero, as the ${\cal O}(\lQ)$ correction diverges logarithmically in $c'$. We also note that there is a value of $c'=c'_{\rm min}$ that makes that $K_X^{(A)}=0$ so that the 
${\cal O}(\lQ)$ correction vanishes. Therefore, we compute $V_A$ for different values of $c'$. For illustration we show some results in Figs. \ref{Fig:BothmethodsVPVnf0} and \ref{Fig:BothmethodsVPVnf3}. We draw lines for $V_{\rm PV}-V_A-K_X^{(A)}\Lambda_X$ at $c'=1$ and $c'=c_{\rm min}$ generating a band. We also explore the dependence on the scheme by comparing the results in the lattice and $\MS$ scheme.  We stress again that in the large $\beta_0$ approximation lattice and $\MS$ schemes just correspond to a redefinition of $\mu$, but quite large indeed. On the other hand the final result is $\mu$ independent. Nevertheless, the way the $\mu \rightarrow \infty$ limit is taken is fixed by $N_A$, as defined in \eq{eq:muinfty}, which is dependent on $\mu$. This explains why different results are obtained. 

In Figs. \ref{Fig:BothmethodsVPVnf0} and \ref{Fig:BothmethodsVPVnf3}, we also compare with results obtained using method 1), more specifically we compare with $V_{\rm PV}-V_P-\frac{1}{r}\Omega_V$, as they both have analogous power accuracy (though method 1) is  parametrically more precise). For $\Omega_V$ we take the exact expression but using its approximated expression does not change the discussion, as the difference is very small. What we see is that the $\MS$ scheme yields more precise predictions than the lattice scheme, and that method 1) yields considerable better results than method 2B). 

Another issue specific to method 2B) is to determine how large we need to take $N$ (and consequently $\mu$) of the truncated sum such that it approximates well $V_A$. For illustrative purposes we show the convergence in Fig. \ref{Fig:Convergencec1nf3} for $n_f=3$ in the lattice and $\MS$ scheme. We find that we have to go to relatively large values of $\mu$ (and $N$) to get it precise. This can be a problem if one wants to go beyond the large 
$\beta_0$. This problem would be less severe if one can use the asymptotic expression for the coefficients beyond certain $n$. Nicely enough, we find that the use of the asymptotic expression for the coefficients for $n>N^*$ ($\sim 3$ in the $\MS$ and $\sim$ 8 in the lattice scheme) is very efficient and basically yields the same results as the exact result.   Finally, we also recall that to approximate well $V_A$ by the truncated sum is more costly for small values of $c'$. 

\section{Conclusion}

We aim to accurately describe observables characterized by having a large scale $Q \gg \lQ$. For those it is believed that the OPE is a good approximation (we do not enter in this paper on the issue of duality violations). We want to make the most of available perturbative expressions of the observable. Our aim is to organize the computation and its associated accuracy within a hyperasymptotic expansion. For this, we carefully study the connection between truncated sums of the perturbative expansions in powers of $\al$ and the associated NP corrections. In practice, we relate those truncated sums with the Borel sum of the perturbative series regulated using the PV prescription. This object has the nice properties of being scale and scheme independent. It may also open the window to connect with studies directly aiming to the NP regime. We then hypothesize that the difference between the Borel sum and the full NP evaluation of the observable complies with the structure of the NP OPE (at least for the first terms of the NP power expansion). Such computational scheme allows us to get a hyperasymptotic expansion of the observable, and, consequently, to unambiguously state the magnitude of the different terms of the hyperasymptotic expansion. 
 
Relating truncated sums of the perturbative expansion with NP definitions of them is not trivial in general. However, this is possible for the case of the PV prescription. We have studied two methods that achieve this goal and explored how reliable they are in practice. We have given analytic formulas (with exponential accuracy) that relate the truncated sum with the PV-regulated Borel sum. We emphasize that these formulas are valid beyond the large-$\beta_0$ approximation.  

These methods allow us to efficiently disentangle the pure perturbative term from the first NP corrections of an arbitrary observable that admits an OPE at large energies. General expressions for arbitrary observables are given (for this paper we neglect ultraviolet renormalons). Nevertheless, the accuracy we achieve for each case is different:
\begin{itemize}
\item
 The method 2B)  (see \eq{Observable2B}) has the handicap that (in principle) needs the perturbative expansion of the observable and the running of $\alpha$ to all orders. On top of that we are only able to obtain the ${\cal O}(e^{-\frac{2\pi d}{\beta_0 \alpha(Q)}}\alpha^{-\frac{d\beta_1}{2\beta_0}}(Q))$ term of the Borel sum, which then sets the precision of the analysis. On the other hand, it has the nice feature that the leading NP power correction of the Borel sum has exactly the same scaling as the NP corrections dictated by the OPE, and that the result is explicitly $\mu$ independent. 
\item
On the other hand, method 1) (see \eq{ObservableA}) shows to be much more powerful. At low orders it is just standard perturbation theory. At high orders (quantified by $N_P$) the series is truncated. This corresponds to the superasymptotic approximation. We can quantify the error committed in summations truncated at the minimal term and state the independence of the result on the scale and scheme used for the perturbative expansion to a given accuracy. This allows us to state the parametric accuracy of determinations of genuine NP power corrections obtained by subtracting the perturbative series from the full observable (the latter being obtained either from lattice simulations or directly from experiment).\\
We then incorporate the NP corrections to the truncated sum associated with the renormalons using the PV regularization prescription. The procedure uses the theory of terminants discussed in \cite{Dingle}. The scale and scheme dependence of this merging is under control in the whole process.  This process is, in principle, systematically improvable. Subleading power corrections can be incorporated in the analysis, reaching  hyperasymptotic accuracy. This analysis also allows us to visualize that truncating the perturbative sum at the minimal term produces, in general, terms that cannot be absorbed in the NP terms of the OPE, because of prefactors proportional to $\sqrt{\al}$. Overall, one obtains an smooth connection between the standard (pure) perturbative computation and the OPE (hyperasymptotic) expansion that includes the NP power corrections.
\end{itemize}

With these methods it is possible to determine the leading difference between the perturbative series truncated at the minimal term with the Borel integral regulated using the PV prescription in terms of the closest singularity to the origin of the Borel transform. This is very good because it allows us to determine such leading NP correction in terms of the normalization of the leading renormalon, $Z_{O_d}^X$, for which approximate determinations can be obtained if the perturbative series is known to high enough orders. It is also worth mentioning that the dependence on $Z_{O_d}^X$ of the hyperasymptotic approximation to the Borel sum is minimal, since it only appears in $\Omega$. Finally note that there is no need of introducing an infrared cutoff $\nu_f$. 

We plan to apply these methods to general observables, but before we want to study the methods in test-objects for which the approximations are under control. In this paper we take the static potential in the large $\beta_0$ approximation, regulating the asymptotic perturbative expansion using the PV prescription, as the observable. It has nice properties: A lot of analytic control is known for it, its Borel transform is known exactly, and it does not have ultraviolet renormalons. 
In this case we know what the genuine NP corrections are. They are zero by construction.

Whereas the general expressions we give in this paper are valid for any scheme, for the specific analysis worked out in this paper (the static potential in the large $\beta_0$ approximation), we use two different schemes: the lattice and the $\MS$ schemes. In the large $\beta_0$ this is equivalent to a redefinition of the renormalization scale. Nevertheless, let us stress that it corresponds to a rather large change in the scale. Different values of $c$ (see \eq{eq:NP}) can also be understood as a change in the renormalization scale. The result is independent on the scheme and factorization scale used for the $\al$ (within the error of the computation). The scheme/scale dependence is a higher order effect. The important thing is that both schemes converge. This does not mean that all schemes converge equally fast. We observe that $\MS$ appears to be more convenient for method 2B). It is also interesting to see the dependence of the observables/methods with $n_f$. Indeed we observe that the range of validity of the hyperasymptotic expansion is sensitive to the value of $n_f$. Changing from $n_f=0$ to $n_f=3$ significantly enlarges the range of validity of the OPE. This is a relevant discussion when trying to determine up to which scale one can apply perturbation theory and the OPE. Concerning how well method 1) and 2B) perform in practice for this observable, we find that both methods converge to the expected result. Method 2B) is not particularly precise though. Method 1) appears to converge faster (besides being systematically improvable).  Finally, and specific to method 2B), one issue that we address is how large the renormalization scale $\mu$ has to be such that the perturbative expansion simulates well the truncated integral in \eq{SAST}. For the case of the static potential in the large $\beta_0$ approximation, we observe that we have to go to relatively high scales. This makes this method not very useful. 

The application of these analyses to QCD observables (beyond the large $\beta_0$ approximation) and the incorporation of ultraviolet renormalons (if necessary) is left to forthcoming papers.

\medskip
 
{\bf Acknowledgments}\\
\noindent
  C.A. thanks the IFAE group at Universitat Aut\`onoma de
Barcelona for warm hospitality during part of this work.
This work was supported in part by the Spanish grants FPA2017-86989-P and SEV-2016-0588 from the ministerio de Ciencia, Innovaci\'on y Universidades, and the grant 2017SGR1069 from the Generalitat de Catalunya; and by the Chileans FONDECYT Postdoctoral Grant No. 3170116, and by FONDECYT Regular Grant No. 1180344.

\begin{appendix}
\appendix
\section{$D_b(-x)$}
\label{Sec:Db}
We define
\be
D_b(-x) \equiv x \int_{0,\rm PV}^{\infty} du e^{-ux}\frac{1}{(1-u)^{1+b}}=\frac{1}{\Gamma(b+1)}\int_{0,\rm PV}^{\infty} 
d \epsilon 
\epsilon^b\frac{1}{1-\frac{\epsilon}{x}}e^{-\epsilon}
\,,
\ee
where $x>0$. 
Note that this integral has a cut in the integration line starting at $u=1$. We have to define how we handle the singularity. 
We demand $D_b(-x)$ to be real for real and positive $x$. 
The first expression can be understood as the analytic continuation in $b$ of the second expression 
(which is first defined for arbitrary positive integer values), and in the second expression we use the PV prescription. Both expressions produce the same asymptotic expansions. Finally, we obtain the following expression
\be
\label{Db}
D_b(-x)=xe^{-x}(-x)^b[\Gamma(-b)-\Gamma(-b,-x)]-\cos(\pi b) \Gamma(-b)x^{1+b}e^{-x}
\,,
\ee
where ($\Gamma(b) \equiv \Gamma(b,0)$)
\be
\Gamma(b,x)=\int_x^{\infty}dt t^{b-1} e^{-t}
\ee
is the incomplete Gamma function. The second term in \eq{Db} is explicitly real, not so for the first term. 
Note that the last term in Eq. (\ref{Db}) is proportional to $\lQ$. From these expressions is difficult to take the $b \rightarrow 0$ limit. It is more convenient to set $b=0$ before computing.

$D_b(-x)$ is long known: $D_b(-x)=\bar \Lambda_b (-x)$, 
where $\bar \Lambda_b (-x)$ is defined in \cite{Dingle}. Variants of that formula read (originally generated with 
$a >0$)
\be
\int_{0}^{\infty} d y e^{-y x} \frac{1}{\left(1+\frac{y}{a}\right)^{1+b}}
=
\frac{a}{\Gamma(b+1)} x^b \int_0^{\infty} dy y^b  e^{-y x} \frac{1}{\left(1+\frac{y}{a}\right)}
\,,
\ee
\be
\int_{0,\rm PV}^{\infty} d \epsilon e^{-\epsilon} \frac{(\alpha \epsilon)^N}{1-\alpha \epsilon}=\alpha^{N-1}\Gamma(N+b+a)
\int_{0,\rm PV}^{\infty} d \epsilon e^{-\epsilon/\alpha} \frac{1}{(1- \epsilon)^{(1+b+N)}}
\,.
\ee 

\end{appendix}

\end{document}